\documentclass[11pt]{article}

\usepackage{geometry} 
\usepackage{graphicx}
\usepackage{amssymb}

\begin{document}

\thispagestyle{empty}

\rightline{JLAB-THY-05-437}
\rightline{RM3-TH/05-10}

\vspace{2cm}

\begin{center}

\LARGE{\bf Leading twist moments of the neutron structure function $F_2^n$\\}

\vspace{2cm}

\Large{M.~Osipenko$^{1,2}$, W.~Melnitchouk$^{3}$, S.~Simula$^{4}$, 
       S.~Kulagin$^{5}$, G.~Ricco$^{1,6}$}

\normalsize

\vspace{1cm}

	$^1$Istituto Nazionale di Fisica Nucleare, Sezione di Genova,
	16146 Genova, Italy \\ [3pt]
$^2$Skobeltsyn Institute of Nuclear Physics, 119992 Moscow, Russia \\ [3pt]
$^3$Jefferson Lab, Newport News, Virginia 23606, USA \\ [3pt]
$^4$Istituto Nazionale di Fisica Nucleare, Sezione Roma III,
	00146 Roma, Italy \\ [3pt]
$^5$Institute for Nuclear Research of Russian Academy of Science,
	117312 Moscow, Russia \\ [3pt]
$^6$Dipartimento di Fisica dell'Universit\`a di Genova, 16146 Genova, Italy \\

\vspace{2cm}

{\bf Abstract} 

\end{center} 

We perform a global analysis of neutron $F_2^n$ structure function data,
obtained by combining proton and deuteron measurements over a large
range of kinematics. From these data the lowest moments ($n \leq 10$) of 
the leading twist neutron $F_2^n$ structure function are extracted.
Particular attention is paid to nuclear effects in the deuteron,   
which become increasingly important for the higher moments. Our results 
for the nonsinglet, isovector $p - n$ combination of the leading twist 
moments are used to test recent lattice simulations.
We also determine the lowest few moments of the higher twist
contributions, and find these to be approximately isospin independent, 
suggesting the possible dominance of $ud$ correlations over $uu$ and 
$dd$ in the nucleon.

\newpage

\section{\label{sec:intro} Introduction}

Inclusive lepton scattering experiments have for some time been the
standard tool with which to study the internal quark-gluon structure 
of nucleons and nuclei.
In unpolarized scattering, this structure is represented through two
structure functions, $F_1$ and $F_2$, which parameterize the differential
cross section,  
\begin{eqnarray}
\frac{ d^2\sigma }{ d\Omega dE' }
&=& \sigma_{\rm Mott}
\left[ \frac{ 2 F_1(x,Q^2) }{ M } \tan^2\frac{ \theta }{ 2 }
       + \frac{ F_2(x,Q^2) }{ \nu }
\right]\ ,
\end{eqnarray}
where $\sigma_{\rm Mott} = (4\alpha^2 E'^2/Q^4) \cos^2\frac{\theta}{2}$
is the Mott cross section for the scattering from point particles,
with $\Omega$ the scattered lepton solid angle ($\theta$ is the polar
scattering angle), and $E'$ the recoil lepton energy.
Here $\nu$ is the energy transfer and $M$ is the nucleon mass.
The structure functions $F_{1,2}$ are functions of the four-momentum
transfer squared $q^2 \equiv -Q^2$, and the Bjorken scaling variable
$x = Q^2/2M\nu$.
In the Bjorken limit ($Q^2, \nu \to \infty$, $x$ fixed), the structure
functions can be related to parton distribution functions, which give the
light-cone momentum distributions of quarks (and gluons, or partons in
general) in the hadron.
At leading order, one has
$F_2 = 2xF_1
     = x \sum_i e_i^2 \left( q_i(x,Q^2) + \bar q_i(x,Q^2) \right)$.
At finite $Q^2$ radiative corrections and higher twist effects give 
rise to a $Q^2$ dependence in the structure functions, leading to 
scaling violations.

The connection between the partonic structure of hadrons and QCD can
be most readily made through {\em moments}, or $x$-weighted integrals,
of structure functions, which through the operator product expansion
(OPE) can be related to hadronic matrix elements of local operators.
In the case of the $F_2$ structure function, to which we restrict
ourselves in the following, the moment is defined by
\begin{eqnarray}
M_n(Q^2) &=& \int_0^1 dx\ x^{n-2}\ F_2(x,Q^2)\ .
\label{eq:momdef_CN}
\end{eqnarray}
While in the Bjorken limit the moments are given by real numbers, 
independent of $Q^2$, at finite $Q^2$ the renormalization group 
equations lead to a specific $Q^2$ dependence of the moments that can
be calculated from perturbative QCD.
In fact, observation of this $Q^2$ dependence was instrumental in
establishing QCD as the theory of the strong nuclear interactions.
The moments can also be calculated nonperturbatively using lattice QCD 
techniques.

A wealth of information has been accumulated over several decades on  
the structure of the proton, deuteron, as well as heavier nuclei.
To completely describe the structure of the nucleon, however, requires
knowledge of both the proton {\em and} neutron structure functions.
Typically, the neutron structure function is obtained by combining
data on the deuteron and proton.
In most analyses the deuteron structure function is assumed to be
given by a simple sum of free proton and neutron structure functions,
$F_2^D \to F_2^p + F_2^n$, which is a good approximation for some
regions of kinematics ($0.1 \lesssim x \lesssim 0.4$).
However, it is well known that at small $x$ ($x \lesssim 0.1$) nuclear 
shadowing reduces the deuteron structure function relative to the free
nucleon \cite{D_SHAD,PILLER,MTshad,KPW94}, while at large $x$
($x \gtrsim 0.4$) nuclear binding and Fermi motion corrections are 
expected to give rise to a small ``nuclear EMC effect'' in the deuteron.

In terms of moments, any nuclear effect at large $x$ will manifest
itself in high moments (large $n$). It is crucial, therefore, that 
a moment analysis of the neutron structure function includes a careful 
treatment of nuclear corrections. In this paper we for the first time 
perform such an analysis, utilizing the recent high-precision data from 
Jefferson Lab together with earlier data 
on $F_2^p$ and $F_2^D$ (see Refs.~\cite{osipenko_f2p,osipenko_f2d} 
and references therein).

The earlier world data on the $F_2$ structure function cover a wide
range of kinematics, from $Q^2 \sim 0$ to $\sim 100$~(GeV/c)$^2$.
However, the large-$x$ domain has been almost completely unexplored,
due to either experimental limitations or intepretation difficulties.
New, high precision data from Jefferson Lab have allowed one to fill
this gap which has in the past prevented a complete moment based
analysis of data.
Jefferson Lab data cover the entire resonance region starting from the
elastic peak (quasi-elastic for the deuteron) and extending to invariant
masses of the produced hadronic system of about 3.5~(GeV/c).
This region is of particular importance for higher moments.
In fact, already for the $n = 4$ moment the contribution of this kinematic
region is significant for $Q^2 < 5$~(GeV/c)$^2$.
As a consequence for $Q^2 < 5$~(GeV/c)$^2$ and $n>2$ the uncertainties
come mostly from Jefferson Lab data.

The lowest moment ($n = 2$) is also sensitive to the small-$x$ region,
where much data, in particular for $Q^2<5$~(GeV/c)$^2$, have been
accumulated in previous experiments.
For the proton $F_2$ structure function the lowest $x$ were achieved
with the HERA collider data which reached down to $x \sim 10^{-5}$.
For the deuteron $F_2$ structure function, the minimum $x$ values are
somewhat larger, $x \sim 10^{-4}$, and were obtained with muon beams in
fixed target experiments at CERN and Fermilab.
The extrapolation to $x = 0$ has been carefully studied in
Refs.~\cite{osipenko_f2p,osipenko_f2d} and the corresponding systematic 
uncertainties have been estimated.
At $Q^2$ values above those currently accessible at Jefferson Lab
($Q^2 > 5$~(GeV/c)$^2$), the large-$x$ extrapolation was also studied
by a comparison of various models and the corresponding systematic
uncertainties were included in the moment analyses.

Unlike previous studies, which have focussed on the $x$ dependence,
and have ignored or underestimated the nuclear corrections, we extract 
the lowest few moments of the neutron $F_2^n$ taking into account the 
systematic uncertainties associated with our knowledge of nuclear effects 
in the deuteron.
Furthermore, this analysis directly uses the leading twist parts of the
moments extracted from the $F_2^p$ and $F_2^D$ data, which allows a more
straightforward implementation of the impulse approximation.

The general framework for analyzing the data is presented in 
Sec.~\ref{sec:moms}, where we discuss various models of nucleon 
light-cone momentum distributions in the deuteron.
The extracted leading twist neutron moments are presented in
Sec.~\ref{sec:Mn}, where we also discuss corrections to the simple
one-dimensional convolution formalism, in the form of nucleon off-shell
effects, as well as nuclear shadowing and meson exchange currents.
The results for the extracted moments are compared with those obtained
from several global parton distribution functions. Moreover we construct the 
nonsinglet, isovector $p - n$ combination of the leading twist moments and 
compare our results with available lattice ones.
As an interesting application of our extraction, in
Sec.~\ref{sec:ht_isospin} we identify the isospin dependence of the
higher twist contributions to the moments, and speculate about possible
interpretations of our findings within a diquark picture.
Finally, in Sec.~\ref{sec:con} we summarize our results and briefly
outline some future directions.


\section{\label{sec:moms}Neutron structure and nuclear effects in the deuteron}

In the absence of free neutron targets, information on the structure
function of the neutron is typically extracted from measurements using
deuterons, together with knowledge of the corresponding proton structure 
function.
Since the deuteron is a nucleus, its nucleon constituents are bound,
albeit weakly, and in general the structure of the bound nucleons will 
differ from the free nucleon structure.
In this section we describe the various nuclear corrections which
enter in the extraction of the free neutron $F_2$ structure function,
and in particular its moments, from the structure function of the
deuteron.

We begin with the standard nuclear impulse approximation, in which
the virtual photon scatters incoherently from individual nucleons in
the deuteron, and discuss several implementations of the convolution
formula.
Following this we describe corrections to convolution, associated with
nucleon off-shell effects, nuclear shadowing, and meson exchange currents.

\subsection{Impulse approximation}
\label{ssec:ia}

For moderate to large values of $x$ ($x \gtrsim 0.1$) inclusive scattering
off a deuteron can be described, using the optical theorem, in terms of 
the amplitudes $\gamma^* N^* \to \gamma^* N^*$ and $N^* d \to N^* d$, 
where $N^*$ refers to the fact that the nucleon is off its mass or
energy shell.
In this approximation interactions between the intermediate states of
the $\gamma^* N^*$ and $N^* d$ subprocesses are neglected.
Formally, the deuteron $F_2$ structure function can then be written as
\begin{eqnarray}
F_2^D(x,Q^2)
&=& \int d^4p\
{\rm Tr}\left[ \widehat{\cal W}(p,q) \cdot \widehat{\cal A}(p,p_D)
	\right]\ ,
\label{eq:F2d_gen}
\end{eqnarray}
where $p$, $p_D$ and $q$ are the virtual nucleon, deuteron and virtual
photon four-momenta, respectively, the Dirac matrix
$\widehat{\cal W}(p,q)
= I\ {\cal W}_0\ + \not\!p\ {\cal W}_1\ + \not\!q\ {\cal W}_2$
is the truncated (off-shell) nucleon structure function, and
$\widehat{\cal A}(p,p_D)
= I {\cal A}_0 + \gamma_\alpha {\cal A}_1^\alpha$
is the relativistic deuteron spectral function.
The amplitudes ${\cal A}_0$ and ${\cal A}_1^\alpha$ depend
on the deuteron wave functions.

For an on-shell nucleon, the $F_2$ structure function is proportional
to $M\ {\cal W}_0 + M^2\ {\cal W}_1 + p\cdot q\ {\cal W}_2$.
In general the presence of several Dirac structures and corresponding
``structure functions'' ${\cal W}_i$ in the trace in the integrand in 
Eq.~(\ref{eq:F2d_gen}) complicates the relation between the deuteron and 
nucleon structure functions.
There are various ways which the formal expression in
Eq.~(\ref{eq:F2d_gen}) can be simplified, however, as we now discuss.

\subsubsection{Nonrelativistic expansion}
\label{sssec:nr}

The formal expression for the deuteron structure function $F_2^D$ in
Eq.~(\ref{eq:F2d_gen}) is Lorentz-invariant and can be evaluated in
any reference frame.
Without loss of generality, however, one can choose to work in the
deuteron rest frame, and take the $z$-axis such that $q_z=-|\bf{q}|$.
Performing a nonrelativistic reduction of the Dirac traces in
Eq.~(\ref{eq:F2d_gen}), one can show that to order ${\bf p}^2/M^2$ in
the Bjorken limit the product of the Dirac structures reduces to a 
single structure proportional to a generalized (off-shell) nucleon
structure function \cite{KPW94},
\begin{equation}
F_2^D(x,Q^2)
= \int\frac{d^3{\bf p}}{(2\pi)^3}
  \left(1+\frac{p_z}{M}\right)
  \left|\Psi_D({\bf p})\right|^2
  F_2^N\left(\frac{x}{z},Q^2,p^2\right)\ .
\label{eq:F2D}
\end{equation}
Here $F_2^N=\frac12(F_2^p+F_2^n)$ is the isoscalar nucleon structure
function, $z = (p_0 + p_z)/M$ is the light-cone momentum fraction of 
the deuteron carried by the interacting nucleon, where $p_0 = M + 
\varepsilon_D - {\bf p}^2/(2M)$ is the energy of the struck nucleon,
and $\varepsilon_D=-2.22$~MeV is the deuteron binding energy.
The deuteron wave function $\Psi_D$ contains the usual nonrelativistic
$S$- and $D$-wave components, $\psi_0$ and $\psi_2$,
\begin{eqnarray}
\frac{1}{2\pi^2}\ |\Psi_D({\bf p})|^2
&=& \psi_0^2(|{\bf p}|) + \psi_2^2(|{\bf p}|)\ ,
\label{eq:PsiD}
\end{eqnarray}
normalized such that 
\begin{equation}
\frac{1}{2 \pi^2} \int_0^\infty d|{\bf p}|\ {\bf p}^2 |\Psi_D({\bf p})|^2 = 1\ .
\label{eq:normalization}
\end{equation}
Note that in addition to the Bjorken variable and $Q^2$, the nucleon 
structure function in (\ref{eq:F2D}) also depends on the nucleon 
off-shellness $p^2 \equiv p_0^2 - {\bf p}^2$.
If one neglects the dependence of $F_2^N$ on $p^2$, then
Eq.~(\ref{eq:F2D}) leads to the simple one-dimensional convolution 
formula in terms of the free nucleon structure function \cite{WEST,JAFFE},
\begin{equation}
F_2^{D ({\rm conv})}(x,Q^2)
= \int_x^{M_D/M} dz\ f^D_{\rm nr}(z)\ F_2^N\left(\frac{x}{z},Q^2\right)\ ,
\label{eq:conv}
\end{equation}
where $M_D$ is the deuteron mass, and $f^D_{\rm nr}(z)$ is the
nonrelativistic nucleon (light-cone) momentum distribution in the
deuteron,
\begin{eqnarray}
f^D_{\rm nr}(z)
&=& \int\frac{d^3{\bf p}}{(2\pi)^3}
    \left(1+\frac{p_z}{M}\right)
    \left|\Psi_D({\bf p})\right|^2
    \delta\left( z - \frac{p_0+p_z}{M}\right)\ .
\label{eq:fDnr}
\end{eqnarray}
Physically the function $f^D_{\rm nr}(z)$ describes the Fermi motion
and binding of nucleons in the deuteron.
Note that some implementations of the convolution formula (\ref{eq:conv})
do not include the ``flux factor'' $(1+p_z/M)$ \cite{FLUX}.
However, since this appears at ${\cal O}({\bf p}/M)$, consistency to
order ${\bf p}^2/M^2$ requires its inclusion \cite{KPW94}.
The wave function normalization (\ref{eq:normalization}) ensures that
\begin{eqnarray}
\int_0^{M_D/M} dz\ f^D_{\rm nr}(z) &=& 1\ ,
\label{eq:fDnorm}
\end{eqnarray}
which guarantees that (for the valence component) the convolution term 
alone preserves the baryon number of the deuteron.

\subsubsection{Relativistic effects}
\label{sssec:rel}

Because the deuteron is a weakly bound nucleus, in most calculations
of deuteron structure the nonrelativistic approximation is adequate.
In the large-$x$ region, however, one is more sensitive to the 
large-momentum components of the deuteron wave function.
In this case relativistic effects are expected to become more important.

Since the distribution function $f^D_{\rm nr}$ in Eq.~(\ref{eq:fDnr})
was obtained in the nonrelativistic approximation, it applies in the 
region $|z-1| \ll 1$.
Outside of this region relativistic corrections may be important.
In order to test the sensitivity to these effects we also consider a 
relativistic version of the distribution in Eq.~(\ref{eq:fDnr}) which
is calculated with the same nonrelativistic wave function but using
relativistic kinematics, namely $p_0 = M_D - E_p$, where
$E_p = \sqrt{M^2 + {\bf p}^2}$ is the energy of the spectator nucleon.
Futhermore, one also replaces the ratio $p_z/M$ in the flux factor with
$p_z/E_p$.

While most deuteron wave functions calculated to date have been 
nonrelativistic, there have been some models which have treated the 
two-nucleon system relativistically \cite{TJON,BG79,GVH92}.
In order to consistently utilize the relativistic wave functions,
one needs to work with the full expression for the structure function
in Eq.~(\ref{eq:F2d_gen}) directly.
This necessarily involves modeling the off-shell nucleon structure
functions, which introduces additional model dependence in the
calculation \cite{MST94}.

One can show, however, that the exact expression in
Eq.~(\ref{eq:F2d_gen}) can be reduced to an on-shell convolution part, 
similar to that in Eq.~(\ref{eq:conv}), plus nuclear off-shell and 
relativistic corrections \cite{MSTplb,KRAK},
\begin{eqnarray}
F_2^D(x,Q^2)
&=& F_2^{D ({\rm conv})}(x,Q^2) + \delta^{\rm (off)} F_2^D(x,Q^2)\ .
\label{eq:F2Drel}
\end{eqnarray}
The convolution component here now involves a relativistic nucleon 
light-cone distribution function \cite{MSTplb},
\begin{equation}
f^D_{\rm rel}(z)
= z M_D \int_{-\infty}^{p_{\rm max}} \frac{dp^2}{16\pi^2}\
  \frac{E_p}{p_0}\ |\widetilde{\Psi}_D({\bf p})|^2\ ,
\label{eq:fDrel}
\end{equation}
where $E_p = \sqrt{M^2+{\bf p}^2} = (M^2 - p^2 + M_D^2)/2 M_D$
is the energy of the (on-shell) nucleon spectator, and the upper limit
of the nucleon virtuality $p^2$ in the integration is given by
$p^2_{\rm max} = z M_D^2/2 - z M^2/(2-z)$.
%
%
In addition to the usual $S$- and $D$-state wave functions $u$ and $w$,
the wave function $\widetilde{\Psi}_D({\bf p})$ also includes the
relativistic $P$-states \cite{BG79,GVH92}
\begin{eqnarray}
\frac{1}{2\pi^2}\ |\widetilde{\Psi}_D({\bf p})|^2
&=& \psi_0^2(|{\bf p}|)\ +\ \psi_2^2(|{\bf p}|)\
 +\ \psi_{1t}^2(|{\bf p}|)\ +\ \psi_{1s}^2(|{\bf p}|)\ ,
\end{eqnarray}
where $\psi_{1s}$ and $\psi_{1t}$ correspond to the singlet and triplet
$P$-states, respectively.
Note that the function $f^D_{\rm rel}$ also includes the relativistic
flux factor, $(E_p/p_0)\ z$ \cite{FLUX}, and therefore satisfies the
same normalization condition as in Eq.~(\ref{eq:fDnorm}), 
$\int dz f^D_{\rm rel} = 1$.
One should point out, however, that the separation of $F_2^D$ into the 
two terms in Eq.~(\ref{eq:F2Drel}) is not unique (the sum, of course,
is unique).
However, the choice (\ref{eq:fDrel}) is the most natural one, and
corresponds to that which has been followed in many phenomenological
treatments.

\subsubsection{Light-cone formulation}
\label{sssec:lc}

An alternative to the covariant (instant-form) formulations of
scattering from the deuteron is the Hamiltonian light-cone (LC) approach.
Here the intermediate state nucleons are on-mass-shell, but
off-energy-shell, so that one avoids the explicit $p^2$ dependence
in the bound nucleon structure function in Eq.~(\ref{eq:F2D}).

Because the nucleons are on-mass-shell, the derivation of the
one-dimensional convolution from Eq.~(\ref{eq:F2d_gen}) is trivial
since each of the three terms in the trace factor become proportional
\cite{MST94}.
The LC nucleon momentum distribution in the deuteron is given by 
\cite{FS81}
\begin{equation}
f^D_{\rm LC}(z)
= \int_{|{\bf p}|_{\rm min}} d|{\bf p}|\ |{\bf p}|\
  \sqrt{M^2+{\bf p}^2}\ |\Psi_D^{\rm LC}({\bf p})|^2\ ,
\label{eq:fDlc}
\end{equation}
where $|{\bf p}|_{\rm min}= M |z-1| / \sqrt{z (2-z)}$,
and $\Psi_D^{\rm LC}$ is the LC deuteron wave function.
Since most of the two-nucleon phenomenology has traditionally been 
formulated within the instant-form approach, in practice for
$\Psi_D^{\rm LC}$ one has usually employed the $S$- and $D$-state wave
functions as calculated in nonrelativistic potential models 
[see Eq.~(\ref{eq:PsiD})].

Note that in the limit where the nucleons are free (i.e. not bound),
the deuteron can be described in terms of an $NN$ Fock state component.
Inclusion of binding requires the addition of $NN+$meson Fock states.
Technically, binding or off-shell effects in the LC approach can be 
included by constructing a wave function which explicitly depends on an 
additional light-like four-vector related to the orientation of the 
null-plane \cite{KARMANOV}.
This introduces a number of additional wave functions, which would need
to be included in $f^D_{\rm LC}$.
Recently calculations of deuteron LC wave functions have been performed
within this framework, and some of the additional wave functions found
to be important at large momentum.
While it is interesting to explore the phenomenology using these wave
functions, in the present study we shall compare the results of the LC
distribution $f^D_{\rm LC}$ using the nonrelativistic $S$- and $D$-state 
wave functions as used in the literature.

In the next section we compare the various light-cone momentum
distributions discussed above, for different model deuteron wave 
function models.

\subsection{Nucleon momentum distributions}
\label{ssec:fz}

Within the convolution approximation there are two sources of nuclear
model dependence in the calculation of the deuteron structure function.
Firstly, as discussed above, different approaches are used to
formulate scattering from the two-nucleon system -- nonrelativistic,
relativistic, or light-cone.
In addition, within each approach there is dependence on the deuteron 
wave function or $NN$ potential.
It is beyond the scope of this work to critically analyze the merits
of the various models; instead we survey a reasonably large sample of
models and estimate the systematic uncertainty from the spread in their
predictions.

\begin{figure}[!ht]
\begin{center}
\includegraphics[bb=0cm 5cm 20cm 25cm, scale=0.365]{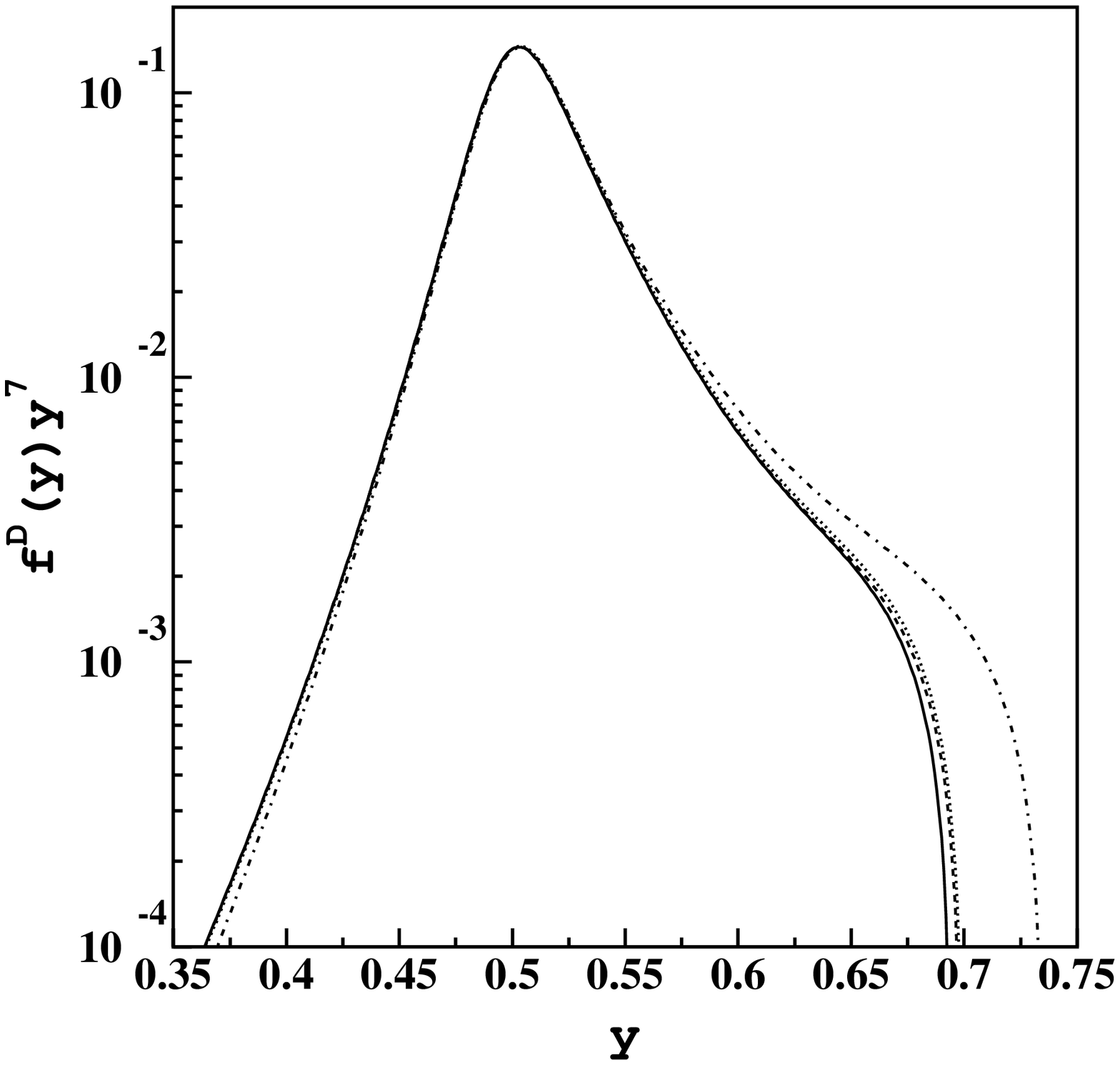}~%
\includegraphics[bb=0cm 5cm 20cm 25cm, scale=0.365]{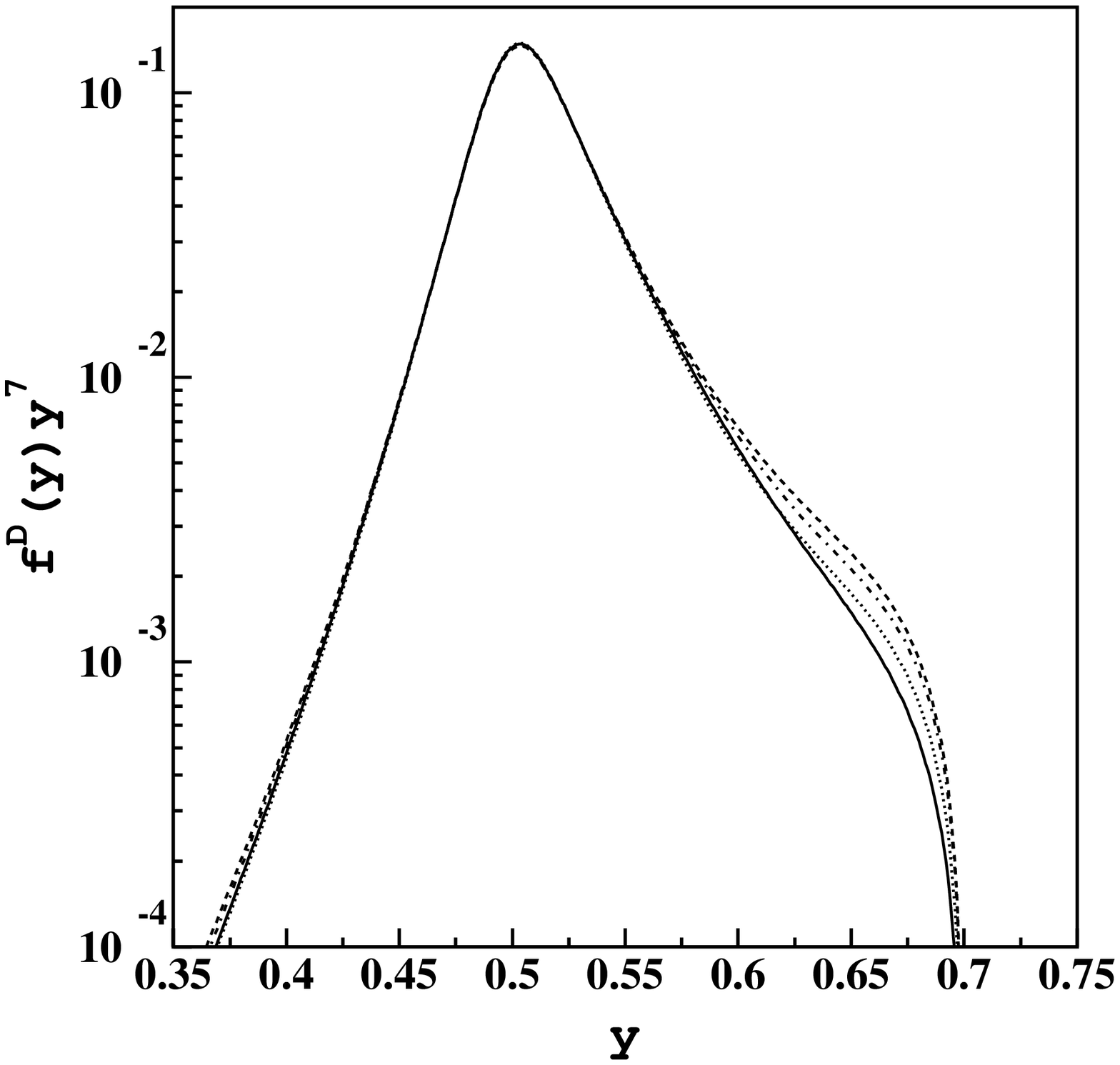}
\caption{\label{fig:wfs_integrands} \it \small
	Nucleon light-cone momentum distribution $f^D$ weighted
	by $y^7$, where $y=z/2$, for various models:
    (a) nonrelativistic distribution (\ref{eq:fDnr}) (solid),
	distribution (\ref{eq:fDnr}) with relativistic kinematics
	(dashed), relativistic model (\ref{eq:fDrel}) (dotted),
	light-cone distribution (\ref{eq:fDlc}), all using the
	Paris wave function \cite{PARIS};
    (b) relativistic distribution (\ref{eq:fDrel}) using the
	Bonn \cite{BONN} (solid), Paris \cite{PARIS} (dashed), and
	relativistic Gross wave functions from Refs.~\cite{BG79}
	(dotted) and \cite{GVH92} (dot-dashed).}
\end{center}
\end{figure}

In practice the deuteron wave functions are determined by fitting $NN$
scattering data.
Since most of the data are at low momentum, the wave functions are
relatively well constrained at small $p$, but the large-$p$ tail of
the wave function is more model dependent.
In terms of the light-cone variable $z$, the large-$p$ region
corresponds to large values of $z$.
To emphasize the large-$z$ region of $f^D$ we plot in
Fig.~\ref{fig:wfs_integrands} the various nucleon momentum distributions,
for different wave functions, weighted by $y^7$, where $y \equiv z/2$.
Note that the momentum integration in each of the curves has been
limited to $|{\bf p}| = 0.5$~GeV (see below).
As expected the model dependence is relatively weak for moderate $y$,
and only becomes significant for $y \gtrsim 0.6$.

\begin{figure}[!ht]
\begin{center}
\includegraphics[bb=0cm 5cm 20cm 25cm, scale=0.365]{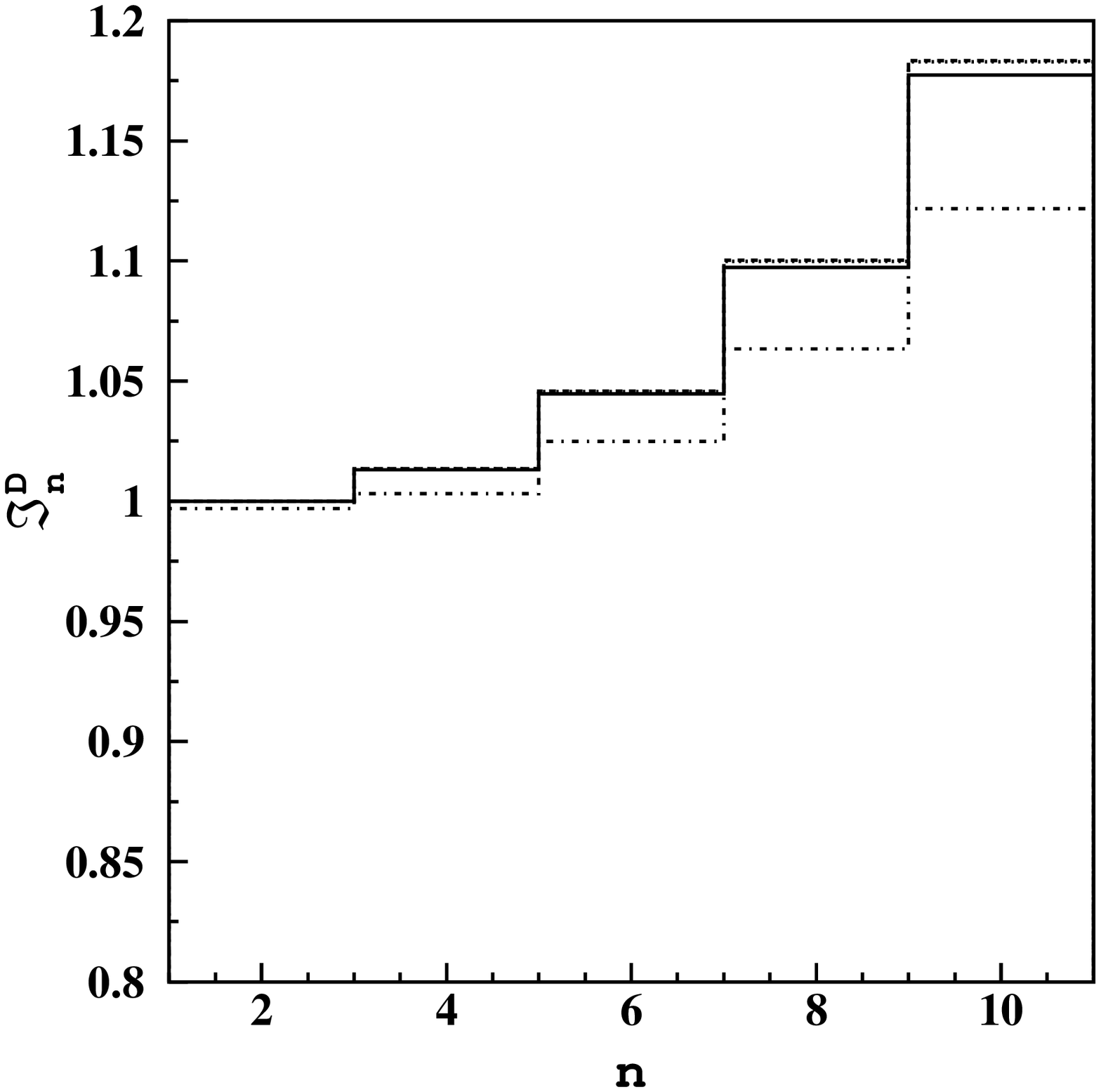}~%
\includegraphics[bb=0cm 5cm 20cm 25cm, scale=0.365]{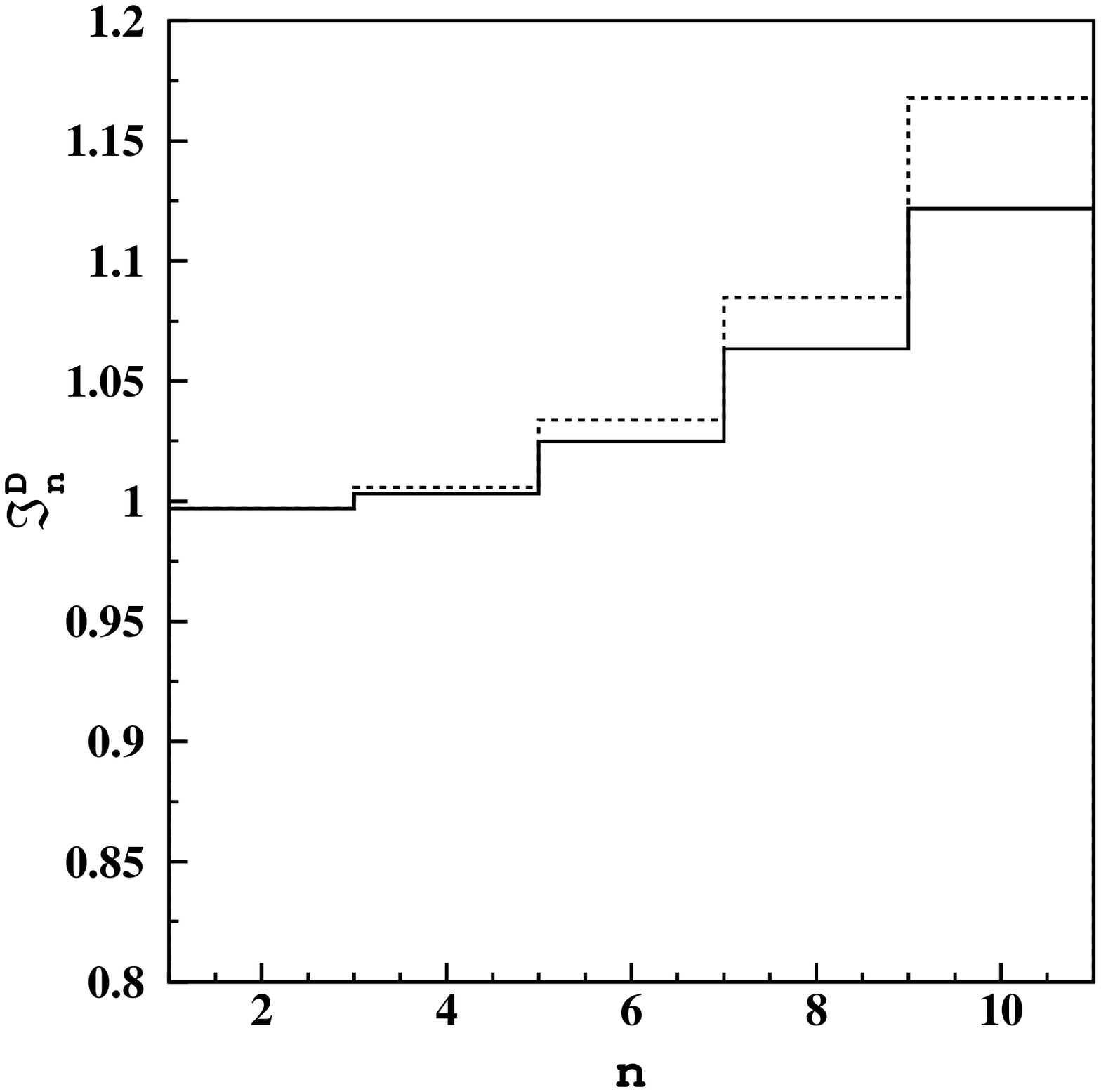}
\caption{\label{fig:wf_moments} \it \small
	Moments ${\cal F}_n^D$ ($n=2-10$) of the nucleon light-cone
	momentum  distribution $f^D(z)$ in Eq.~(\ref{eq:fDrel})
    (a) for different deuteron wave functions:
	BonnCD \cite{BONNCD} (solid),
	Nijmegen~93 \cite{NIJM} (dashed),
	Argonne $v_{18}$ \cite{AV18} (dotted),
	Gross \cite{BG79} (dot-dashed);
    (b) for the Gross wave function \cite{GVH92} with the momentum
	integration restricted to $p_{\rm max}=0.5$~GeV (solid) and
	1~GeV (dashed).}
\end{center}
\end{figure}

Uncertainties due to the nuclear models can also be quantified in terms
of moments.
This is actually more relevant for the present analysis since ultimately
we extract the moments of the neutron $F_2^n$ structure function.
If we define the $n$-th moment of the distribution $f^D(z)$ by
\begin{equation}
{\cal F}_n^D = \int_0^{M_D/M} dz\ z^{n-1}\ f^D(z)\ ,
\label{eq:fz_mom}
\end{equation}
then clearly the uncertainties in the large-$z$ part of the distribution
will be reflected in the uncertainties of the moments of $f^D$ at large-$n$,
which are illustrated in Fig.~\ref{fig:wf_moments}.

As mentioned above, the differences between the momentum distributions
at large $z$ arise because the deuteron wave function is not well 
constrained at large momentum.
In fact, $NN$ scattering data which have been used to constrain the
deuteron wave function typically extend only to $p \approx 0.4$~GeV.
At large momentum ($p \gtrsim 0.3-0.4$~GeV) the inter-nucleon distances
probed becomes smaller than the nucleon radius, so that here a
description in terms of nucleon degrees of freedom alone may be 
questionable.
To estimate the uncertainty arising from the large-$p$ tail of
the deuteron wave function, we examine the effect of truncating
the momentum integration at $p=p_{\rm max}$.
The moments ${\cal F}_n^D$ of the distribution $f^D(z)$ with
$p_{\rm max}=0.5$ and 1~GeV are plotted in Fig.~\ref{fig:wf_moments}(b), 
where the distributions have been renormalized to satisfy
Eq.~(\ref{eq:fDnorm}).
For low moments ($n=2, 4$) the effect is negligible, while for the
$n = 10$ moment the effect of the $p_{\rm max}=0.5$ (1)~GeV cuts is
6\% (1\%).
We estimate that the uncertainty introduced by our poor knowledge of
the large-$p$ components of the deuteron wave function is of the same
order of magnitude as the wave function model dependence.
The maximum deviation among the moments obtained with different deuteron
wave functions will be taken as an estimate of the systematic error
due to the wave function model dependence.

An independent test of the nucleon distribution functions may be made
by considering quasi-elastic scattering.
Within the impulse approximation and neglecting finite-$Q^2$ effects, 
the quasi-elastic contribution to the deuteron structure function, 
$F_2^{D(QE)}$, can be written in factorized form, in terms of the 
nucleon distribution function $f^D$ and the nucleon elastic form factors,
\begin{equation}
F_2^{D(QE)}(x_D,Q^2)
\to \frac{1}{2} x_D f^D(x_D)
  \Bigr\{ \frac{G_E^{p\mbox{ }2}(Q^2)+G_E^{n\mbox{ }2}(Q^2)
	+ \tau(G_M^{p\mbox{ }2}(Q^2)+G_M^{n\mbox{ }2}(Q^2))}
	{1+\tau}
  \Bigl\} ~~,
\label{eq:conv_qe}
\end{equation}
where $x_D \equiv Q^2 / 2M_D \nu \simeq x / 2$. Thus, one can define 
{\em phenomenological} nucleon distribution moments 
$\widetilde{\cal F}_n^D(Q^2)$ in terms of the moments $M_n^{D(QE)}(Q^2)$
of the deuteron quasi-elastic structure function $F_2^{D(QE)}$ 
and the nucleon form factors,
\begin{equation}
\label{eq:nqe}
\widetilde{\cal F}_n^D(Q^2)
= \frac{(1+\tau) M_n^{D(QE)}(Q^2)}
       {G_E^{p\mbox{ }2}(Q^2)+G_E^{n\mbox{ }2}(Q^2)
        + \tau(G_M^{p\mbox{ }2}(Q^2)+G_M^{n\mbox{ }2}(Q^2))}\ .
\end{equation}
We expect that $\widetilde{\cal F}_n^D(Q^2) \simeq {\cal F}_n^D$ at 
large enough $Q^2$. Note that for the moment $M_n^{D(QE)}(Q^2)$ we use 
the Nachtmann definition \cite{Nachtmann} to eliminate the finite-$Q^2$ 
effects due to the nonzero target mass (i.e.~the deuteron mass),
\begin{equation}
M_n^{D(QE)}(Q^2)
\to \int_0^1 dx_D \frac{\xi_D^{n+1}}{x_D^3} F_2^{D(QE)}(x_D,Q^2)
		\frac{3+3(n+1)r_D+n(n+2)r_D^2}{(n+2)(n+3)}\ ,
\end{equation}
where $r_D = \sqrt{1 + 4M_D^2x_D^2/Q^2}$ and $\xi_D = 2x_D/(1+r_D)$.

\begin{figure}[!ht]
\begin{center}
\includegraphics[bb=0cm 5cm 20cm 25cm, scale=0.365]{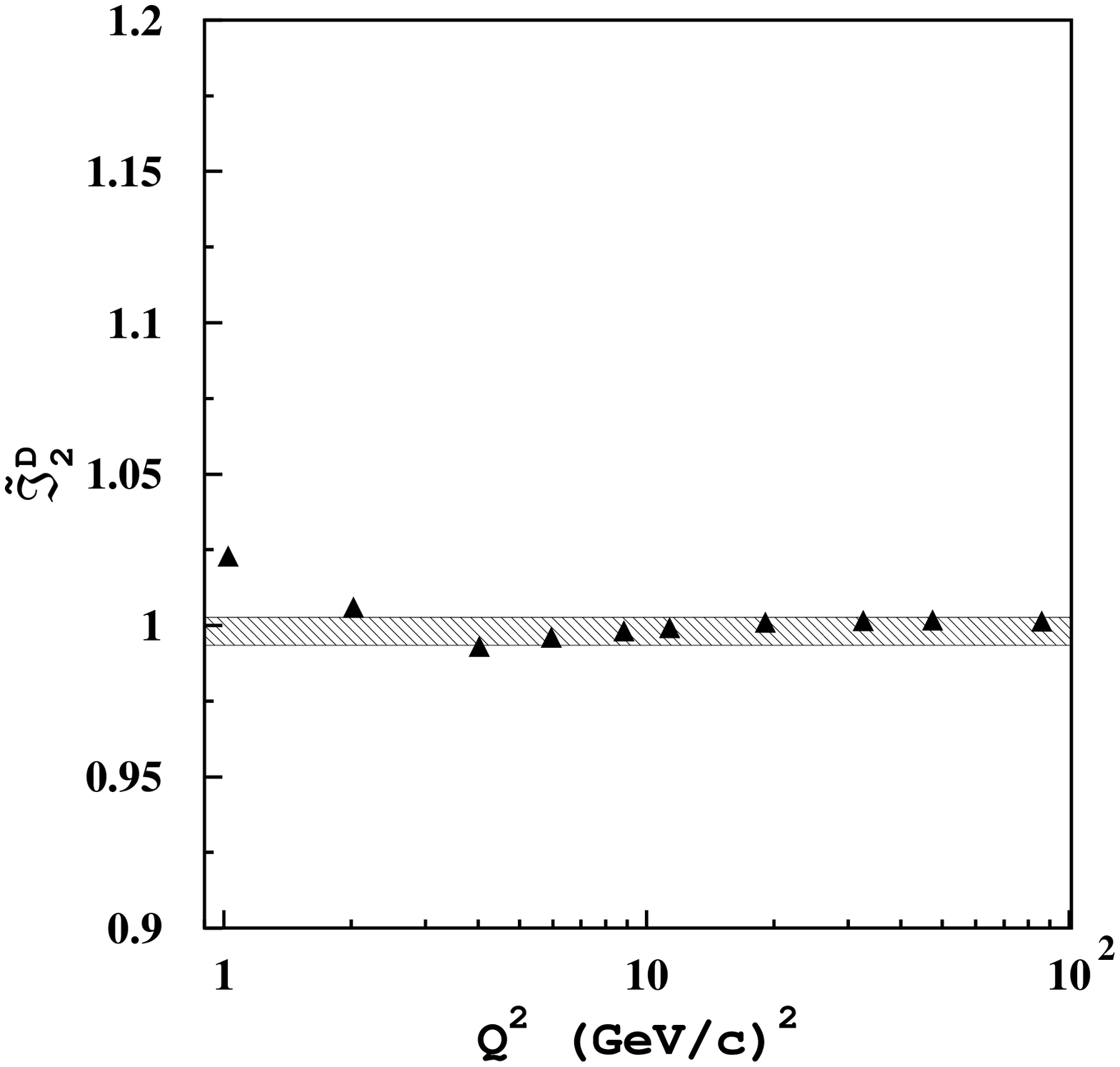}~%
\includegraphics[bb=0cm 5cm 20cm 25cm, scale=0.365]{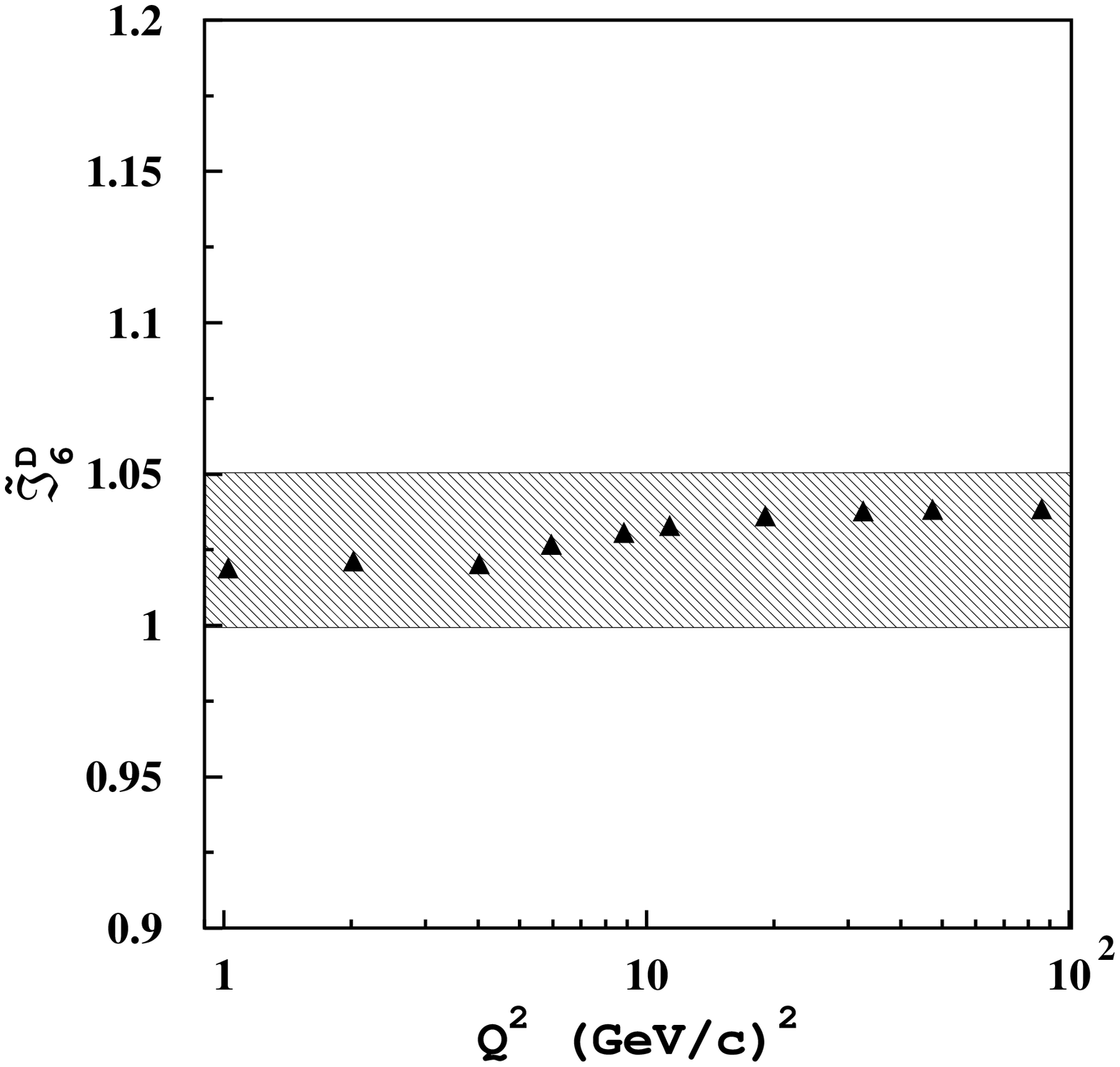}
\caption{\label{fig:neutron_moments_qe} \it \small
	Moments $\widetilde{\cal F}_2^D(Q^2)$ (a) and $\widetilde{\cal F}_6^D(Q^2)$ (b)
	extracted from the quasi-elastic peak moments defined in Eq.~(\ref{eq:nqe})
	(triangles). The hatched areas represent the corresponding moments 
	${\cal F}_2^D$ and ${\cal F}_6^D$ calculated from Eq.~(\ref{eq:fz_mom}), 
	including the nuclear model uncertainties shown in Fig.~\ref{fig:wf_moments}.}
\end{center}
\end{figure}

The deuteron quasi-elastic contribution to the structure function $F_2^{D(QE)}$ 
has been calculated using the approach developed in Ref.~\cite{Simula_qe}, 
which includes final state interactions and has been tested
against the recent deuteron experimental data of Ref.~\cite{osipenko_f2d}.
As an illustration, the extracted $\widetilde{\cal F}_2^D(Q^2)$ and 
$\widetilde{\cal F}_6^D(Q^2)$ moments are shown in
Fig.~\ref{fig:neutron_moments_qe} as a function of $Q^2$, and
compared with the corresponding moments ${\cal F}_2^D$ and ${\cal F}_6^D$ 
calculated from the distribution function $f^D(z)$ in Eq.~(\ref{eq:fz_mom}).
The results agree well within the nuclear model uncertainties.

\subsection{Off-shell effects}
\label{ssec:offshell}

The expressions for the deuteron structure function thus far have
been in the framework of the one-dimensional convolution formula,
in which nucleon off-shell, relativistic, and other effects beyond
the impulse approximation are neglected.
While this constitutes a reasonable approximation at moderate $x$,
or for low moments, at large (and very small) $x$ values these
effects will become more important.

Within the nonrelativistic approach discussed in Sec.~\ref{sssec:nr},
the fact that the structure of bound nucleons can differ from that
of free nucleons is incorporated in the dependence of the bound
nucleon structure function in Eq.~(\ref{eq:F2D}) on the nucleon
virtuality, or off-shellness, $p^2$.
Since the deuteron is treated as a nonrelativistic system, the 
off-shellness parameter $v \equiv (p^2 - M^2)/M^2$ is on average a 
small number.
Expanding the leading twist structure function in $v$ and keeping only
the first order correction in $v$, one finds
\begin{eqnarray}
\label{eq:SF_OS}
F_2^N(x,Q^2,p^2)
& = & F_2^N(x,Q^2)\left( 1+\delta f_2(x)\,v \right)\ ,\\
\label{eq:delf}
\delta f_2 & = & {\partial\ln F_2^N}/{\partial\ln p^2}\ ,
\end{eqnarray}
where the first term on the right in Eq.~(\ref{eq:SF_OS}) is the
structure function of the on-mass-shell nucleon, and the derivative
is evaluated at $p^2=M^2$.
The deuteron structure function $F_2^D$ can then be written as in
Eq.~(\ref{eq:F2Drel}), with
\begin{eqnarray}
\delta^{\rm (off)} F_2^D(x,Q^2)
&=& \int\frac{d^3{\bf p}}{(2\pi)^3}
    \left(1+\frac{p_z}{M}\right)\
    \left|\Psi_D(\bf{p})\right|^2\
    \delta f_2(x/z,Q^2)\ v\ F_2^N(x/z,Q^2)\ .
\label{eq:delF2D_nr}
\end{eqnarray}
The off-shell correction $\delta f_2$ was studied phenomenologically
in Ref.~\cite{KP04} by analyzing data on the ratios of structure
functions of different nuclei (EMC ratios).
It was found that to a good approximation $\delta f_2$ was independent
of $Q^2$, and could be parameterized as
\begin{equation}
\label{eq:delf_fit}
\delta f_2 = C_N (x - x_1)(x - x_0)(1 + x_0 - x)\ ,
\end{equation}
with the best fit parameters $C_N = 8.1 \pm 0.3_{\rm stat} \pm 0.5_{\rm syst}$, 
$x_1 = 0.05$, and $x_0 = 0.448 \pm 0.005_{\rm stat} \pm 0.007_{\rm syst}$.
Futhermore, the off-shell correction to the structure function $F_3$
is given by the same function $\delta f_2$, which is consistent with
the normalization of nuclear valence quark distribution
(for more details see Ref.~\cite{KP04}).

Expressing the off-shell correction to $F_2^D$ in terms of moments,
one can change the order of the ${\bf p}$ and $x$ integrations to obtain
\begin{equation}
\label{eq:MnD}
M_n^D(Q^2) = {\cal F}_n^D\ M_n^N(Q^2)\ +\ \delta C_n\ \delta M_n^N(Q^2)\ ,
\end{equation}
where the first term on the right hand side is the usual convolution
term and the second is the off-shell correction, with
\begin{equation}
\delta C_n
= \int\frac{d^3{\bf p}}{(2\pi)^3}
  \left(1+\frac{p_z}{M}\right)\
  z^{n-1}\ v\
  \left|\Psi_D(\bf{p})\right|^2\
\label{eq:delCn}
\end{equation}
and
\begin{equation}
\delta M_n^N(Q^2)
= \int_0^1 dx\, x^{n-2}\ \delta f_2(x)\ F_2^N(x,Q^2)\ .
\label{eq:delMnN}
\end{equation}
Writing the nucleon moment in terms of the proton and neutron moments
explicitly, $M_n^N = (M_n^p + M_n^n)/2$, one can solve
Eq.~(\ref{eq:MnD}) to express the neutron moment in terms of the
experimental proton and deuteron moments, and the theoretical
${\cal F}_n^D$ and off-shell corrections,
\begin{equation}
\label{eq:nmom_offshell}
M_n^n(Q^2)
= 2 M_n^D(Q^2)
  \left( \frac{ 1 - \Delta^{({\rm off})}_n }{ {\cal F}_n^D } \right)
- M_n^p(Q^2)\ ,
\end{equation}
where the off-shell ratio $\Delta^{({\rm off})}_n$ is given by
\begin{eqnarray}
\label{eq:delta_offshell}
\Delta^{(\rm off)}_n(Q^2)
&=& \frac{1}{M_n^D}
    \int dx\ x^{n-2}\ \delta^{\rm (off)}(x)\ F_2^D(x,Q^2)\ ,
\end{eqnarray}
and $\delta^{\rm (off)}$ is defined as
\begin{eqnarray}
\delta^{\rm (off)}(x,Q^2)
&=& \frac{ \delta^{\rm (off)} F_2^D(x,Q^2) }{ F_2^D(x,Q^2) }\ .
\end{eqnarray}
From Eqs.~(\ref{eq:delF2D_nr}), (\ref{eq:delCn}) and (\ref{eq:delMnN})
one can also write the moment correction as
\begin{eqnarray}
\Delta^{(\rm off)}_n(Q^2)
&=& \frac{ \delta C_n\ \delta M_n^N }{ 2 M_n^D }\ .
\end{eqnarray}

For the relativistic model of Sec.~\ref{sssec:rel}, the off-shell
correction $\delta^{\rm (off)} F_2^D$ receives two contributions:
one which depends on the relativistic $P$-state wave functions
$\psi_{1s}$ and $\psi_{1t}$, and one which is given by the off-shell
part of the truncated nucleon structure functions ${\cal W}_{0,1,2}$.
The former was found to be of the order 1\% and negative over the
whole $x$ range, while the latter was somewhat smaller in magnitude.
The total off-shell ratio $\delta^{\rm (off)}$ was parameterized
at $Q^2 \approx 5$~GeV$^2$ as \cite{MSTplb}
\begin{eqnarray}
\delta^{\rm (off)}(x,Q^2)
&=& a_0 (1 + a_1 x^{a_2}) \left( 1 - (a_3-x^{a_4})^{a_5} \right)\ ,
\label{eq:del_rel}
\end{eqnarray}
with the parameters $a_0 = -0.014$, $a_1 = 3.0$, $a_2 = 20.0$,
$a_3 = 1.067$, $a_4 = 1.5$, and $a_5 = 18.0$.
The $Q^2$ dependence of this ratio was found to be weak.
Note that the estimate of $\delta^{\rm (off)}$ in Ref.~\cite{MSTplb}
was based on the set of wave functions in Ref.~\cite{BG79} derived from 
an $NN$ potential with a pseudoscalar $\pi NN$ coupling, which are known 
to give larger $P$-state contributions than more modern treatments 
\cite{TJON,GVH92}.
The correction $\delta^{\rm (off)}$ in Eq.~(\ref{eq:del_rel}) should
therefore serve as an upper limit on the size of the off-shell 
correction in this model.

\begin{figure}[!ht]
\begin{center}
\includegraphics[bb=0cm 5cm 20cm 25cm, scale=0.365]{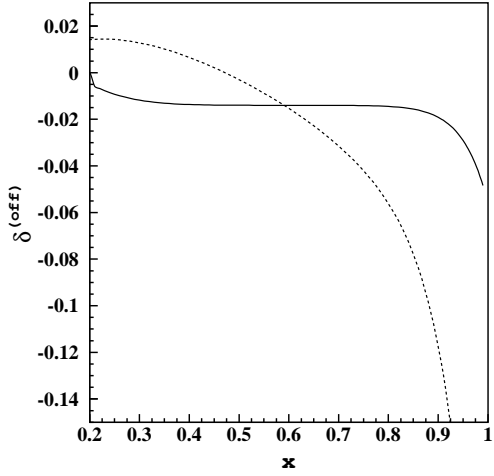}
\caption{\label{fig:delta_r} \it \small
	Nucleon off-shell correction $\delta^{(\rm off)}(x,Q^2)$ as a
	function of $x$ at $Q^2=10$~GeV$^2$, in the nonrelativistic
	\cite{KP04} (dashed) and relativistic \cite{MSTplb} (solid)
	models.}
\end{center}
\end{figure}

\begin{table*}[!ht]
\begin{center}
\caption{\label{table:delta_r} \it \small
	Moments of the non-convolution corrections to the deuteron
	$F_2^D$ structure function for $n=2-10$: the first two columns
	give off-shell corrections $\Delta^{(\rm off)}_n$ in the
	relativistic \cite{MSTplb} and nonrelativistic \cite{KP04}
	models, respectively; 
	the third and fourth columns represent the contributions from meson
	exchange currents and nuclear
	shadowing \cite{MTshad}, respectively.} \vspace{5mm}
\begin{tabular}{|c|c|c|c|c|} \hline
$n$ & $\Delta^{(\rm off)}_n$ \cite{MSTplb}
    & $\Delta^{(\rm off)}_n$ \cite{KP04}
    & $\Delta^{(\rm MEC)}_n$  \cite{MTshad}
    & $\Delta^{(\rm shad)}_n$ \cite{MTshad} \\ \hline
 2  & -5.67$\times 10^{-3}$ &  5.08$\times 10^{-3}$
    &  2.6   $\times 10^{-4}$ & -1.0   $\times 10^{-4}$ \\ \hline
 4  & -1.21$\times 10^{-2}$ & -3.34$\times 10^{-3}$
    & 7.6    $\times 10^{-6}$ &  7.6   $\times 10^{-6}$ \\ \hline
 6  & -1.39$\times 10^{-2}$ & -1.68$\times 10^{-2}$
    & 8.4    $\times 10^{-7}$ &  1.5   $\times 10^{-6}$ \\ \hline
 8  & -1.46$\times 10^{-2}$ & -3.01$\times 10^{-2}$
    & 1.6    $\times 10^{-7}$ &  4.1$\times 10^{-7}$ \\ \hline
10  & -1.53$\times 10^{-2}$ & -3.97$\times 10^{-2}$
    & 4.4    $\times 10^{-8}$ &  1.4$\times 10^{-7}$ \\ \hline
\end{tabular}
\end{center}
\end{table*}

The $x$ dependence of the corrections $\delta^{\rm (off)}$ in the
models of Refs.~\cite{MSTplb,KP04}, evaluated using the leading twist
parameterization \cite{osipenko_f2d} of $F_2^D$ at $Q^2 = 10$~GeV$^2$,
is displayed in Fig.~\ref{fig:delta_r}.
Both of the corrections are around 1--2\% for $x \lesssim 0.6-0.7$,
but increase significantly in magnitude for $x \gtrsim 0.8$.
The correction at large $x$ is negative in both cases, and is
especially dramatic for the model of Ref.~\cite{KP04}.
The resulting moments $\Delta^{(\rm off)}_n$ are listed in Table~\ref{table:delta_r} 
for $n=2-10$. The magnitude of the correction for the lowest moment is similar
in the two models, but is of opposite sign.
For higher $n$, the relative importance of the off-shell correction
increases.
Because of the steep decrease of $\delta^{\rm (off)}$ in the
nonrelativistic model \cite{KP04}, the magnitude of the correction
$\Delta^{(\rm off)}_n$ at $n = 10$ is about 2-3 times larger than the 
one in the model of Ref.~\cite{MSTplb}.

Note that part of the off-shell corrections in the model of
Ref.~\cite{MSTplb} are included in the relativistic distribution
function (\ref{eq:fDrel}), which in the model of Ref.~\cite{KP04}
are contained only in the non-convolution term.
Therefore the definition of off-shell corrections can differ between
various models, and one should ensure that off-shell effects in general
are calculated self-consistently.
The difference between the two models provides an
estimate of the systematic error associated with the poor knowledge
of the change in the nucleon structure when the nucleon is bound in
deuteron (see also Ref.~\cite{Cothran}).

\subsection{Small-$x$ effects: shadowing and meson exchange currents}
\label{ssec:shad}

For completeness, in this section we briefly discuss the nuclear
corrections in the deuteron at small values of $x$.
From the definition of the moments of the $F_2$ structure function,
it is clear that small-$x$ effects will mostly be relevant for the
lowest moment, and will be negligible for $n \geq 4$.

At small values of $x$, the coherence length of a virtual $q\bar q$
fluctuation of the photon probe in the laboratory frame is $\sim 1/Mx$,
which for $x \lesssim 0.1$ becomes larger than the inter-nucleon separation.
Consequently at small $x$ the scattering from the deuteron can take
place via double scattering from both nucleons, which gives rise to the 
phenomenon of nuclear shadowing (see e.g. Refs.~\cite{KPW94,PILLER}).

At low $Q^2$ the $q\bar q$ fluctuations are usually represented in
the form of the vector mesons, $\rho$, $\omega$ and $\phi$.
Their contributions scale as $1/Q^2$, however, and are formally of
higher twist.
Leading twist shadowing contributions arise from diffractive scattering
of the virtual photon from the bound nucleons, which is described
through Pomeron exchange, as well as contributions in which the
exchanged quanta are mesons (most notably the pion).

\begin{figure}[!ht]
\begin{center}
\includegraphics[bb=0cm 5cm 20cm 25cm, scale=0.365]{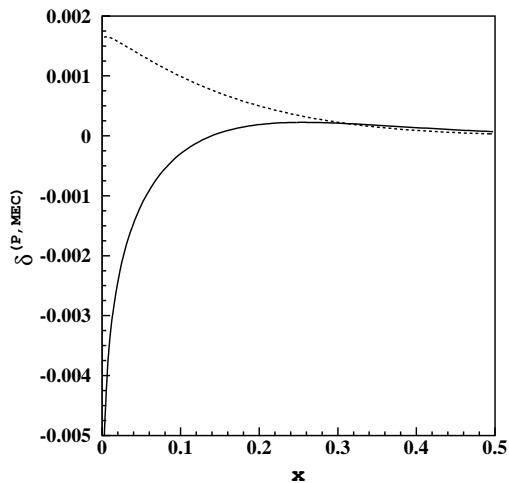}
\caption{\label{fig:delta} \it \small
	Shadowing (Pomeron exchange) and meson exchange current corrections
	$\delta^{(P)}(x,Q^2)$ (solid) and $\delta^{(\rm MEC)}(x,Q^2)$
	(dotted) as a function of $x$ at $Q^2=10$~GeV$^2$ \cite{MTshad}.}
\end{center}
\end{figure}

Both the Pomeron ($P$) exchange and meson exchange current (MEC)
contributions to the deuteron $F_2^D$ structure function have been 
estimated by several authors \cite{D_SHAD,PILLER,MTshad,KPW94,KP04,UMNIKOV},
and enter as additive corrections to $F_2^D$,
\begin{equation}
F_2^D(x,Q^2) \to F_2^D(x,Q^2)\ +\ \delta^{(P, {\rm MEC})} F_2^D(x,Q^2)\ .
\end{equation}
The Pomeron shadowing contribution $\delta^{(P)} F_2^D$ is negative, 
while the meson (pion) exchange contribution $\delta^{(\rm MEC)} F_2^D$
is positive and hence corresponds to an antishadowing effect.
To estimate their effect on the moment analysis, we use the results
from Ref.~\cite{MTshad}, which can be conveniently parameterized at
$Q^2 \sim 5$~GeV$^2$ as
\begin{equation}
\delta^{(P)} F_2^D
= b_0\ x^{b_1} (1-x)^{b_2} (1 + b_3 x^{b_4})\ ,
\end{equation}
for the $P$-exchange, where $b_0 = -0.003$, $b_1 = -0.13$, $b_2 = 5.0$,
$b_3 = -2.2$ and $b_4 = 0.4$, and
\begin{equation}
\delta^{(\rm MEC)} F_2^D
= c_0\ x^{c_1} (1-x)^{c_2}
\end{equation}
for meson exchange, with $c_0 = 0.002$, $c_1 = 0.03$ and $c_2 = 6.0$.
As a fraction of the empirical leading twist $F_2^D$, the ratios
\begin{equation}
\delta^{(P, {\rm MEC})}(x,Q^2)
= \frac{ \delta^{(P, {\rm MEC})} F_2^D(x,Q^2) }{ F_2^D(x,Q^2) }
\end{equation}
are plotted in Fig.~\ref{fig:delta}.
The corresponding moments, defined in analogy with 
Eq.~(\ref{eq:delta_offshell}),
\begin{equation}
\Delta_n^{(P, {\rm MEC})}(Q^2)
= \frac{1}{M_n^D}
  \int dx\ x^{n-2}\ \delta^{(P, {\rm MEC})}(x,Q^2)\ F_2^D(x,Q^2)\ ,
\end{equation}
are listed in Table~\ref{table:delta_r}, where it is clear that these corrections
are negligible for $n \geq 4$.

\section{\label{sec:Mn} Moments of the neutron structure function}

In this section we present the extraction of the lowest few moments
of the neutron structure function $F_2^n$, taking into account the
systematic uncertainties associated with our incomplete knowledge
of nuclear effects in the deuteron.
We shall focus on the leading twist (LT) part of the structure function,
and make use of the previous determination of the LT moments of the proton 
and deuteron structure functions obtained by the twist analyses of
Refs.~\cite{osipenko_f2p,osipenko_f2d}.
Higher twist contributions to the moments, in the form of subleading
$1/Q^2$ power corrections, will be discussed in Sec.~\ref{sec:ht_isospin}.
Moreover, since the twist extraction performed in 
Refs.~\cite{osipenko_f2p,osipenko_f2d} is based on the Nachtmann 
definition of the moments \cite{Nachtmann}, our LT terms are determined 
without any {\em a priori} knowledge of the $x$-shape of parton
distribution functions (PDFs).

The moments of the LT $F_2$ structure function of the neutron are
related to those of proton and deuteron by
\begin{equation}
\label{eq:neutron_LT}
 M_n^n(Q^2) = 2 M_n^D(Q^2) \frac{ 1 - \Delta^{({\rm off})}_n(Q^2) - 
 \Delta_n^{(P, {\rm MEC})}(Q^2) }{ {\cal F}_n^D } - M_n^p(Q^2)\ ,
\end{equation}
where
${\cal F}_n^D / [1 - \Delta^{({\rm off})}_n(Q^2) - 
\Delta_n^{(P, {\rm MEC})}(Q^2)]$ represents the nuclear correction factor, 
or in other words the EMC effect of the deuteron in moment space.
From the values reported in Fig.~\ref{fig:wf_moments} and in Table~\ref{table:delta_r},
the EMC effect on the second moment is expected to be below 1\% [see 
also Fig.~\ref{fig:neutron_moments_qe}(a)]. 

\begin{figure}[!ht]
\begin{center}
\includegraphics[bb=0cm 5cm 20cm 25cm, scale=0.365]{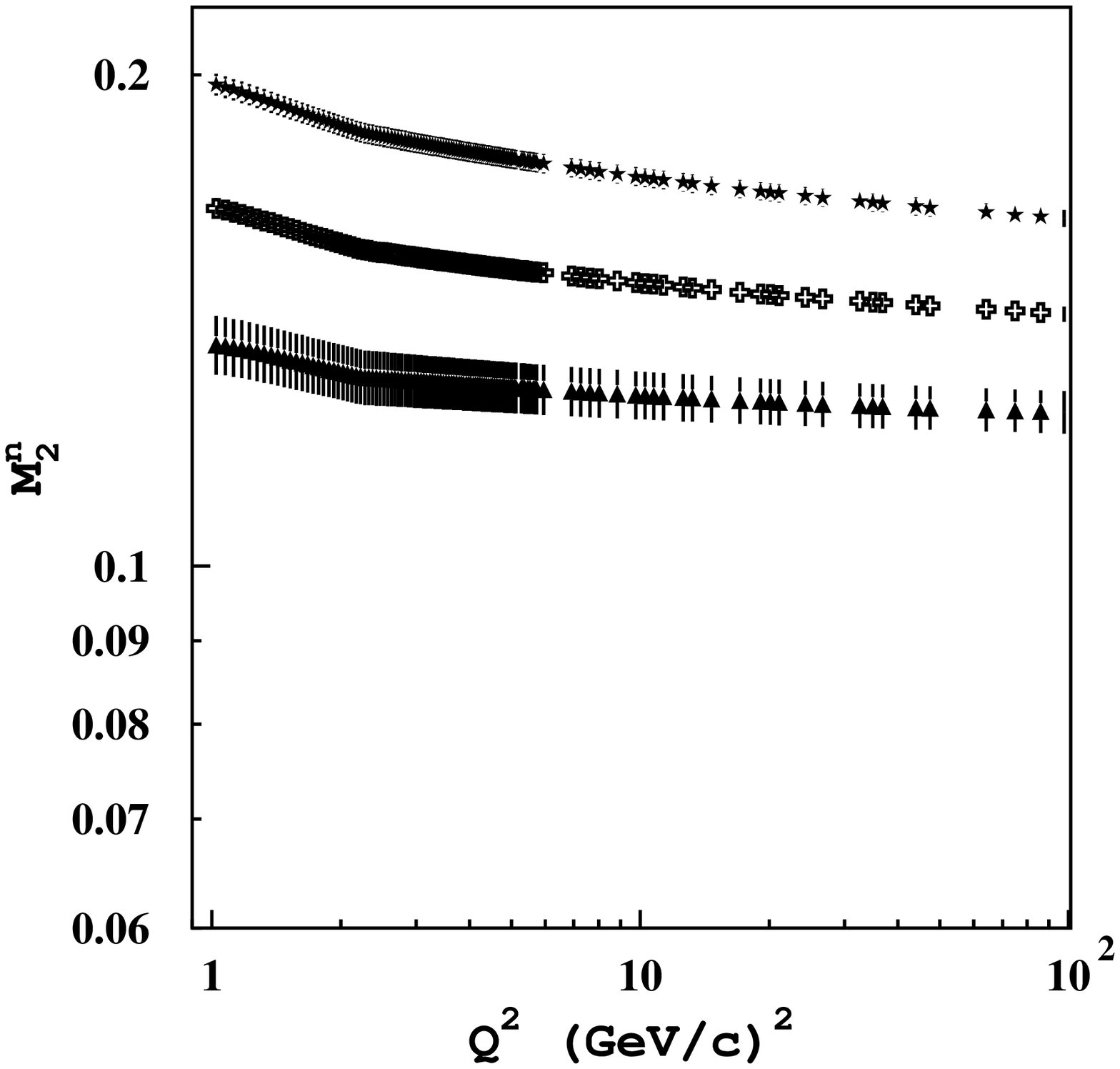}~%
\includegraphics[bb=0cm 5cm 20cm 25cm, scale=0.365]{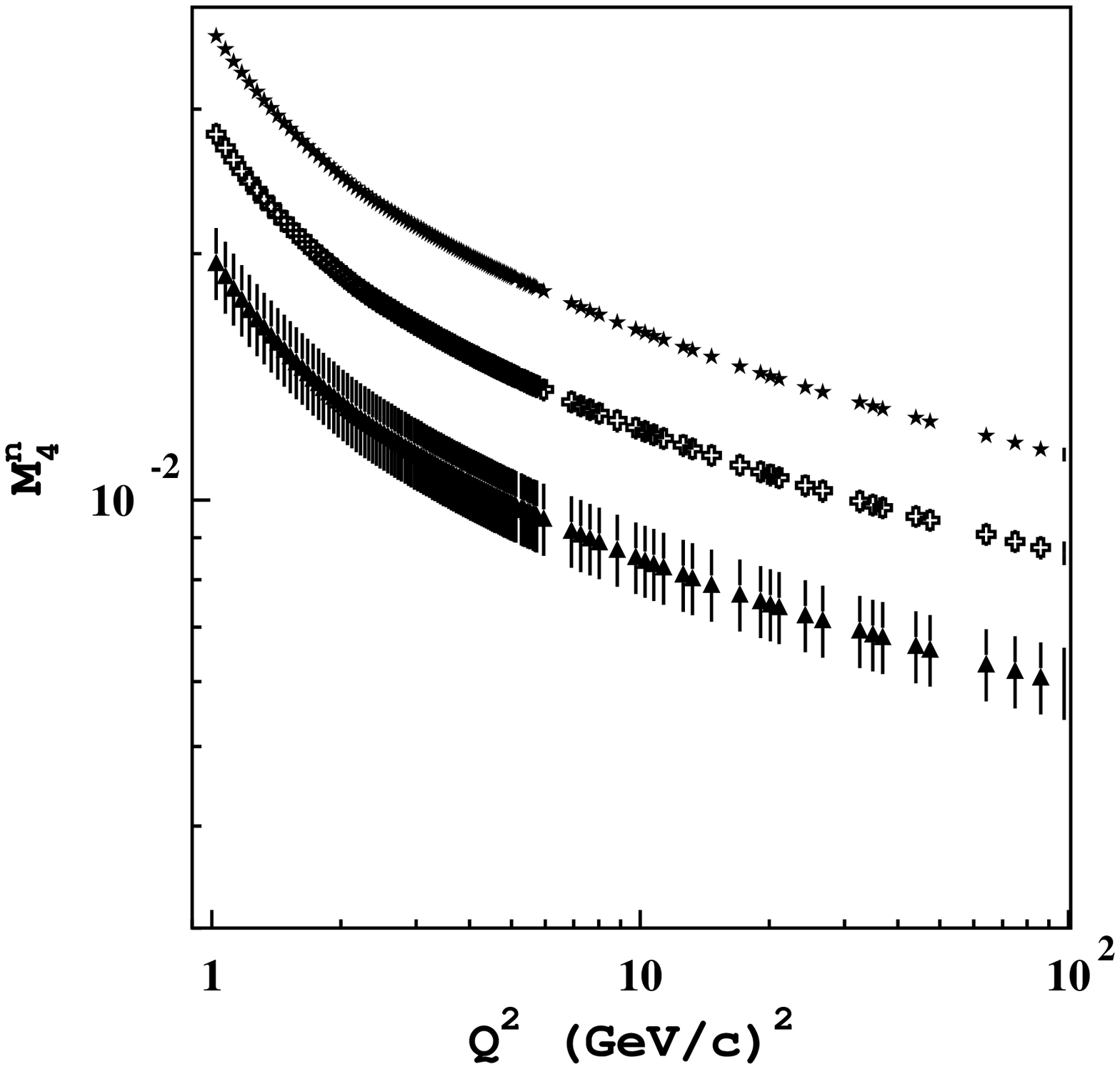}
\includegraphics[bb=0cm 5cm 20cm 25cm, scale=0.365]{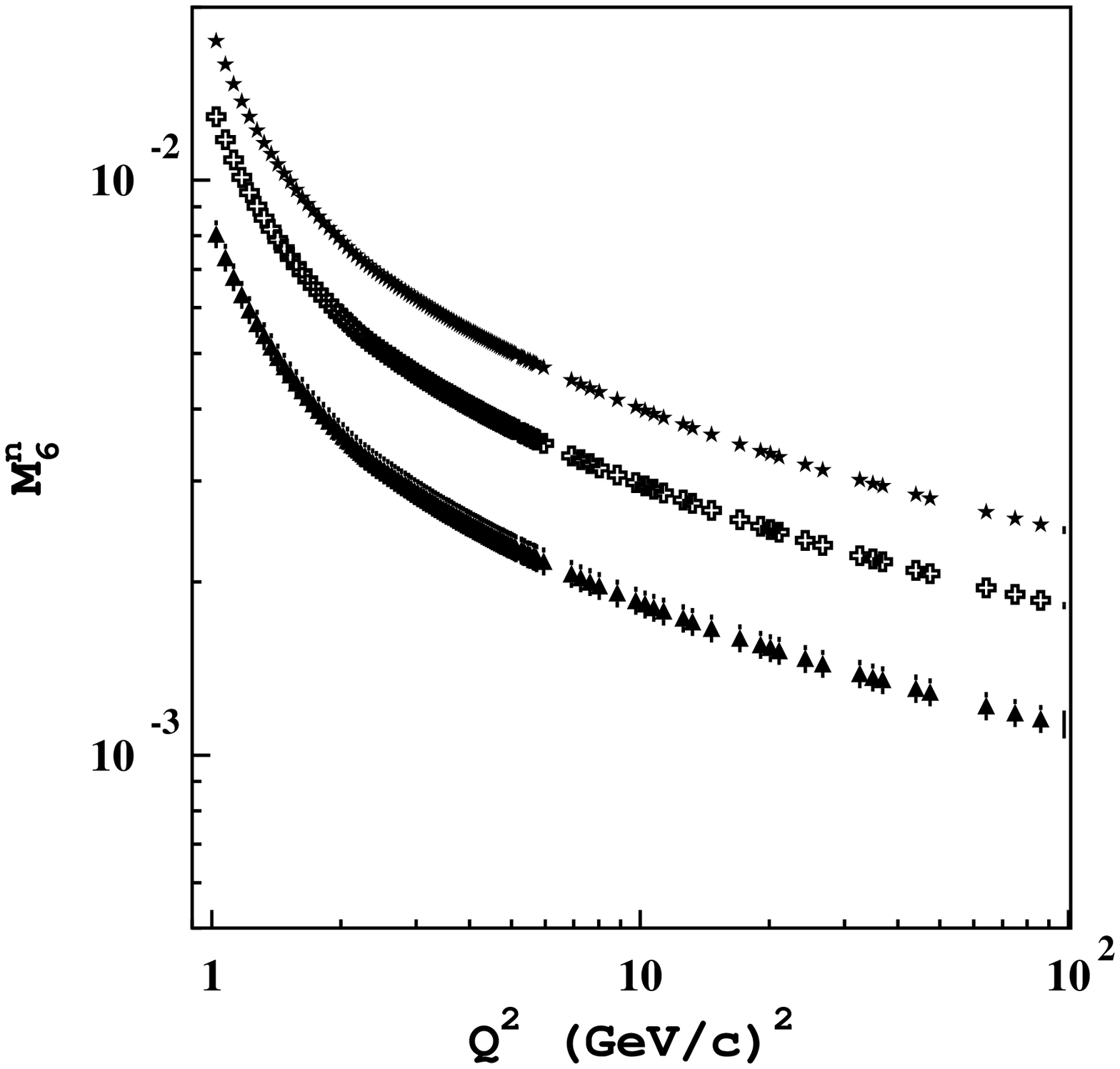}~%
\includegraphics[bb=0cm 5cm 20cm 25cm, scale=0.365]{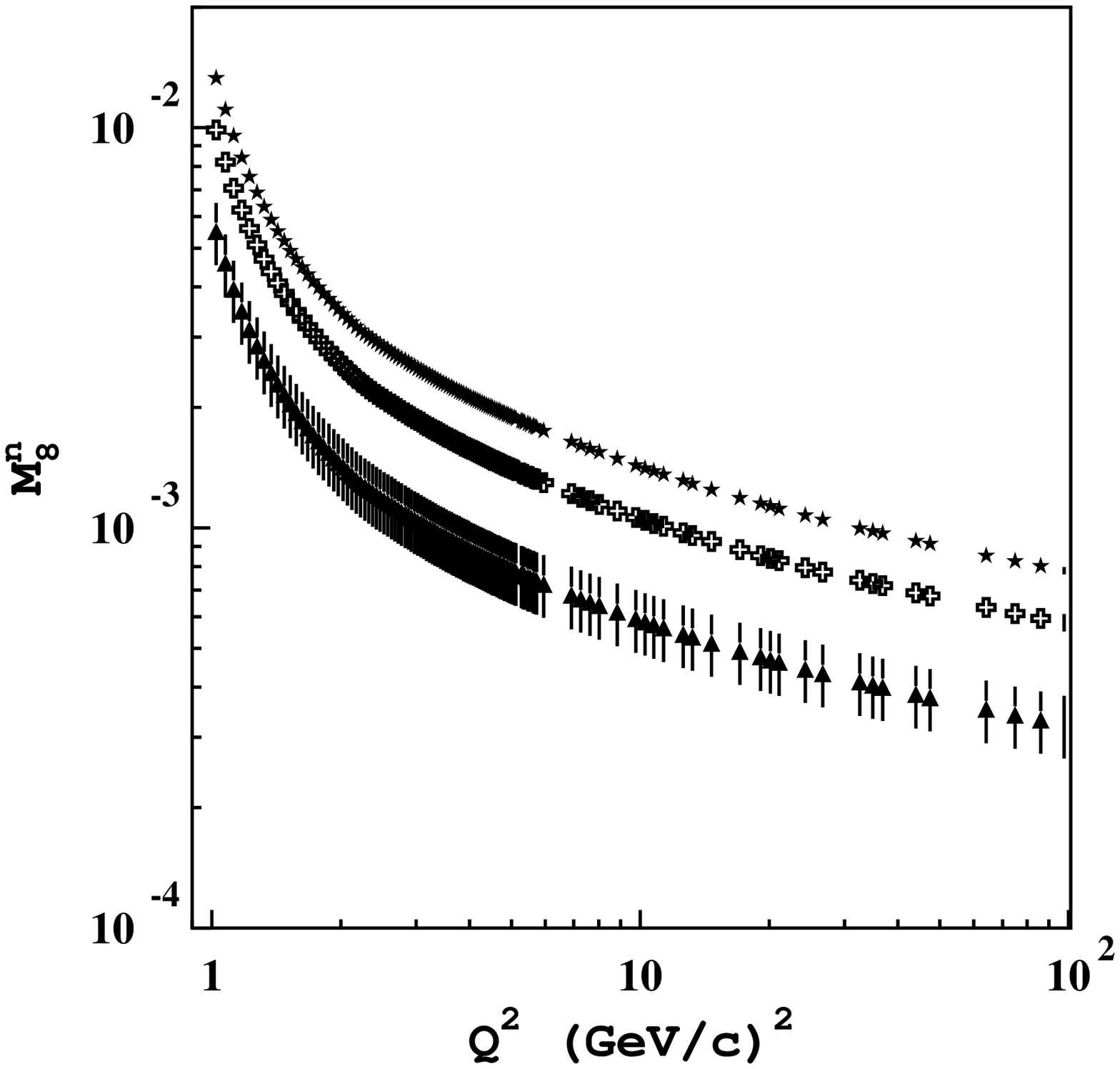}
\caption{\label{fig:neutron_moments} \it \small
	Leading twist moments of the neutron structure function for 
	$n = 2, 4, 6$ and 8 as a function of $Q^2$ (triangles), 
	extracted using the relativistic distribution in Eq.~(\ref{eq:fDrel}) 
	with the deuteron wave function from Ref.~\cite{GVH92}.
	For comparison we also show the proton (stars) and deuteron
	(open crosses) moments.
	The errors shown are statistical.}
\end{center}
\end{figure}

\begin{figure}[!ht]
\begin{center}
\includegraphics[bb=0cm 5cm 20cm 25cm, scale=0.365]{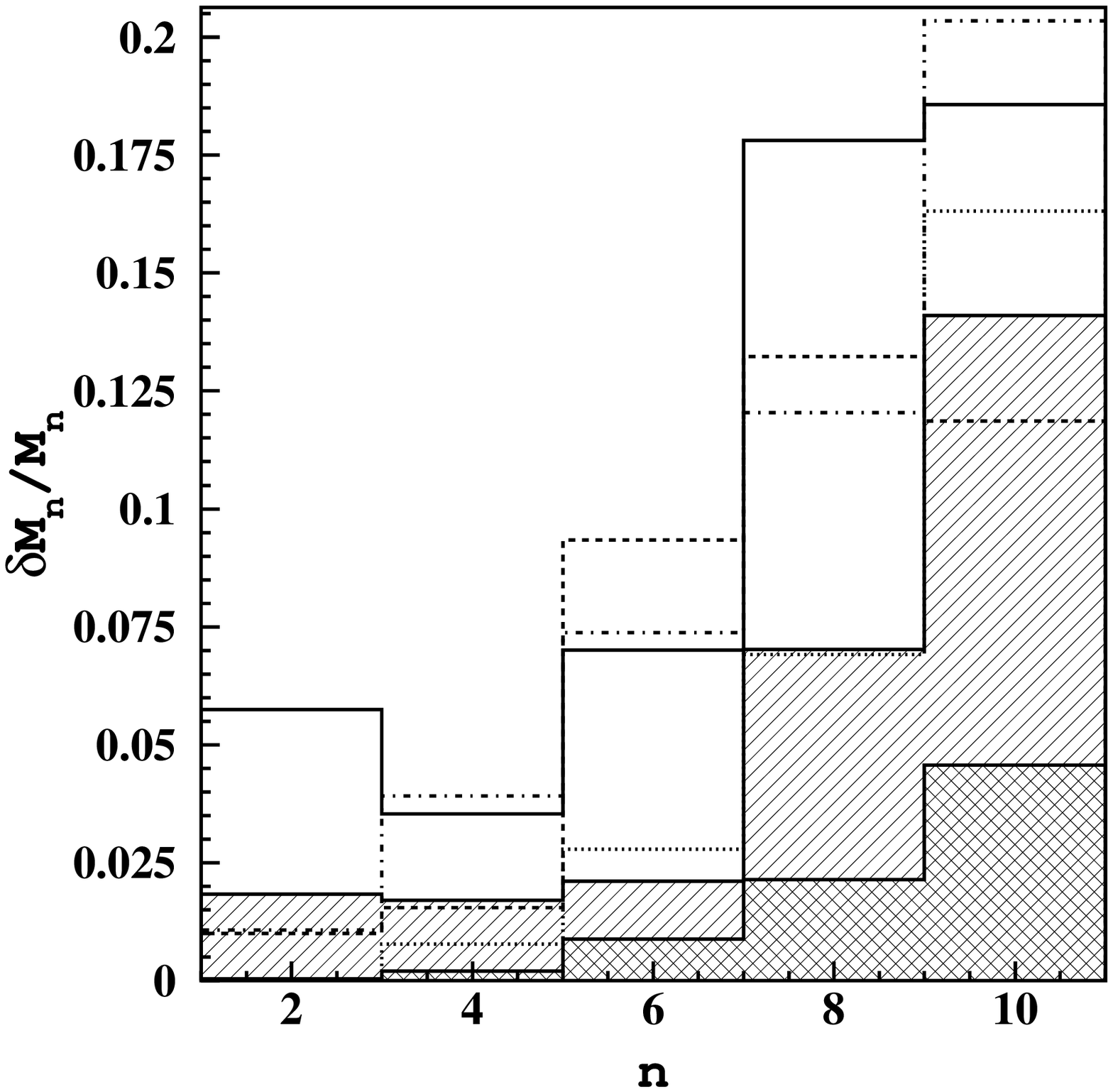}
\caption{\label{fig:sys_err} \it \small
	Various contributions to the systematic errors on the LT
	neutron moments for $n = 2 - 10$:
	deuteron LT moment error (solid);
	proton LT moment error (dashed);
	error due to the momentum cut (dotted);
	dependence on the relativistic treatment (dot-dashed);
	offshellness uncertainty (left-hatched);
	wave function model dependence (cross-hatched).}
\end{center}
\end{figure}

Using the nuclear ingredients discussed in the previous Section, the extracted 
LT moments $M_n^n(Q^2)$ are listed in Table~\ref{table:ltw} and shown in 
Fig.~\ref{fig:neutron_moments}. The central values refer to the case of the 
relativistic distribution (\ref{eq:fDrel}) with the deuteron wave function
from Ref.~\cite{GVH92}. For comparison we also plot the corresponding proton 
(stars) and deuteron (open crosses) LT moments. It turns out that the 
neutron LT moments with $n \leq 8$ can be determined with quite good precision, 
namely $\leq 18 \%$ (statistical) and $\leq 30 \%$ (systematic). Different 
contributions to the total systematic error are shown in Fig.~\ref{fig:sys_err}. 
It can be seen that all the uncertainties related to the nuclear correction 
factor steeply rise with $n$, so that the total errors due to nuclear effects 
become comparable with the experimental ones around $n = 10$. Thus, 
an improvement in the measurements of the proton and deuteron structure 
functions at large $x$ would be beneficial and also the moments at $n = 12$ 
may be still reasonably accessible.

\begin{table*}
\begin{center}
\caption{\label{table:ltw} \it \small
	Extracted leading twist moments of the neutron $F_2$ structure
	functions, $M_n^n(Q^2)$, for $n = 2, 4, 6, 8$, shown in 
	Fig.~\ref{fig:neutron_moments}, together with statistical and
	systematic uncertainties.} \vspace{3mm}
{\scriptsize
\begin{tabular}{|c|c|c|c|c|} \cline{1-5}
$Q^2~[$(GeV/c)$^2]$
 & $M_2^n(Q^2) \times 10^{-1}$ & $M_4^n(Q^2) \times 10^{-2}$ 
 & $M_6^n(Q^2) \times 10^{-3}$ & $M_8^n(Q^2) \times 10^{-3}$ \\ \cline{1-5}
  1.025 & 1.37  $\pm$ 0.06  $\pm$ 0.12 & 1.95  $\pm$ 0.20  $\pm$ 0.11 & 8.06  $\pm$ 0.45  $\pm$ 1.15 & 5.51  $\pm$ 0.98  $\pm$ 1.49 \\ \cline{1-5}
  1.075 & 1.36  $\pm$ 0.06  $\pm$ 0.12 & 1.88  $\pm$ 0.19  $\pm$ 0.11 & 7.35  $\pm$ 0.41  $\pm$ 1.04 & 4.59  $\pm$ 0.81  $\pm$ 1.24 \\ \cline{1-5}
  1.125 & 1.36  $\pm$ 0.06  $\pm$ 0.12 & 1.82  $\pm$ 0.18  $\pm$ 0.11 & 6.78  $\pm$ 0.38  $\pm$ 0.96 & 3.95  $\pm$ 0.70  $\pm$ 1.07 \\ \cline{1-5}
  1.175 & 1.36  $\pm$ 0.06  $\pm$ 0.12 & 1.76  $\pm$ 0.18  $\pm$ 0.10 & 6.33  $\pm$ 0.35  $\pm$ 0.90 & 3.48  $\pm$ 0.62  $\pm$ 0.94 \\ \cline{1-5}
  1.225 & 1.36  $\pm$ 0.05  $\pm$ 0.11 & 1.71  $\pm$ 0.17  $\pm$ 0.10 & 5.95  $\pm$ 0.33  $\pm$ 0.85 & 3.13  $\pm$ 0.56  $\pm$ 0.85 \\ \cline{1-5}
  1.275 & 1.35  $\pm$ 0.05  $\pm$ 0.11 & 1.67  $\pm$ 0.17  $\pm$ 0.10 & 5.63  $\pm$ 0.31  $\pm$ 0.80 & 2.85  $\pm$ 0.51  $\pm$ 0.77 \\ \cline{1-5}
  1.325 & 1.35  $\pm$ 0.05  $\pm$ 0.11 & 1.63  $\pm$ 0.16  $\pm$ 0.09 & 5.36  $\pm$ 0.30  $\pm$ 0.76 & 2.63  $\pm$ 0.47  $\pm$ 0.71 \\ \cline{1-5}
  1.375 & 1.35  $\pm$ 0.05  $\pm$ 0.11 & 1.59  $\pm$ 0.16  $\pm$ 0.09 & 5.13  $\pm$ 0.28  $\pm$ 0.73 & 2.44  $\pm$ 0.43  $\pm$ 0.66 \\ \cline{1-5}
  1.425 & 1.34  $\pm$ 0.05  $\pm$ 0.11 & 1.56  $\pm$ 0.16  $\pm$ 0.09 & 4.93  $\pm$ 0.27  $\pm$ 0.70 & 2.29  $\pm$ 0.41  $\pm$ 0.62 \\ \cline{1-5}
  1.475 & 1.34  $\pm$ 0.05  $\pm$ 0.11 & 1.53  $\pm$ 0.15  $\pm$ 0.09 & 4.74  $\pm$ 0.26  $\pm$ 0.68 & 2.16  $\pm$ 0.38  $\pm$ 0.58 \\ \cline{1-5}
  1.525 & 1.34  $\pm$ 0.05  $\pm$ 0.11 & 1.50  $\pm$ 0.15  $\pm$ 0.09 & 4.58  $\pm$ 0.25  $\pm$ 0.65 & 2.04  $\pm$ 0.36  $\pm$ 0.55 \\ \cline{1-5}
  1.575 & 1.34  $\pm$ 0.05  $\pm$ 0.11 & 1.48  $\pm$ 0.15  $\pm$ 0.09 & 4.44  $\pm$ 0.25  $\pm$ 0.63 & 1.94  $\pm$ 0.35  $\pm$ 0.53 \\ \cline{1-5}
  1.625 & 1.33  $\pm$ 0.05  $\pm$ 0.11 & 1.45  $\pm$ 0.15  $\pm$ 0.08 & 4.31  $\pm$ 0.24  $\pm$ 0.61 & 1.86  $\pm$ 0.33  $\pm$ 0.50 \\ \cline{1-5}
  1.675 & 1.33  $\pm$ 0.05  $\pm$ 0.11 & 1.43  $\pm$ 0.14  $\pm$ 0.08 & 4.19  $\pm$ 0.23  $\pm$ 0.60 & 1.78  $\pm$ 0.32  $\pm$ 0.48 \\ \cline{1-5}
  1.725 & 1.33  $\pm$ 0.05  $\pm$ 0.11 & 1.41  $\pm$ 0.14  $\pm$ 0.08 & 4.08  $\pm$ 0.23  $\pm$ 0.58 & 1.71  $\pm$ 0.30  $\pm$ 0.46 \\ \cline{1-5}
  1.775 & 1.32  $\pm$ 0.05  $\pm$ 0.11 & 1.39  $\pm$ 0.14  $\pm$ 0.08 & 3.99  $\pm$ 0.22  $\pm$ 0.57 & 1.65  $\pm$ 0.29  $\pm$ 0.45 \\ \cline{1-5}
  1.825 & 1.32  $\pm$ 0.05  $\pm$ 0.11 & 1.37  $\pm$ 0.14  $\pm$ 0.08 & 3.90  $\pm$ 0.22  $\pm$ 0.55 & 1.60  $\pm$ 0.28  $\pm$ 0.43 \\ \cline{1-5}
  1.875 & 1.32  $\pm$ 0.05  $\pm$ 0.11 & 1.36  $\pm$ 0.14  $\pm$ 0.08 & 3.81  $\pm$ 0.21  $\pm$ 0.54 & 1.55  $\pm$ 0.27  $\pm$ 0.42 \\ \cline{1-5}
  1.925 & 1.32  $\pm$ 0.05  $\pm$ 0.11 & 1.34  $\pm$ 0.13  $\pm$ 0.08 & 3.73  $\pm$ 0.21  $\pm$ 0.53 & 1.50  $\pm$ 0.27  $\pm$ 0.41 \\ \cline{1-5}
  1.975 & 1.32  $\pm$ 0.05  $\pm$ 0.11 & 1.33  $\pm$ 0.13  $\pm$ 0.08 & 3.66  $\pm$ 0.20  $\pm$ 0.52 & 1.46  $\pm$ 0.26  $\pm$ 0.40 \\ \cline{1-5}
  2.025 & 1.31  $\pm$ 0.05  $\pm$ 0.11 & 1.31  $\pm$ 0.13  $\pm$ 0.08 & 3.59  $\pm$ 0.20  $\pm$ 0.51 & 1.42  $\pm$ 0.25  $\pm$ 0.38 \\ \cline{1-5}
  2.075 & 1.31  $\pm$ 0.05  $\pm$ 0.11 & 1.30  $\pm$ 0.13  $\pm$ 0.08 & 3.53  $\pm$ 0.20  $\pm$ 0.50 & 1.38  $\pm$ 0.25  $\pm$ 0.38 \\ \cline{1-5}
  2.125 & 1.31  $\pm$ 0.05  $\pm$ 0.11 & 1.29  $\pm$ 0.13  $\pm$ 0.07 & 3.47  $\pm$ 0.19  $\pm$ 0.49 & 1.35  $\pm$ 0.24  $\pm$ 0.37 \\ \cline{1-5}
  2.175 & 1.31  $\pm$ 0.05  $\pm$ 0.10 & 1.27  $\pm$ 0.13  $\pm$ 0.07 & 3.42  $\pm$ 0.19  $\pm$ 0.49 & 1.32  $\pm$ 0.23  $\pm$ 0.36 \\ \cline{1-5}
  2.225 & 1.31  $\pm$ 0.05  $\pm$ 0.10 & 1.26  $\pm$ 0.13  $\pm$ 0.07 & 3.36  $\pm$ 0.19  $\pm$ 0.48 & 1.29  $\pm$ 0.23  $\pm$ 0.35 \\ \cline{1-5}
  2.275 & 1.31  $\pm$ 0.05  $\pm$ 0.10 & 1.25  $\pm$ 0.13  $\pm$ 0.07 & 3.32  $\pm$ 0.18  $\pm$ 0.47 & 1.27  $\pm$ 0.22  $\pm$ 0.34 \\ \cline{1-5}
  2.325 & 1.31  $\pm$ 0.05  $\pm$ 0.10 & 1.24  $\pm$ 0.12  $\pm$ 0.07 & 3.28  $\pm$ 0.18  $\pm$ 0.47 & 1.25  $\pm$ 0.22  $\pm$ 0.34 \\ \cline{1-5}
  2.375 & 1.30  $\pm$ 0.05  $\pm$ 0.10 & 1.23  $\pm$ 0.12  $\pm$ 0.07 & 3.24  $\pm$ 0.18  $\pm$ 0.46 & 1.23  $\pm$ 0.22  $\pm$ 0.33 \\ \cline{1-5}
  2.425 & 1.30  $\pm$ 0.05  $\pm$ 0.10 & 1.22  $\pm$ 0.12  $\pm$ 0.07 & 3.20  $\pm$ 0.18  $\pm$ 0.46 & 1.21  $\pm$ 0.21  $\pm$ 0.33 \\ \cline{1-5}
  2.475 & 1.30  $\pm$ 0.05  $\pm$ 0.10 & 1.22  $\pm$ 0.12  $\pm$ 0.07 & 3.17  $\pm$ 0.18  $\pm$ 0.45 & 1.19  $\pm$ 0.21  $\pm$ 0.32 \\ \cline{1-5}
  2.525 & 1.30  $\pm$ 0.05  $\pm$ 0.10 & 1.21  $\pm$ 0.12  $\pm$ 0.07 & 3.13  $\pm$ 0.17  $\pm$ 0.45 & 1.17  $\pm$ 0.21  $\pm$ 0.32 \\ \cline{1-5}
  2.575 & 1.30  $\pm$ 0.05  $\pm$ 0.10 & 1.20  $\pm$ 0.12  $\pm$ 0.07 & 3.10  $\pm$ 0.17  $\pm$ 0.44 & 1.15  $\pm$ 0.21  $\pm$ 0.31 \\ \cline{1-5}
  2.625 & 1.30  $\pm$ 0.05  $\pm$ 0.10 & 1.19  $\pm$ 0.12  $\pm$ 0.07 & 3.07  $\pm$ 0.17  $\pm$ 0.44 & 1.14  $\pm$ 0.20  $\pm$ 0.31 \\ \cline{1-5}
  2.675 & 1.30  $\pm$ 0.05  $\pm$ 0.10 & 1.18  $\pm$ 0.12  $\pm$ 0.07 & 3.04  $\pm$ 0.17  $\pm$ 0.43 & 1.12  $\pm$ 0.20  $\pm$ 0.30 \\ \cline{1-5}
  2.725 & 1.30  $\pm$ 0.05  $\pm$ 0.10 & 1.18  $\pm$ 0.12  $\pm$ 0.07 & 3.01  $\pm$ 0.17  $\pm$ 0.43 & 1.11  $\pm$ 0.20  $\pm$ 0.30 \\ \cline{1-5}
  2.775 & 1.30  $\pm$ 0.05  $\pm$ 0.10 & 1.17  $\pm$ 0.12  $\pm$ 0.07 & 2.98  $\pm$ 0.17  $\pm$ 0.42 & 1.10  $\pm$ 0.19  $\pm$ 0.30 \\ \cline{1-5}
  2.825 & 1.30  $\pm$ 0.05  $\pm$ 0.10 & 1.16  $\pm$ 0.12  $\pm$ 0.07 & 2.96  $\pm$ 0.16  $\pm$ 0.42 & 1.08  $\pm$ 0.19  $\pm$ 0.29 \\ \cline{1-5}
  2.875 & 1.30  $\pm$ 0.05  $\pm$ 0.10 & 1.16  $\pm$ 0.12  $\pm$ 0.07 & 2.93  $\pm$ 0.16  $\pm$ 0.42 & 1.07  $\pm$ 0.19  $\pm$ 0.29 \\ \cline{1-5}
  2.925 & 1.30  $\pm$ 0.05  $\pm$ 0.10 & 1.15  $\pm$ 0.12  $\pm$ 0.07 & 2.91  $\pm$ 0.16  $\pm$ 0.41 & 1.06  $\pm$ 0.19  $\pm$ 0.29 \\ \cline{1-5}
  2.975 & 1.30  $\pm$ 0.05  $\pm$ 0.10 & 1.15  $\pm$ 0.12  $\pm$ 0.07 & 2.88  $\pm$ 0.16  $\pm$ 0.41 & 1.05  $\pm$ 0.19  $\pm$ 0.28 \\ \cline{1-5}
  3.025 & 1.30  $\pm$ 0.05  $\pm$ 0.10 & 1.14  $\pm$ 0.11  $\pm$ 0.07 & 2.86  $\pm$ 0.16  $\pm$ 0.41 & 1.04  $\pm$ 0.18  $\pm$ 0.28 \\ \cline{1-5}
  3.075 & 1.30  $\pm$ 0.05  $\pm$ 0.10 & 1.13  $\pm$ 0.11  $\pm$ 0.07 & 2.84  $\pm$ 0.16  $\pm$ 0.40 & 1.03  $\pm$ 0.18  $\pm$ 0.28 \\ \cline{1-5}
  3.125 & 1.30  $\pm$ 0.05  $\pm$ 0.10 & 1.13  $\pm$ 0.11  $\pm$ 0.07 & 2.82  $\pm$ 0.16  $\pm$ 0.40 & 1.02  $\pm$ 0.18  $\pm$ 0.28 \\ \cline{1-5}
  3.175 & 1.30  $\pm$ 0.05  $\pm$ 0.10 & 1.12  $\pm$ 0.11  $\pm$ 0.07 & 2.80  $\pm$ 0.16  $\pm$ 0.40 & 1.01  $\pm$ 0.18  $\pm$ 0.27 \\ \cline{1-5}
  3.225 & 1.30  $\pm$ 0.05  $\pm$ 0.10 & 1.12  $\pm$ 0.11  $\pm$ 0.06 & 2.78  $\pm$ 0.15  $\pm$ 0.40 & 1.00  $\pm$ 0.18  $\pm$ 0.27 \\ \cline{1-5}
  3.275 & 1.30  $\pm$ 0.05  $\pm$ 0.10 & 1.11  $\pm$ 0.11  $\pm$ 0.06 & 2.76  $\pm$ 0.15  $\pm$ 0.39 & 0.99  $\pm$ 0.18  $\pm$ 0.27 \\ \cline{1-5}
  3.325 & 1.30  $\pm$ 0.05  $\pm$ 0.10 & 1.11  $\pm$ 0.11  $\pm$ 0.06 & 2.74  $\pm$ 0.15  $\pm$ 0.39 & 0.98  $\pm$ 0.17  $\pm$ 0.27 \\ \cline{1-5}
  3.375 & 1.30  $\pm$ 0.05  $\pm$ 0.10 & 1.10  $\pm$ 0.11  $\pm$ 0.06 & 2.72  $\pm$ 0.15  $\pm$ 0.39 & 0.97  $\pm$ 0.17  $\pm$ 0.26 \\ \cline{1-5}
  3.425 & 1.30  $\pm$ 0.05  $\pm$ 0.10 & 1.10  $\pm$ 0.11  $\pm$ 0.06 & 2.70  $\pm$ 0.15  $\pm$ 0.38 & 0.96  $\pm$ 0.17  $\pm$ 0.26 \\ \cline{1-5}
  3.475 & 1.30  $\pm$ 0.05  $\pm$ 0.10 & 1.09  $\pm$ 0.11  $\pm$ 0.06 & 2.68  $\pm$ 0.15  $\pm$ 0.38 & 0.95  $\pm$ 0.17  $\pm$ 0.26 \\ \cline{1-5}
  3.525 & 1.29  $\pm$ 0.05  $\pm$ 0.10 & 1.09  $\pm$ 0.11  $\pm$ 0.06 & 2.67  $\pm$ 0.15  $\pm$ 0.38 & 0.94  $\pm$ 0.17  $\pm$ 0.26 \\ \cline{1-5}
  3.575 & 1.29  $\pm$ 0.05  $\pm$ 0.10 & 1.08  $\pm$ 0.11  $\pm$ 0.06 & 2.65  $\pm$ 0.15  $\pm$ 0.38 & 0.94  $\pm$ 0.17  $\pm$ 0.25 \\ \cline{1-5}
  3.625 & 1.29  $\pm$ 0.05  $\pm$ 0.10 & 1.08  $\pm$ 0.11  $\pm$ 0.06 & 2.64  $\pm$ 0.15  $\pm$ 0.38 & 0.93  $\pm$ 0.17  $\pm$ 0.25 \\ \cline{1-5}
  3.675 & 1.29  $\pm$ 0.05  $\pm$ 0.10 & 1.08  $\pm$ 0.11  $\pm$ 0.06 & 2.62  $\pm$ 0.15  $\pm$ 0.37 & 0.92  $\pm$ 0.16  $\pm$ 0.25 \\ \cline{1-5}
  3.725 & 1.29  $\pm$ 0.05  $\pm$ 0.10 & 1.07  $\pm$ 0.11  $\pm$ 0.06 & 2.60  $\pm$ 0.14  $\pm$ 0.37 & 0.92  $\pm$ 0.16  $\pm$ 0.25 \\ \cline{1-5}
  3.775 & 1.29  $\pm$ 0.05  $\pm$ 0.10 & 1.07  $\pm$ 0.11  $\pm$ 0.06 & 2.59  $\pm$ 0.14  $\pm$ 0.37 & 0.91  $\pm$ 0.16  $\pm$ 0.25 \\ \cline{1-5}
  3.825 & 1.29  $\pm$ 0.05  $\pm$ 0.10 & 1.06  $\pm$ 0.11  $\pm$ 0.06 & 2.58  $\pm$ 0.14  $\pm$ 0.37 & 0.90  $\pm$ 0.16  $\pm$ 0.24 \\ \cline{1-5}
\end{tabular}
}
\end{center}
\end{table*}
\begin{table*}
\begin{center}
{\scriptsize
\begin{tabular}{|c|c|c|c|c|} \cline{1-5}
$Q^2~[$(GeV/c)$^2]$
& $M_2^n(Q^2) \times 10^{-1}$ & $M_4^n(Q^2) \times 10^{-2}$
& $M_6^n(Q^2) \times 10^{-3}$ & $M_8^n(Q^2) \times 10^{-3}$ \\ \cline{1-5}
  3.875 & 1.29  $\pm$ 0.05  $\pm$ 0.10 & 1.06  $\pm$ 0.11  $\pm$ 0.06 & 2.56  $\pm$ 0.14  $\pm$ 0.36 & 0.90  $\pm$ 0.16  $\pm$ 0.24 \\ \cline{1-5}
  3.925 & 1.29  $\pm$ 0.05  $\pm$ 0.10 & 1.06  $\pm$ 0.11  $\pm$ 0.06 & 2.55  $\pm$ 0.14  $\pm$ 0.36 & 0.89  $\pm$ 0.16  $\pm$ 0.24 \\ \cline{1-5}
  3.975 & 1.29  $\pm$ 0.05  $\pm$ 0.10 & 1.05  $\pm$ 0.11  $\pm$ 0.06 & 2.53  $\pm$ 0.14  $\pm$ 0.36 & 0.88  $\pm$ 0.16  $\pm$ 0.24 \\ \cline{1-5}
  4.025 & 1.29  $\pm$ 0.05  $\pm$ 0.10 & 1.05  $\pm$ 0.11  $\pm$ 0.06 & 2.52  $\pm$ 0.14  $\pm$ 0.36 & 0.88  $\pm$ 0.16  $\pm$ 0.24 \\ \cline{1-5}
  4.075 & 1.29  $\pm$ 0.05  $\pm$ 0.10 & 1.05  $\pm$ 0.11  $\pm$ 0.06 & 2.51  $\pm$ 0.14  $\pm$ 0.36 & 0.87  $\pm$ 0.15  $\pm$ 0.24 \\ \cline{1-5}
  4.125 & 1.29  $\pm$ 0.05  $\pm$ 0.10 & 1.04  $\pm$ 0.10  $\pm$ 0.06 & 2.50  $\pm$ 0.14  $\pm$ 0.36 & 0.87  $\pm$ 0.15  $\pm$ 0.23 \\ \cline{1-5}
  4.175 & 1.29  $\pm$ 0.05  $\pm$ 0.10 & 1.04  $\pm$ 0.10  $\pm$ 0.06 & 2.48  $\pm$ 0.14  $\pm$ 0.35 & 0.86  $\pm$ 0.15  $\pm$ 0.23 \\ \cline{1-5}
  4.225 & 1.29  $\pm$ 0.05  $\pm$ 0.10 & 1.04  $\pm$ 0.10  $\pm$ 0.06 & 2.47  $\pm$ 0.14  $\pm$ 0.35 & 0.86  $\pm$ 0.15  $\pm$ 0.23 \\ \cline{1-5}
  4.275 & 1.29  $\pm$ 0.05  $\pm$ 0.10 & 1.03  $\pm$ 0.10  $\pm$ 0.06 & 2.46  $\pm$ 0.14  $\pm$ 0.35 & 0.85  $\pm$ 0.15  $\pm$ 0.23 \\ \cline{1-5}
  4.325 & 1.29  $\pm$ 0.05  $\pm$ 0.10 & 1.03  $\pm$ 0.10  $\pm$ 0.06 & 2.45  $\pm$ 0.14  $\pm$ 0.35 & 0.84  $\pm$ 0.15  $\pm$ 0.23 \\ \cline{1-5}
  4.375 & 1.29  $\pm$ 0.05  $\pm$ 0.10 & 1.03  $\pm$ 0.10  $\pm$ 0.06 & 2.44  $\pm$ 0.14  $\pm$ 0.35 & 0.84  $\pm$ 0.15  $\pm$ 0.23 \\ \cline{1-5}
  4.425 & 1.29  $\pm$ 0.05  $\pm$ 0.10 & 1.02  $\pm$ 0.10  $\pm$ 0.06 & 2.43  $\pm$ 0.13  $\pm$ 0.35 & 0.84  $\pm$ 0.15  $\pm$ 0.23 \\ \cline{1-5}
  4.475 & 1.29  $\pm$ 0.05  $\pm$ 0.10 & 1.02  $\pm$ 0.10  $\pm$ 0.06 & 2.42  $\pm$ 0.13  $\pm$ 0.34 & 0.83  $\pm$ 0.15  $\pm$ 0.23 \\ \cline{1-5}
  4.525 & 1.29  $\pm$ 0.05  $\pm$ 0.10 & 1.02  $\pm$ 0.10  $\pm$ 0.06 & 2.41  $\pm$ 0.13  $\pm$ 0.34 & 0.83  $\pm$ 0.15  $\pm$ 0.22 \\ \cline{1-5}
  4.575 & 1.29  $\pm$ 0.05  $\pm$ 0.10 & 1.01  $\pm$ 0.10  $\pm$ 0.06 & 2.40  $\pm$ 0.13  $\pm$ 0.34 & 0.82  $\pm$ 0.15  $\pm$ 0.22 \\ \cline{1-5}
  4.625 & 1.29  $\pm$ 0.05  $\pm$ 0.10 & 1.01  $\pm$ 0.10  $\pm$ 0.06 & 2.39  $\pm$ 0.13  $\pm$ 0.34 & 0.82  $\pm$ 0.15  $\pm$ 0.22 \\ \cline{1-5}
  4.675 & 1.29  $\pm$ 0.05  $\pm$ 0.10 & 1.01  $\pm$ 0.10  $\pm$ 0.06 & 2.38  $\pm$ 0.13  $\pm$ 0.34 & 0.81  $\pm$ 0.14  $\pm$ 0.22 \\ \cline{1-5}
  4.725 & 1.29  $\pm$ 0.05  $\pm$ 0.10 & 1.01  $\pm$ 0.10  $\pm$ 0.06 & 2.37  $\pm$ 0.13  $\pm$ 0.34 & 0.81  $\pm$ 0.14  $\pm$ 0.22 \\ \cline{1-5}
  4.775 & 1.29  $\pm$ 0.05  $\pm$ 0.10 & 1.00  $\pm$ 0.10  $\pm$ 0.06 & 2.36  $\pm$ 0.13  $\pm$ 0.34 & 0.80  $\pm$ 0.14  $\pm$ 0.22 \\ \cline{1-5}
  4.825 & 1.29  $\pm$ 0.05  $\pm$ 0.10 & 1.00  $\pm$ 0.10  $\pm$ 0.06 & 2.35  $\pm$ 0.13  $\pm$ 0.33 & 0.80  $\pm$ 0.14  $\pm$ 0.22 \\ \cline{1-5}
  4.875 & 1.29  $\pm$ 0.05  $\pm$ 0.10 & 1.00  $\pm$ 0.10  $\pm$ 0.06 & 2.34  $\pm$ 0.13  $\pm$ 0.33 & 0.80  $\pm$ 0.14  $\pm$ 0.22 \\ \cline{1-5}
  4.925 & 1.29  $\pm$ 0.05  $\pm$ 0.10 & 1.00  $\pm$ 0.10  $\pm$ 0.06 & 2.33  $\pm$ 0.13  $\pm$ 0.33 & 0.79  $\pm$ 0.14  $\pm$ 0.21 \\ \cline{1-5}
  4.975 & 1.29  $\pm$ 0.05  $\pm$ 0.10 & 0.99  $\pm$ 0.10  $\pm$ 0.06 & 2.32  $\pm$ 0.13  $\pm$ 0.33 & 0.79  $\pm$ 0.14  $\pm$ 0.21 \\ \cline{1-5}
  5.025 & 1.29  $\pm$ 0.05  $\pm$ 0.10 & 0.99  $\pm$ 0.10  $\pm$ 0.06 & 2.31  $\pm$ 0.13  $\pm$ 0.33 & 0.78  $\pm$ 0.14  $\pm$ 0.21 \\ \cline{1-5}
  5.075 & 1.29  $\pm$ 0.05  $\pm$ 0.10 & 0.99  $\pm$ 0.10  $\pm$ 0.06 & 2.30  $\pm$ 0.13  $\pm$ 0.33 & 0.78  $\pm$ 0.14  $\pm$ 0.21 \\ \cline{1-5}
  5.125 & 1.29  $\pm$ 0.05  $\pm$ 0.10 & 0.99  $\pm$ 0.10  $\pm$ 0.06 & 2.30  $\pm$ 0.13  $\pm$ 0.33 & 0.78  $\pm$ 0.14  $\pm$ 0.21 \\ \cline{1-5}
  5.275 & 1.29  $\pm$ 0.05  $\pm$ 0.10 & 0.98  $\pm$ 0.10  $\pm$ 0.06 & 2.27  $\pm$ 0.13  $\pm$ 0.32 & 0.77  $\pm$ 0.14  $\pm$ 0.21 \\ \cline{1-5}
  5.325 & 1.29  $\pm$ 0.05  $\pm$ 0.10 & 0.98  $\pm$ 0.10  $\pm$ 0.06 & 2.26  $\pm$ 0.13  $\pm$ 0.32 & 0.76  $\pm$ 0.14  $\pm$ 0.21 \\ \cline{1-5}
  5.375 & 1.29  $\pm$ 0.05  $\pm$ 0.10 & 0.97  $\pm$ 0.10  $\pm$ 0.06 & 2.26  $\pm$ 0.13  $\pm$ 0.32 & 0.76  $\pm$ 0.13  $\pm$ 0.21 \\ \cline{1-5}
  5.475 & 1.28  $\pm$ 0.05  $\pm$ 0.09 & 0.97  $\pm$ 0.10  $\pm$ 0.06 & 2.24  $\pm$ 0.12  $\pm$ 0.32 & 0.75  $\pm$ 0.13  $\pm$ 0.20 \\ \cline{1-5}
  5.525 & 1.28  $\pm$ 0.05  $\pm$ 0.09 & 0.97  $\pm$ 0.10  $\pm$ 0.06 & 2.23  $\pm$ 0.12  $\pm$ 0.32 & 0.75  $\pm$ 0.13  $\pm$ 0.20 \\ \cline{1-5}
  5.625 & 1.28  $\pm$ 0.05  $\pm$ 0.09 & 0.96  $\pm$ 0.10  $\pm$ 0.06 & 2.22  $\pm$ 0.12  $\pm$ 0.32 & 0.74  $\pm$ 0.13  $\pm$ 0.20 \\ \cline{1-5}
  5.675 & 1.28  $\pm$ 0.05  $\pm$ 0.09 & 0.96  $\pm$ 0.10  $\pm$ 0.06 & 2.21  $\pm$ 0.12  $\pm$ 0.31 & 0.74  $\pm$ 0.13  $\pm$ 0.20 \\ \cline{1-5}
  5.725 & 1.28  $\pm$ 0.05  $\pm$ 0.09 & 0.96  $\pm$ 0.10  $\pm$ 0.06 & 2.21  $\pm$ 0.12  $\pm$ 0.31 & 0.74  $\pm$ 0.13  $\pm$ 0.20 \\ \cline{1-5}
  5.955 & 1.28  $\pm$ 0.05  $\pm$ 0.09 & 0.95  $\pm$ 0.10  $\pm$ 0.06 & 2.18  $\pm$ 0.12  $\pm$ 0.31 & 0.72  $\pm$ 0.13  $\pm$ 0.20 \\ \cline{1-5}
  6.915 & 1.28  $\pm$ 0.04  $\pm$ 0.09 & 0.92  $\pm$ 0.09  $\pm$ 0.05 & 2.07  $\pm$ 0.12  $\pm$ 0.29 & 0.68  $\pm$ 0.12  $\pm$ 0.18 \\ \cline{1-5}
  7.267 & 1.28  $\pm$ 0.04  $\pm$ 0.09 & 0.91  $\pm$ 0.09  $\pm$ 0.05 & 2.04  $\pm$ 0.11  $\pm$ 0.29 & 0.67  $\pm$ 0.12  $\pm$ 0.18 \\ \cline{1-5}
  7.630 & 1.28  $\pm$ 0.04  $\pm$ 0.09 & 0.90  $\pm$ 0.09  $\pm$ 0.05 & 2.00  $\pm$ 0.11  $\pm$ 0.29 & 0.65  $\pm$ 0.12  $\pm$ 0.18 \\ \cline{1-5}
  8.021 & 1.28  $\pm$ 0.04  $\pm$ 0.09 & 0.89  $\pm$ 0.09  $\pm$ 0.05 & 1.97  $\pm$ 0.11  $\pm$ 0.28 & 0.64  $\pm$ 0.11  $\pm$ 0.17 \\ \cline{1-5}
  8.847 & 1.28  $\pm$ 0.04  $\pm$ 0.09 & 0.87  $\pm$ 0.09  $\pm$ 0.05 & 1.91  $\pm$ 0.11  $\pm$ 0.27 & 0.62  $\pm$ 0.11  $\pm$ 0.17 \\ \cline{1-5}
  9.775 & 1.27  $\pm$ 0.04  $\pm$ 0.09 & 0.85  $\pm$ 0.09  $\pm$ 0.05 & 1.86  $\pm$ 0.10  $\pm$ 0.26 & 0.59  $\pm$ 0.11  $\pm$ 0.16 \\ \cline{1-5}
 10.267 & 1.27  $\pm$ 0.04  $\pm$ 0.09 & 0.85  $\pm$ 0.09  $\pm$ 0.05 & 1.83  $\pm$ 0.10  $\pm$ 0.26 & 0.58  $\pm$ 0.10  $\pm$ 0.16 \\ \cline{1-5}
 10.762 & 1.27  $\pm$ 0.04  $\pm$ 0.09 & 0.84  $\pm$ 0.08  $\pm$ 0.05 & 1.81  $\pm$ 0.10  $\pm$ 0.26 & 0.57  $\pm$ 0.10  $\pm$ 0.16 \\ \cline{1-5}
 11.344 & 1.27  $\pm$ 0.04  $\pm$ 0.09 & 0.83  $\pm$ 0.08  $\pm$ 0.05 & 1.78  $\pm$ 0.10  $\pm$ 0.25 & 0.56  $\pm$ 0.10  $\pm$ 0.15 \\ \cline{1-5}
 12.580 & 1.27  $\pm$ 0.04  $\pm$ 0.09 & 0.81  $\pm$ 0.08  $\pm$ 0.05 & 1.73  $\pm$ 0.10  $\pm$ 0.25 & 0.54  $\pm$ 0.10  $\pm$ 0.15 \\ \cline{1-5}
 13.238 & 1.27  $\pm$ 0.04  $\pm$ 0.09 & 0.80  $\pm$ 0.08  $\pm$ 0.05 & 1.71  $\pm$ 0.10  $\pm$ 0.24 & 0.53  $\pm$ 0.09  $\pm$ 0.15 \\ \cline{1-5}
 14.689 & 1.27  $\pm$ 0.04  $\pm$ 0.09 & 0.79  $\pm$ 0.08  $\pm$ 0.05 & 1.66  $\pm$ 0.09  $\pm$ 0.24 & 0.52  $\pm$ 0.09  $\pm$ 0.14 \\ \cline{1-5}
 17.108 & 1.26  $\pm$ 0.04  $\pm$ 0.09 & 0.77  $\pm$ 0.08  $\pm$ 0.04 & 1.60  $\pm$ 0.09  $\pm$ 0.23 & 0.49  $\pm$ 0.09  $\pm$ 0.13 \\ \cline{1-5}
 19.072 & 1.26  $\pm$ 0.04  $\pm$ 0.09 & 0.75  $\pm$ 0.08  $\pm$ 0.04 & 1.56  $\pm$ 0.09  $\pm$ 0.22 & 0.48  $\pm$ 0.08  $\pm$ 0.13 \\ \cline{1-5}
 20.108 & 1.26  $\pm$ 0.04  $\pm$ 0.09 & 0.75  $\pm$ 0.08  $\pm$ 0.04 & 1.54  $\pm$ 0.09  $\pm$ 0.22 & 0.47  $\pm$ 0.08  $\pm$ 0.13 \\ \cline{1-5}
 21.097 & 1.26  $\pm$ 0.04  $\pm$ 0.09 & 0.74  $\pm$ 0.07  $\pm$ 0.04 & 1.52  $\pm$ 0.08  $\pm$ 0.22 & 0.46  $\pm$ 0.08  $\pm$ 0.13 \\ \cline{1-5}
 24.259 & 1.26  $\pm$ 0.04  $\pm$ 0.08 & 0.73  $\pm$ 0.07  $\pm$ 0.04 & 1.47  $\pm$ 0.08  $\pm$ 0.21 & 0.44  $\pm$ 0.08  $\pm$ 0.12 \\ \cline{1-5}
 26.680 & 1.26  $\pm$ 0.04  $\pm$ 0.08 & 0.71  $\pm$ 0.07  $\pm$ 0.04 & 1.44  $\pm$ 0.08  $\pm$ 0.21 & 0.43  $\pm$ 0.08  $\pm$ 0.12 \\ \cline{1-5}
 32.500 & 1.25  $\pm$ 0.04  $\pm$ 0.08 & 0.69  $\pm$ 0.07  $\pm$ 0.04 & 1.39  $\pm$ 0.08  $\pm$ 0.20 & 0.41  $\pm$ 0.07  $\pm$ 0.11 \\ \cline{1-5}
 34.932 & 1.25  $\pm$ 0.04  $\pm$ 0.08 & 0.69  $\pm$ 0.07  $\pm$ 0.04 & 1.37  $\pm$ 0.08  $\pm$ 0.19 & 0.41  $\pm$ 0.07  $\pm$ 0.11 \\ \cline{1-5}
 36.750 & 1.25  $\pm$ 0.04  $\pm$ 0.08 & 0.68  $\pm$ 0.07  $\pm$ 0.04 & 1.35  $\pm$ 0.08  $\pm$ 0.19 & 0.40  $\pm$ 0.07  $\pm$ 0.11 \\ \cline{1-5}
 43.970 & 1.25  $\pm$ 0.04  $\pm$ 0.08 & 0.66  $\pm$ 0.07  $\pm$ 0.04 & 1.31  $\pm$ 0.07  $\pm$ 0.19 & 0.38  $\pm$ 0.07  $\pm$ 0.10 \\ \cline{1-5}
 47.440 & 1.25  $\pm$ 0.04  $\pm$ 0.08 & 0.66  $\pm$ 0.07  $\pm$ 0.04 & 1.29  $\pm$ 0.07  $\pm$ 0.18 & 0.38  $\pm$ 0.07  $\pm$ 0.10 \\ \cline{1-5}
 64.270 & 1.25  $\pm$ 0.04  $\pm$ 0.08 & 0.63  $\pm$ 0.06  $\pm$ 0.04 & 1.22  $\pm$ 0.07  $\pm$ 0.17 & 0.35  $\pm$ 0.06  $\pm$ 0.10 \\ \cline{1-5}
 75.000 & 1.25  $\pm$ 0.04  $\pm$ 0.08 & 0.62  $\pm$ 0.06  $\pm$ 0.04 & 1.19  $\pm$ 0.07  $\pm$ 0.17 & 0.34  $\pm$ 0.06  $\pm$ 0.09 \\ \cline{1-5}
 86.000 & 1.24  $\pm$ 0.04  $\pm$ 0.08 & 0.61  $\pm$ 0.06  $\pm$ 0.04 & 1.16  $\pm$ 0.06  $\pm$ 0.17 & 0.33  $\pm$ 0.06  $\pm$ 0.09 \\ \cline{1-5}
 97.690 & 1.24  $\pm$ 0.04  $\pm$ 0.08 & 0.60  $\pm$ 0.06  $\pm$ 0.03 & 1.13  $\pm$ 0.06  $\pm$ 0.16 & 0.32  $\pm$ 0.06  $\pm$ 0.09 \\ \cline{1-5}
\end{tabular}
}
\end{center}
\end{table*}

\begin{figure}[!ht]
\begin{center}
\includegraphics[bb=0cm 5cm 20cm 25cm, scale=0.365]{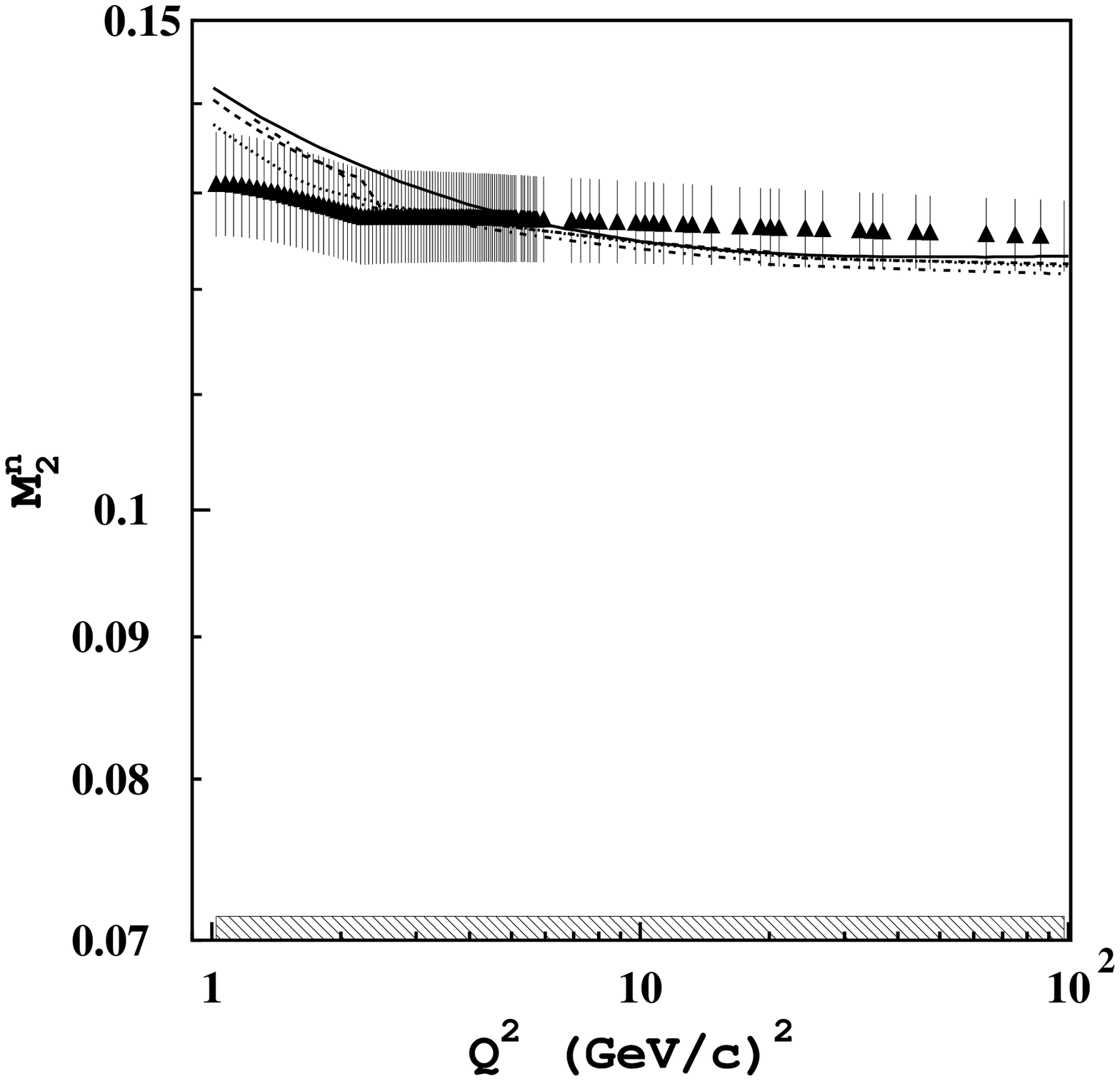}~%
\includegraphics[bb=0cm 5cm 20cm 25cm, scale=0.365]{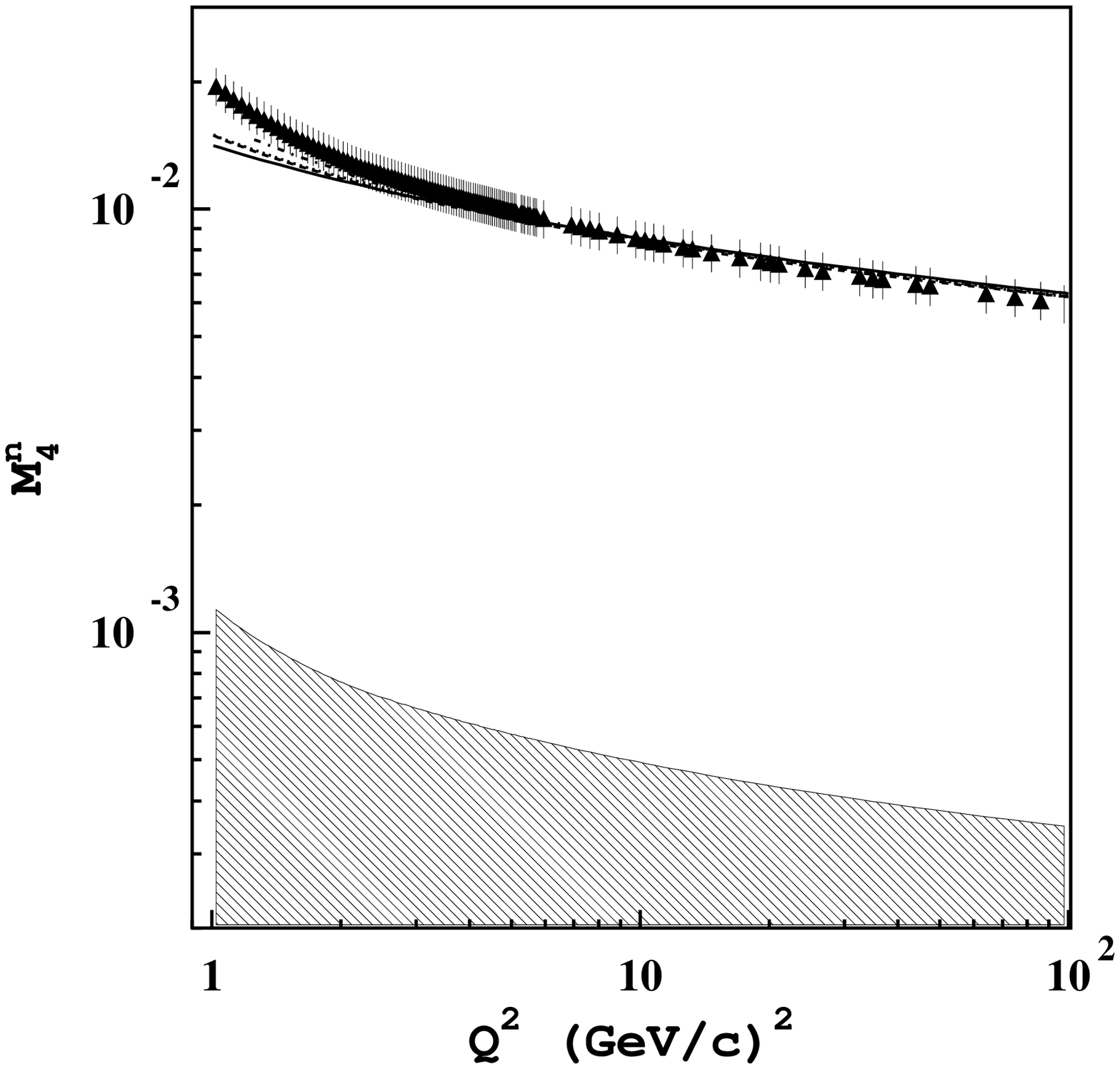}
\includegraphics[bb=0cm 5cm 20cm 25cm, scale=0.365]{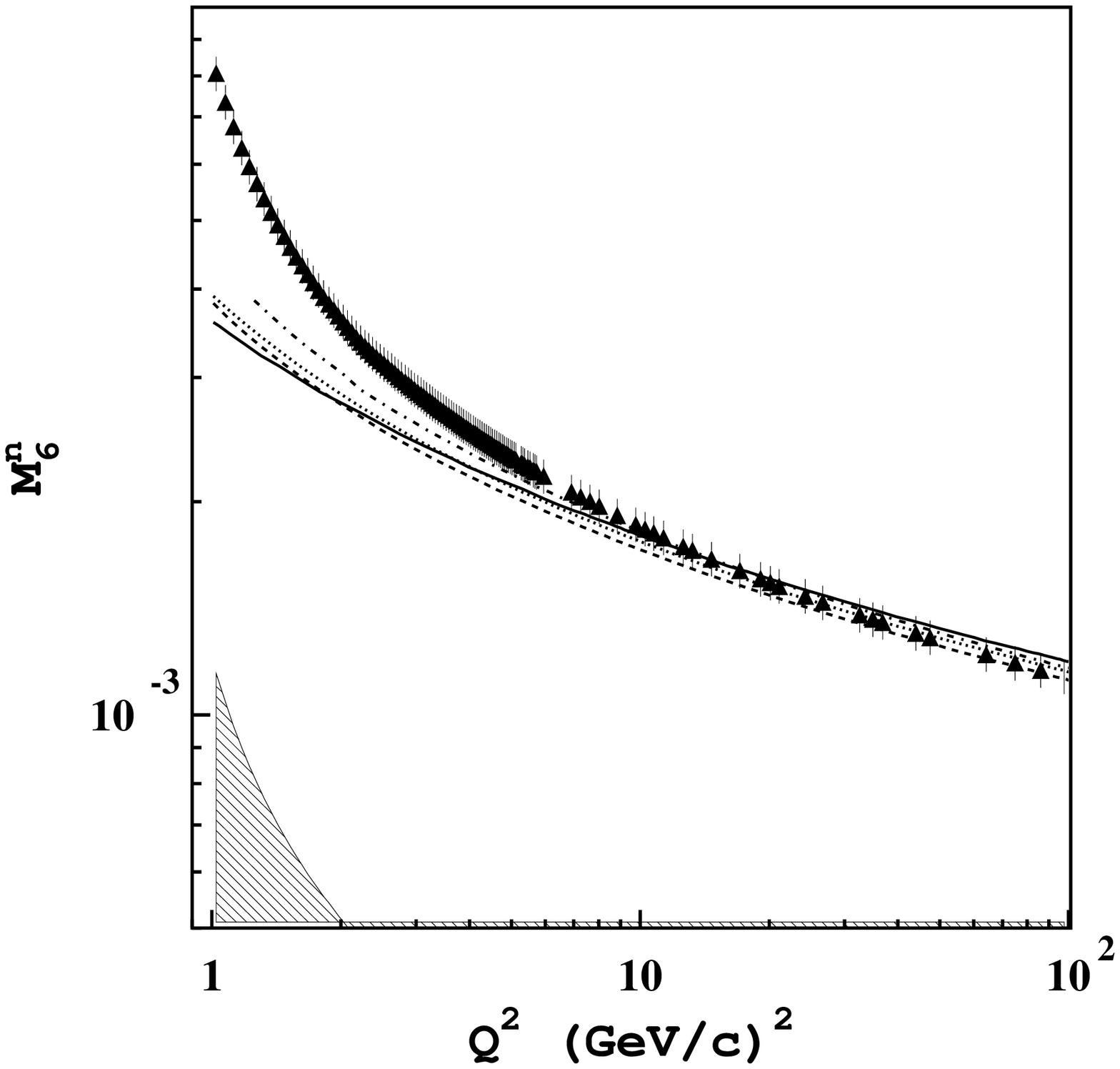}~%
\includegraphics[bb=0cm 5cm 20cm 25cm, scale=0.365]{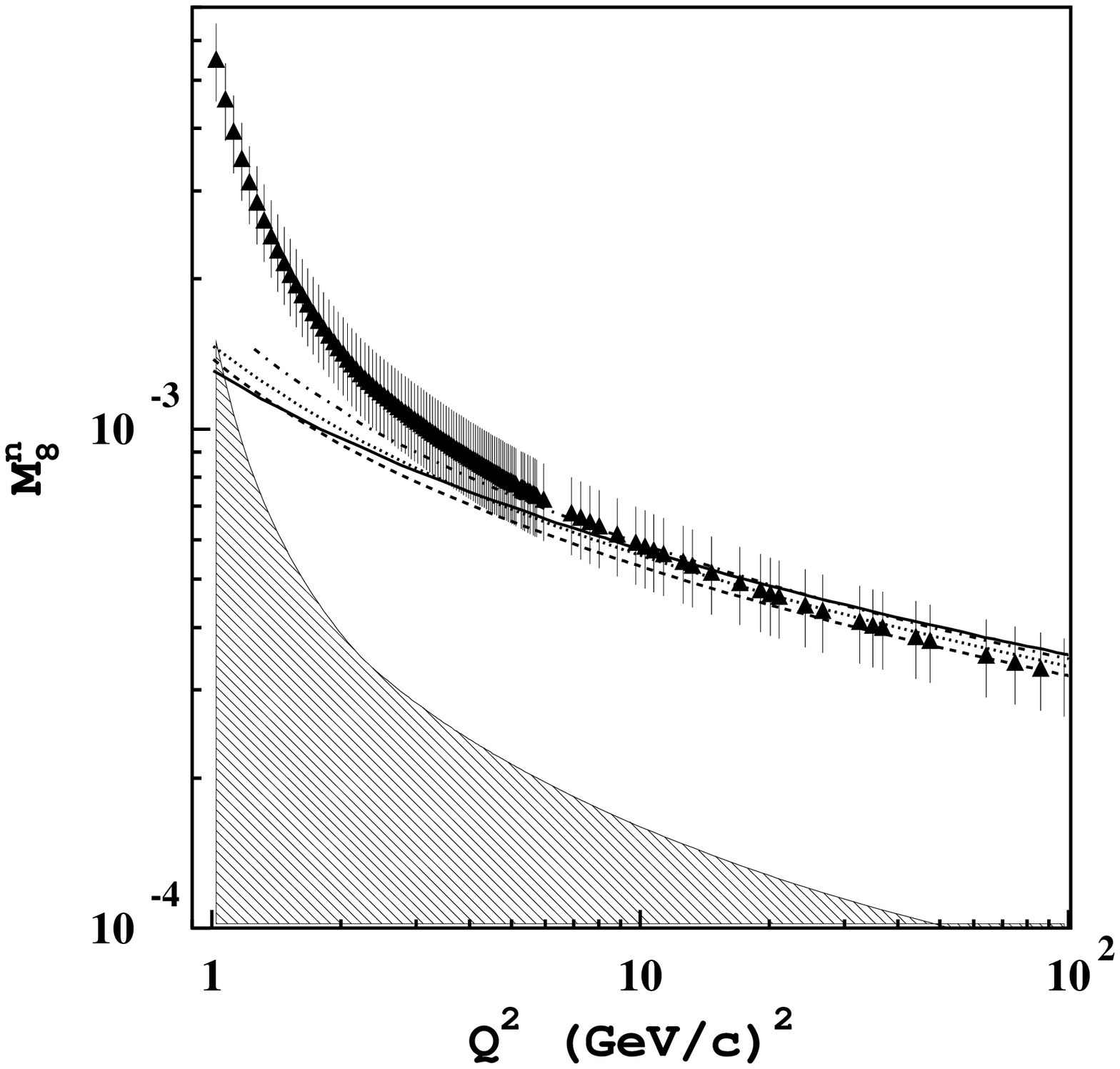}
\caption{\label{fig:neutron_moments_pdfs} \it \small
	Extracted moments of the neutron structure function for $n=2,4,6$ and 8
	as a function of $Q^2$ in comparison to moments of different PDFs:
	solid line represents GRV~\cite{GRV98},
	dashed line represents Alekhin~\cite{AL04},
	dotted line represents CTEQ~\cite{CTEQ04},
	dot-dashed line represents MRST~\cite{MRST04+qed}.
	The hatched area shows systematic errors.}
\end{center}
\end{figure}

In Fig.~\ref{fig:neutron_moments_pdfs} our extracted neutron moments are compared 
with the corresponding moments calculated from various PDF sets available in the 
literature \cite{GRV98,MRST04+qed,CTEQ04,AL04}. For $n > 2$ all PDF sets 
predict significantly lower moments at low $Q^2$. This difference is due to
Sudakov effects at large-$x$; indeed, as was firstly investigated in Ref.~\cite{SIM00}, 
soft gluon resummation plays a relevant role for $n > 2 $ already at $Q^2 = 3-4$~GeV$^2$. 
While resummation effects are taken into account in the analysis of 
Refs.~\cite{osipenko_f2p,osipenko_f2d}, all the PDF sets of
Refs.~\cite{GRV98,MRST04+qed,CTEQ04,AL04} are based on next-to-leading 
(NLO) or NNLO treatments. At higher $Q^2$ the PDF predictions are 
consistent with our results for $n \leq 10$ within the quoted statistical errors. 
For $M_2$ however all PDFs give somehow smaller values at large $Q^2$, 
though within the statistical uncertainties.

The large-$n$ behavior of the moments is also closely related to the
asymptotic $x \to 1$ behavior of the structure functions.
Indeed, inspecting the definition of the moments, Eq.~(\ref{eq:momdef_CN}),
we conclude that at $n \gg 1$ the factor $x^{n-2}$ in the integrand
suppresses the contribution from the region of small $x$, leaving the
moment effectively saturated by a region close to $x=1$.
Furthermore, assuming that the structure functions $F_2^p$ and $F_2^n$ 
vanish as $(1-x)^\beta$ with the same exponent for the proton and 
neutron (as one would expect from perturbative QCD) \cite{LB}, it is
possible to derive the asymptotic relation between the ratios of the
moments and the structure functions
\begin{equation}
\lim_{n\to\infty} \frac{M_n^n}{M_n^p}
= \lim_{x\to 1} \frac{F_2^n(x)}{F_2^p(x)}\ .
\end{equation}
This relation can be useful in obtaining some insights into the large-$x$
asymptotics of the $d/u$ ratio from the large-$n$ limit of the neutron
and proton moments.
Note that most of the PDF fits \cite{GRV98,MRST04+qed,CTEQ04,AL04} are
characterized by the limiting value $\lim_{x\to 1} F_2^n/F_2^p = 1/4$,
which is expected if soft, nonperturbative effects dominate the scattering
at large $x$ (namely, if the energy of the spectator $uu$ system which
accompanies the $d$ quark is increased relative to the $ud$ system which
is relevant for the $u$ quark \cite{Feynman}).
On the other hand, arguments based on helicity conservation in pQCD,
when combined with an unperturbed spin-flavor symmetric wave function,
suggest an asymptotic value of 3/7 \cite{pQCD}.

\begin{figure}[!ht]
\begin{center}
\includegraphics[bb=0cm 5cm 20cm 25cm, scale=0.365]{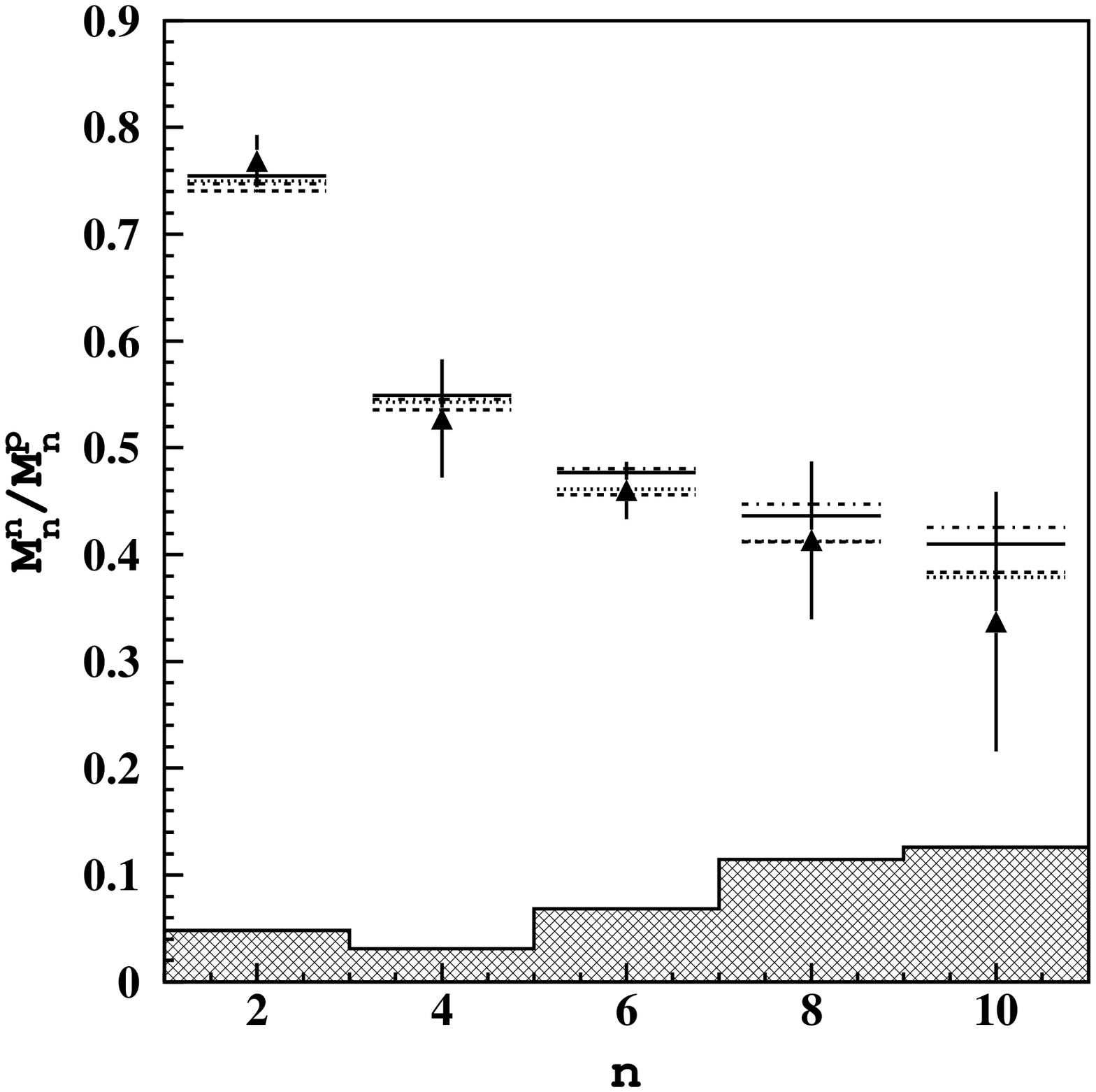}
\caption{\label{fig:np_ratio} \it \small
	Ratio of neutron to proton leading twist moments for various $n$
	(triangles), together with the statistical and systematic 
	uncertainties (hatched area).
	The horizontal lines denote the ratios obtained with different
	PDF sets: GRV \cite{GRV98} (solid), Alekhin \cite{AL04} (dashed),
	CTEQ \cite{CTEQ04} (dotted), MRST \cite{MRST04+qed} (dot-dashed).}
\end{center}
\end{figure}

The ratios of neutron to proton LT moments are shown in
Fig.~\ref{fig:np_ratio} for several $n$, where they are also compared
to various PDF parameterizations \cite{GRV98,MRST04+qed,CTEQ04,AL04}.
The ratio depends only slightly on $Q^2$ for $n = 2$, while it is almost 
completely $Q^2$-independent for larger $n$. 
The latter are in reasonable agreement with our extracted ratios within
both statistical and systematical errors.

The $n/p$ ratio systematically decreases with $n$ and reaches the value
$\simeq 0.34 \pm 0.12_{\rm stat} \pm 0.13_{\rm syst}$ at $n = 10$.
To compare this with the theoretical predictions the mean value of $x$ 
probed by our moments needs to be estimated.
This can be done in the following way: for any calculated $n$-th neutron 
and proton moment we estimate a corresponding average value of $x$
defined as $\langle x \rangle_n
\equiv (M_{n+1}^n + M_{n+1}^p) / (M_n^n + M_n^p)$; 
then the ratio of neutron to proton structure functions is computed at 
the values $\langle x \rangle_n$. 

\begin{figure}[!ht]
\begin{center}
\includegraphics[bb=0cm 5cm 20cm 25cm, scale=0.365]{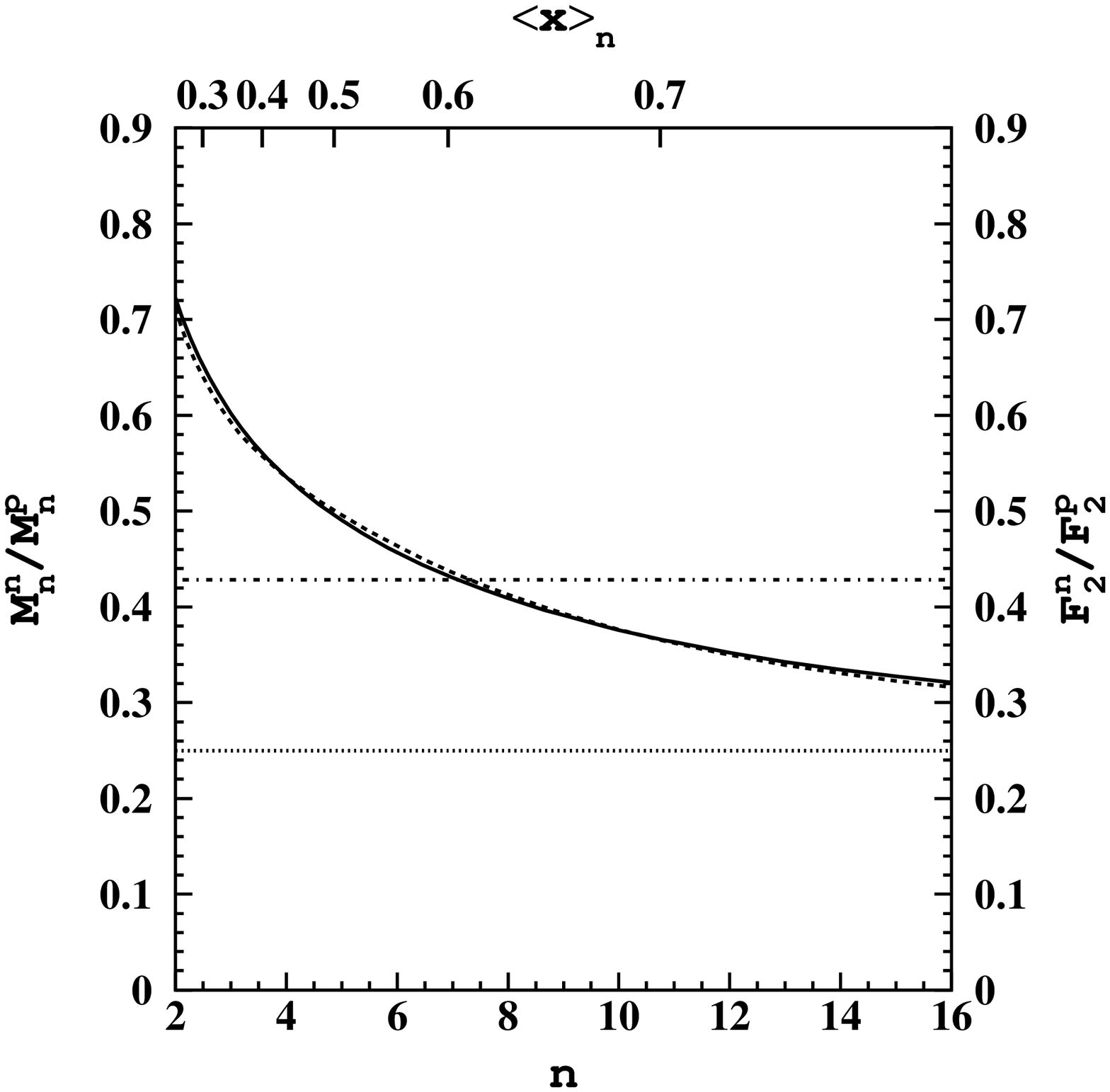}
\caption{\label{fig:np_ratio_pdf} \it \small
	Ratio of extracted neutron to proton leading twist moments as a
	function of $n$ (solid), compared with the structure function 
	ratio (dashed) evaluated using the PDFs of Ref.~\cite{CTEQ04},
	as a function of $\langle x \rangle_n$ (top axis).
	The horizontal lines denote the asymptotic $x\to 1$ values for
	the structure function moments of 1/4 \cite{Feynman} and
	3/7 \cite{pQCD}.}
\end{center}
\end{figure}

The results of this procedure are shown in Fig.~\ref{fig:np_ratio_pdf}
(dashed curve), where they are also compared with the ratio of the
extracted LT moments (solid curve).
The two ratios coincide with good accuracy, suggesting that the extracted 
ratio of moments is similar to the ratio of structure functions taken 
at the average value $\langle x \rangle_n$. 
Note that at $n = 10$ one has $\langle x \rangle_{10} \simeq 0.70$, and the
$F_2^n/F_2^p$ ratio at $x \simeq 0.70$ is equal to
$0.34 \pm 0.12_{\rm stat} \pm 0.13_{\rm syst}$. 
As discussed above (see Fig.~\ref{fig:sys_err}), moments for $n = 12$
may still be accessible by improving measurements of the proton and 
deuteron structure functions at large $x$, and with $n = 12$ one can 
probe values of $x$ as large as 0.75.
However, higher $\langle x \rangle_n$ values are difficult to access
by this method because of the rapid increase of the uncertainties on 
the nuclear corrections with the order $n$.

\begin{figure}[!ht]
\begin{center}
\includegraphics[bb=0cm 5cm 20cm 25cm, scale=0.365]{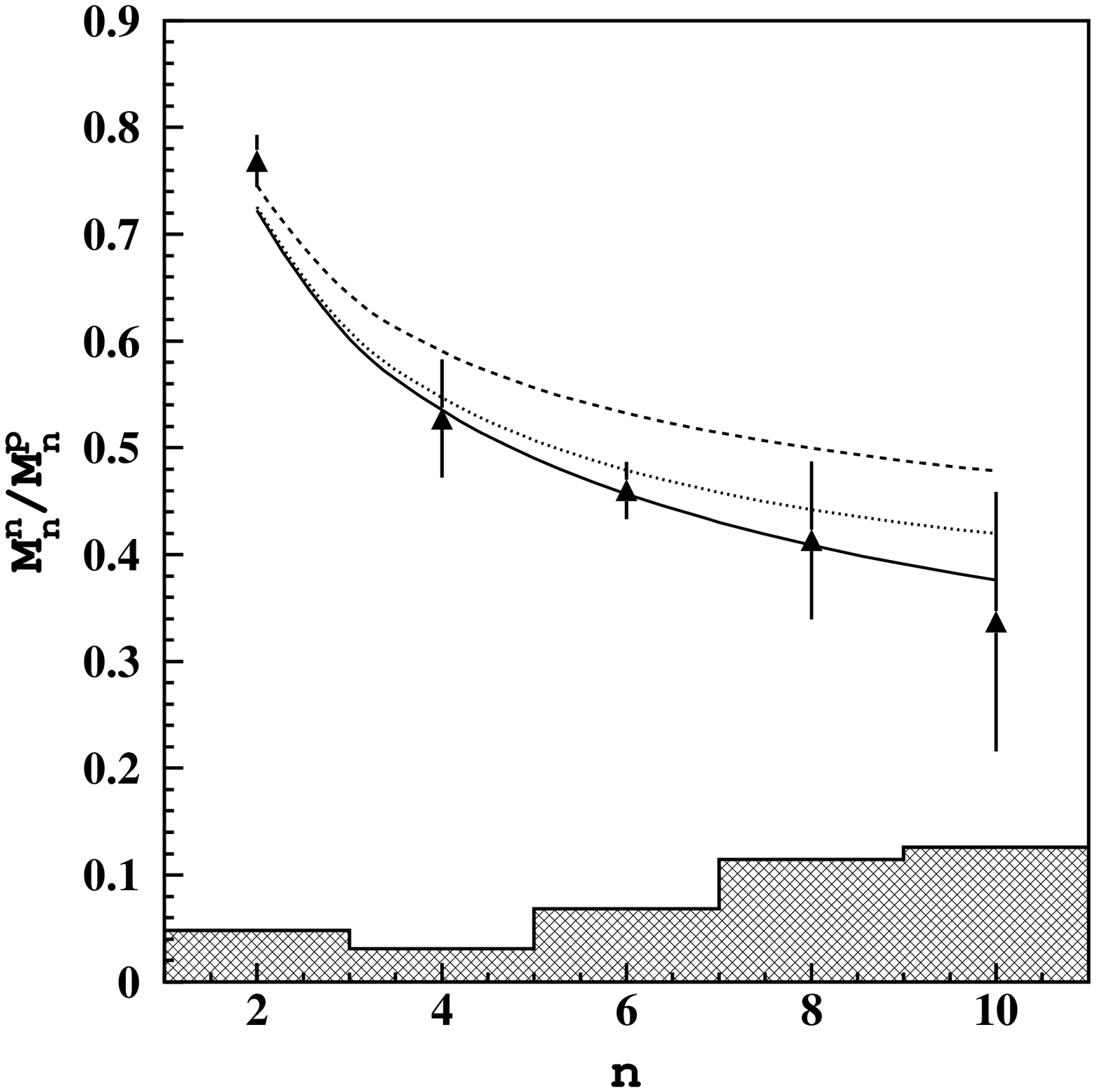}
\caption{\label{fig:d_mod} \it \small
	Ratio of the extracted neutron to proton leading twist moments
	(triangles) as a function of $n$, compared with the corresponding
	PDF parameterization of Ref.~\cite{CTEQ04} (solid).
	The hatched area represents our systematic errors. 
	The dashed and dotted curves correspond to the moment ratio
	evaluated after modifying the $d$-quark distribution according
	to Eq.~(\ref{eq:d_mod}) with $m = 1$ and $m = 4$, respectively.}
\end{center}
\end{figure}

The sensitivity of the moment's ratio to the high-$x$ tail of the PDFs is 
illustrated in Fig.~\ref{fig:d_mod}, where our $n/p$ moment's ratio (triangles)
is compared with the PDF parameterizations from Ref.~\cite{CTEQ04} with 
and without modifying the $d$-quark distribution as: 
\begin{eqnarray}
 d(x) \rightarrow d(x) + 0.1 ~ x^m ~ (1 + x) ~ u(x). 
 \label{eq:d_mod}
\end{eqnarray}
with $m$ a parameter. 
Depending on the value of $m$, the form in Eq.~(\ref{eq:d_mod}) allows
the $x\to 1$ behavior of the $d/u$ ratio to vary between 0 and 1/5,
which correspond respectively to the structure function limits
$\lim_{x\rightarrow 1} F_2^n(x) / F_2^p(x)$ of 1/4 and 3/7.
The dotted line in Fig.~\ref{fig:d_mod} corresponds to the case $m = 1$ 
in Eq.~(\ref{eq:d_mod}), which was advocated in the NNLO analysis of 
both electron and neutrino DIS data of Ref.~\cite{Bodek}.
It appears that our results rule out such an enhancement of the $d$
quark distribution, since the correction is not negligible even at
values of $x$ as low as $x \simeq 0.5$.
A larger value of the parameter $m$ confines the enhancement only to
larger values of $x$, and the results obtained adopting value $m = 4$, 
corresponding to the dashed line in Fig.~\ref{fig:d_mod}, are more 
consistent with our extracted $n/p$ ratio.

Before closing this section, our results for the LT moments of the nonsinglet $p-n$ 
part of the $F_2$ structure function, $M_n^{p-n}(Q^2) = M_n^p(Q^2) - M_n^n(Q^2)$, 
are listed in Table~\ref{table:pn_ltw} together with their statistical and systematic 
uncertainties. We point out that an important fraction of the total systematic error 
for the second moments $M_2^p(Q^2)$ and $M_2^n(Q^2)$ comes from the extrapolation in 
the unmeasured low-$x$ regions. Such systematic uncertainty basically cancels out in 
the isovector $p - n$ combination. This allows us to determine $M_2^{p-n}(Q^2)$ within 
$\approx 20$\% global accuracy.

\begin{table*}
\begin{center}
\caption{\label{table:pn_ltw} \it \small
	Leading twist moments of the nonsinglet $p-n$ part of the $F_2$ structure
	function, $M_n^{p-n}(Q^2) = M_n^p(Q^2) - M_n^n(Q^2)$, together with statistical 
	and systematic uncertainties.} \vspace{3mm}
{\scriptsize
\begin{tabular}{|c|c|c|c|c|} \cline{1-5}
$Q^2~[$(GeV/c)$^2]$
& $M^{p-n}_2(Q^2) \times 10^{-2}$ & $M^{p-n}_4(Q^2) \times 10^{-2}$
& $M^{p-n}_6(Q^2) \times 10^{-3}$ & $M^{p-n}_8(Q^2) \times 10^{-3}$ \\ \cline{1-5}
  1.025 & 6.1  $\pm$ 0.6  $\pm$ 1.0 & 1.74  $\pm$ 0.21  $\pm$ 0.12 & 9.39  $\pm$ 0.53  $\pm$ 0.12 & 7.78  $\pm$ 1.01  $\pm$ 1.66 \\ \cline{1-5}
  1.075 & 6.0  $\pm$ 0.6  $\pm$ 1.0 & 1.67  $\pm$ 0.20  $\pm$ 0.11 & 8.56  $\pm$ 0.48  $\pm$ 0.11 & 6.48  $\pm$ 0.84  $\pm$ 1.38 \\ \cline{1-5}
  1.125 & 6.0  $\pm$ 0.6  $\pm$ 1.0 & 1.62  $\pm$ 0.19  $\pm$ 0.11 & 7.91  $\pm$ 0.44  $\pm$ 0.11 & 5.58  $\pm$ 0.73  $\pm$ 1.19 \\ \cline{1-5}
  1.175 & 5.9  $\pm$ 0.6  $\pm$ 1.0 & 1.57  $\pm$ 0.19  $\pm$ 0.11 & 7.38  $\pm$ 0.41  $\pm$ 0.11 & 4.92  $\pm$ 0.64  $\pm$ 1.05 \\ \cline{1-5}
  1.225 & 5.9  $\pm$ 0.6  $\pm$ 1.0 & 1.52  $\pm$ 0.18  $\pm$ 0.10 & 6.94  $\pm$ 0.39  $\pm$ 0.10 & 4.42  $\pm$ 0.57  $\pm$ 0.94 \\ \cline{1-5}
  1.275 & 5.8  $\pm$ 0.6  $\pm$ 1.0 & 1.48  $\pm$ 0.18  $\pm$ 0.10 & 6.57  $\pm$ 0.37  $\pm$ 0.10 & 4.03  $\pm$ 0.52  $\pm$ 0.86 \\ \cline{1-5}
  1.325 & 5.8  $\pm$ 0.6  $\pm$ 1.0 & 1.45  $\pm$ 0.17  $\pm$ 0.10 & 6.25  $\pm$ 0.35  $\pm$ 0.10 & 3.71  $\pm$ 0.48  $\pm$ 0.79 \\ \cline{1-5}
  1.375 & 5.8  $\pm$ 0.6  $\pm$ 1.0 & 1.42  $\pm$ 0.17  $\pm$ 0.10 & 5.98  $\pm$ 0.34  $\pm$ 0.10 & 3.45  $\pm$ 0.45  $\pm$ 0.74 \\ \cline{1-5}
  1.425 & 5.7  $\pm$ 0.6  $\pm$ 1.0 & 1.39  $\pm$ 0.17  $\pm$ 0.09 & 5.74  $\pm$ 0.32  $\pm$ 0.09 & 3.23  $\pm$ 0.42  $\pm$ 0.69 \\ \cline{1-5}
  1.475 & 5.7  $\pm$ 0.6  $\pm$ 1.0 & 1.36  $\pm$ 0.16  $\pm$ 0.09 & 5.53  $\pm$ 0.31  $\pm$ 0.09 & 3.04  $\pm$ 0.40  $\pm$ 0.65 \\ \cline{1-5}
  1.525 & 5.7  $\pm$ 0.6  $\pm$ 1.0 & 1.34  $\pm$ 0.16  $\pm$ 0.09 & 5.35  $\pm$ 0.30  $\pm$ 0.09 & 2.88  $\pm$ 0.37  $\pm$ 0.62 \\ \cline{1-5}
  1.575 & 5.7  $\pm$ 0.6  $\pm$ 0.9 & 1.31  $\pm$ 0.16  $\pm$ 0.09 & 5.18  $\pm$ 0.29  $\pm$ 0.09 & 2.75  $\pm$ 0.36  $\pm$ 0.59 \\ \cline{1-5}
  1.625 & 5.6  $\pm$ 0.6  $\pm$ 0.9 & 1.29  $\pm$ 0.16  $\pm$ 0.09 & 5.03  $\pm$ 0.28  $\pm$ 0.09 & 2.62  $\pm$ 0.34  $\pm$ 0.56 \\ \cline{1-5}
  1.675 & 5.6  $\pm$ 0.6  $\pm$ 0.9 & 1.27  $\pm$ 0.15  $\pm$ 0.09 & 4.89  $\pm$ 0.27  $\pm$ 0.09 & 2.52  $\pm$ 0.33  $\pm$ 0.54 \\ \cline{1-5}
  1.725 & 5.6  $\pm$ 0.6  $\pm$ 0.9 & 1.26  $\pm$ 0.15  $\pm$ 0.08 & 4.77  $\pm$ 0.27  $\pm$ 0.08 & 2.42  $\pm$ 0.31  $\pm$ 0.52 \\ \cline{1-5}
  1.775 & 5.6  $\pm$ 0.6  $\pm$ 0.9 & 1.24  $\pm$ 0.15  $\pm$ 0.08 & 4.65  $\pm$ 0.26  $\pm$ 0.08 & 2.33  $\pm$ 0.30  $\pm$ 0.50 \\ \cline{1-5}
  1.825 & 5.5  $\pm$ 0.6  $\pm$ 0.9 & 1.22  $\pm$ 0.15  $\pm$ 0.08 & 4.54  $\pm$ 0.26  $\pm$ 0.08 & 2.25  $\pm$ 0.29  $\pm$ 0.48 \\ \cline{1-5}
  1.875 & 5.5  $\pm$ 0.6  $\pm$ 0.9 & 1.21  $\pm$ 0.14  $\pm$ 0.08 & 4.45  $\pm$ 0.25  $\pm$ 0.08 & 2.18  $\pm$ 0.28  $\pm$ 0.47 \\ \cline{1-5}
  1.925 & 5.5  $\pm$ 0.6  $\pm$ 0.9 & 1.19  $\pm$ 0.14  $\pm$ 0.08 & 4.36  $\pm$ 0.24  $\pm$ 0.08 & 2.12  $\pm$ 0.28  $\pm$ 0.45 \\ \cline{1-5}
  1.975 & 5.5  $\pm$ 0.6  $\pm$ 0.9 & 1.18  $\pm$ 0.14  $\pm$ 0.08 & 4.27  $\pm$ 0.24  $\pm$ 0.08 & 2.06  $\pm$ 0.27  $\pm$ 0.44 \\ \cline{1-5}
  2.025 & 5.4  $\pm$ 0.6  $\pm$ 0.9 & 1.17  $\pm$ 0.14  $\pm$ 0.08 & 4.19  $\pm$ 0.24  $\pm$ 0.08 & 2.00  $\pm$ 0.26  $\pm$ 0.43 \\ \cline{1-5}
  2.075 & 5.4  $\pm$ 0.6  $\pm$ 0.9 & 1.16  $\pm$ 0.14  $\pm$ 0.08 & 4.12  $\pm$ 0.23  $\pm$ 0.08 & 1.95  $\pm$ 0.25  $\pm$ 0.42 \\ \cline{1-5}
  2.125 & 5.4  $\pm$ 0.6  $\pm$ 0.9 & 1.15  $\pm$ 0.14  $\pm$ 0.08 & 4.05  $\pm$ 0.23  $\pm$ 0.08 & 1.91  $\pm$ 0.25  $\pm$ 0.41 \\ \cline{1-5}
  2.175 & 5.4  $\pm$ 0.6  $\pm$ 0.9 & 1.13  $\pm$ 0.14  $\pm$ 0.08 & 3.99  $\pm$ 0.22  $\pm$ 0.08 & 1.86  $\pm$ 0.24  $\pm$ 0.40 \\ \cline{1-5}
  2.225 & 5.4  $\pm$ 0.6  $\pm$ 0.9 & 1.12  $\pm$ 0.13  $\pm$ 0.08 & 3.93  $\pm$ 0.22  $\pm$ 0.08 & 1.82  $\pm$ 0.24  $\pm$ 0.39 \\ \cline{1-5}
  2.275 & 5.4  $\pm$ 0.6  $\pm$ 0.9 & 1.11  $\pm$ 0.13  $\pm$ 0.07 & 3.87  $\pm$ 0.22  $\pm$ 0.07 & 1.79  $\pm$ 0.23  $\pm$ 0.38 \\ \cline{1-5}
  2.325 & 5.3  $\pm$ 0.6  $\pm$ 0.9 & 1.11  $\pm$ 0.13  $\pm$ 0.07 & 3.82  $\pm$ 0.21  $\pm$ 0.07 & 1.76  $\pm$ 0.23  $\pm$ 0.38 \\ \cline{1-5}
  2.375 & 5.3  $\pm$ 0.6  $\pm$ 0.9 & 1.10  $\pm$ 0.13  $\pm$ 0.07 & 3.78  $\pm$ 0.21  $\pm$ 0.07 & 1.73  $\pm$ 0.23  $\pm$ 0.37 \\ \cline{1-5}
  2.425 & 5.3  $\pm$ 0.6  $\pm$ 0.9 & 1.09  $\pm$ 0.13  $\pm$ 0.07 & 3.74  $\pm$ 0.21  $\pm$ 0.07 & 1.70  $\pm$ 0.22  $\pm$ 0.36 \\ \cline{1-5}
  2.475 & 5.3  $\pm$ 0.6  $\pm$ 0.9 & 1.08  $\pm$ 0.13  $\pm$ 0.07 & 3.70  $\pm$ 0.21  $\pm$ 0.07 & 1.68  $\pm$ 0.22  $\pm$ 0.36 \\ \cline{1-5}
  2.525 & 5.3  $\pm$ 0.6  $\pm$ 0.9 & 1.08  $\pm$ 0.13  $\pm$ 0.07 & 3.66  $\pm$ 0.21  $\pm$ 0.07 & 1.66  $\pm$ 0.22  $\pm$ 0.35 \\ \cline{1-5}
  2.575 & 5.3  $\pm$ 0.6  $\pm$ 0.9 & 1.07  $\pm$ 0.13  $\pm$ 0.07 & 3.62  $\pm$ 0.20  $\pm$ 0.07 & 1.63  $\pm$ 0.21  $\pm$ 0.35 \\ \cline{1-5}
  2.625 & 5.3  $\pm$ 0.6  $\pm$ 0.9 & 1.06  $\pm$ 0.13  $\pm$ 0.07 & 3.58  $\pm$ 0.20  $\pm$ 0.07 & 1.61  $\pm$ 0.21  $\pm$ 0.34 \\ \cline{1-5}
  2.675 & 5.3  $\pm$ 0.6  $\pm$ 0.9 & 1.06  $\pm$ 0.13  $\pm$ 0.07 & 3.55  $\pm$ 0.20  $\pm$ 0.07 & 1.59  $\pm$ 0.21  $\pm$ 0.34 \\ \cline{1-5}
  2.725 & 5.2  $\pm$ 0.6  $\pm$ 0.9 & 1.05  $\pm$ 0.13  $\pm$ 0.07 & 3.52  $\pm$ 0.20  $\pm$ 0.07 & 1.57  $\pm$ 0.20  $\pm$ 0.33 \\ \cline{1-5}
  2.775 & 5.2  $\pm$ 0.6  $\pm$ 0.9 & 1.04  $\pm$ 0.13  $\pm$ 0.07 & 3.48  $\pm$ 0.20  $\pm$ 0.07 & 1.55  $\pm$ 0.20  $\pm$ 0.33 \\ \cline{1-5}
  2.825 & 5.2  $\pm$ 0.6  $\pm$ 0.9 & 1.04  $\pm$ 0.12  $\pm$ 0.07 & 3.45  $\pm$ 0.19  $\pm$ 0.07 & 1.53  $\pm$ 0.20  $\pm$ 0.33 \\ \cline{1-5}
  2.875 & 5.2  $\pm$ 0.6  $\pm$ 0.9 & 1.03  $\pm$ 0.12  $\pm$ 0.07 & 3.42  $\pm$ 0.19  $\pm$ 0.07 & 1.51  $\pm$ 0.20  $\pm$ 0.32 \\ \cline{1-5}
  2.925 & 5.2  $\pm$ 0.6  $\pm$ 0.9 & 1.03  $\pm$ 0.12  $\pm$ 0.07 & 3.40  $\pm$ 0.19  $\pm$ 0.07 & 1.50  $\pm$ 0.19  $\pm$ 0.32 \\ \cline{1-5}
  2.975 & 5.2  $\pm$ 0.6  $\pm$ 0.9 & 1.02  $\pm$ 0.12  $\pm$ 0.07 & 3.37  $\pm$ 0.19  $\pm$ 0.07 & 1.48  $\pm$ 0.19  $\pm$ 0.32 \\ \cline{1-5}
  3.025 & 5.2  $\pm$ 0.5  $\pm$ 0.9 & 1.02  $\pm$ 0.12  $\pm$ 0.07 & 3.34  $\pm$ 0.19  $\pm$ 0.07 & 1.47  $\pm$ 0.19  $\pm$ 0.31 \\ \cline{1-5}
  3.075 & 5.2  $\pm$ 0.5  $\pm$ 0.9 & 1.01  $\pm$ 0.12  $\pm$ 0.07 & 3.32  $\pm$ 0.19  $\pm$ 0.07 & 1.45  $\pm$ 0.19  $\pm$ 0.31 \\ \cline{1-5}
  3.125 & 5.2  $\pm$ 0.5  $\pm$ 0.9 & 1.01  $\pm$ 0.12  $\pm$ 0.07 & 3.29  $\pm$ 0.18  $\pm$ 0.07 & 1.44  $\pm$ 0.19  $\pm$ 0.31 \\ \cline{1-5}
  3.175 & 5.1  $\pm$ 0.5  $\pm$ 0.9 & 1.00  $\pm$ 0.12  $\pm$ 0.07 & 3.27  $\pm$ 0.18  $\pm$ 0.07 & 1.42  $\pm$ 0.18  $\pm$ 0.30 \\ \cline{1-5}
  3.225 & 5.1  $\pm$ 0.5  $\pm$ 0.9 & 1.00  $\pm$ 0.12  $\pm$ 0.07 & 3.24  $\pm$ 0.18  $\pm$ 0.07 & 1.41  $\pm$ 0.18  $\pm$ 0.30 \\ \cline{1-5}
  3.275 & 5.1  $\pm$ 0.5  $\pm$ 0.9 & 0.99  $\pm$ 0.12  $\pm$ 0.07 & 3.22  $\pm$ 0.18  $\pm$ 0.07 & 1.40  $\pm$ 0.18  $\pm$ 0.30 \\ \cline{1-5}
  3.325 & 5.1  $\pm$ 0.5  $\pm$ 0.9 & 0.99  $\pm$ 0.12  $\pm$ 0.07 & 3.20  $\pm$ 0.18  $\pm$ 0.07 & 1.38  $\pm$ 0.18  $\pm$ 0.29 \\ \cline{1-5}
  3.375 & 5.1  $\pm$ 0.5  $\pm$ 0.8 & 0.98  $\pm$ 0.12  $\pm$ 0.07 & 3.18  $\pm$ 0.18  $\pm$ 0.07 & 1.37  $\pm$ 0.18  $\pm$ 0.29 \\ \cline{1-5}
  3.425 & 5.1  $\pm$ 0.5  $\pm$ 0.8 & 0.98  $\pm$ 0.12  $\pm$ 0.07 & 3.16  $\pm$ 0.18  $\pm$ 0.07 & 1.36  $\pm$ 0.18  $\pm$ 0.29 \\ \cline{1-5}
  3.475 & 5.1  $\pm$ 0.5  $\pm$ 0.8 & 0.98  $\pm$ 0.12  $\pm$ 0.07 & 3.14  $\pm$ 0.18  $\pm$ 0.07 & 1.35  $\pm$ 0.18  $\pm$ 0.29 \\ \cline{1-5}
  3.525 & 5.1  $\pm$ 0.5  $\pm$ 0.8 & 0.97  $\pm$ 0.12  $\pm$ 0.07 & 3.12  $\pm$ 0.17  $\pm$ 0.07 & 1.34  $\pm$ 0.17  $\pm$ 0.28 \\ \cline{1-5}
  3.575 & 5.1  $\pm$ 0.5  $\pm$ 0.8 & 0.97  $\pm$ 0.12  $\pm$ 0.07 & 3.10  $\pm$ 0.17  $\pm$ 0.07 & 1.32  $\pm$ 0.17  $\pm$ 0.28 \\ \cline{1-5}
  3.625 & 5.1  $\pm$ 0.5  $\pm$ 0.8 & 0.96  $\pm$ 0.12  $\pm$ 0.06 & 3.08  $\pm$ 0.17  $\pm$ 0.06 & 1.31  $\pm$ 0.17  $\pm$ 0.28 \\ \cline{1-5}
  3.675 & 5.1  $\pm$ 0.5  $\pm$ 0.8 & 0.96  $\pm$ 0.11  $\pm$ 0.06 & 3.06  $\pm$ 0.17  $\pm$ 0.06 & 1.30  $\pm$ 0.17  $\pm$ 0.28 \\ \cline{1-5}
  3.725 & 5.1  $\pm$ 0.5  $\pm$ 0.8 & 0.96  $\pm$ 0.11  $\pm$ 0.06 & 3.04  $\pm$ 0.17  $\pm$ 0.06 & 1.29  $\pm$ 0.17  $\pm$ 0.28 \\ \cline{1-5}
  3.775 & 5.0  $\pm$ 0.5  $\pm$ 0.8 & 0.95  $\pm$ 0.11  $\pm$ 0.06 & 3.03  $\pm$ 0.17  $\pm$ 0.06 & 1.28  $\pm$ 0.17  $\pm$ 0.27 \\ \cline{1-5}
  3.825 & 5.0  $\pm$ 0.5  $\pm$ 0.8 & 0.95  $\pm$ 0.11  $\pm$ 0.06 & 3.01  $\pm$ 0.17  $\pm$ 0.06 & 1.28  $\pm$ 0.17  $\pm$ 0.27 \\ \cline{1-5}
\end{tabular}
}
\end{center}
\end{table*}
\begin{table*}
\begin{center}
{\scriptsize
\begin{tabular}{|c|c|c|c|c|} \cline{1-5}
$Q^2~[$(GeV/c)$^2]$
& $M^{p-n}_2(Q^2) \times 10^{-2}$ & $M^{p-n}_4(Q^2) \times 10^{-2}$
& $M^{p-n}_6(Q^2) \times 10^{-3}$ & $M^{p-n}_8(Q^2) \times 10^{-3}$ \\ \cline{1-5}
  3.875 & 5.0  $\pm$ 0.5  $\pm$ 0.8 & 0.95  $\pm$ 0.11  $\pm$ 0.06 & 2.99  $\pm$ 0.17  $\pm$ 0.06 & 1.27  $\pm$ 0.16  $\pm$ 0.27 \\ \cline{1-5}
  3.925 & 5.0  $\pm$ 0.5  $\pm$ 0.8 & 0.94  $\pm$ 0.11  $\pm$ 0.06 & 2.98  $\pm$ 0.17  $\pm$ 0.06 & 1.26  $\pm$ 0.16  $\pm$ 0.27 \\ \cline{1-5}
  3.975 & 5.0  $\pm$ 0.5  $\pm$ 0.8 & 0.94  $\pm$ 0.11  $\pm$ 0.06 & 2.96  $\pm$ 0.17  $\pm$ 0.06 & 1.25  $\pm$ 0.16  $\pm$ 0.27 \\ \cline{1-5}
  4.025 & 5.0  $\pm$ 0.5  $\pm$ 0.8 & 0.94  $\pm$ 0.11  $\pm$ 0.06 & 2.95  $\pm$ 0.17  $\pm$ 0.06 & 1.24  $\pm$ 0.16  $\pm$ 0.26 \\ \cline{1-5}
  4.075 & 5.0  $\pm$ 0.5  $\pm$ 0.8 & 0.93  $\pm$ 0.11  $\pm$ 0.06 & 2.93  $\pm$ 0.16  $\pm$ 0.06 & 1.23  $\pm$ 0.16  $\pm$ 0.26 \\ \cline{1-5}
  4.125 & 5.0  $\pm$ 0.5  $\pm$ 0.8 & 0.93  $\pm$ 0.11  $\pm$ 0.06 & 2.92  $\pm$ 0.16  $\pm$ 0.06 & 1.22  $\pm$ 0.16  $\pm$ 0.26 \\ \cline{1-5}
  4.175 & 5.0  $\pm$ 0.5  $\pm$ 0.8 & 0.93  $\pm$ 0.11  $\pm$ 0.06 & 2.90  $\pm$ 0.16  $\pm$ 0.06 & 1.22  $\pm$ 0.16  $\pm$ 0.26 \\ \cline{1-5}
  4.225 & 5.0  $\pm$ 0.5  $\pm$ 0.8 & 0.92  $\pm$ 0.11  $\pm$ 0.06 & 2.89  $\pm$ 0.16  $\pm$ 0.06 & 1.21  $\pm$ 0.16  $\pm$ 0.26 \\ \cline{1-5}
  4.275 & 5.0  $\pm$ 0.5  $\pm$ 0.8 & 0.92  $\pm$ 0.11  $\pm$ 0.06 & 2.87  $\pm$ 0.16  $\pm$ 0.06 & 1.20  $\pm$ 0.16  $\pm$ 0.26 \\ \cline{1-5}
  4.325 & 5.0  $\pm$ 0.5  $\pm$ 0.8 & 0.92  $\pm$ 0.11  $\pm$ 0.06 & 2.86  $\pm$ 0.16  $\pm$ 0.06 & 1.19  $\pm$ 0.16  $\pm$ 0.25 \\ \cline{1-5}
  4.375 & 5.0  $\pm$ 0.5  $\pm$ 0.8 & 0.92  $\pm$ 0.11  $\pm$ 0.06 & 2.85  $\pm$ 0.16  $\pm$ 0.06 & 1.19  $\pm$ 0.15  $\pm$ 0.25 \\ \cline{1-5}
  4.425 & 5.0  $\pm$ 0.5  $\pm$ 0.8 & 0.91  $\pm$ 0.11  $\pm$ 0.06 & 2.84  $\pm$ 0.16  $\pm$ 0.06 & 1.18  $\pm$ 0.15  $\pm$ 0.25 \\ \cline{1-5}
  4.475 & 5.0  $\pm$ 0.5  $\pm$ 0.8 & 0.91  $\pm$ 0.11  $\pm$ 0.06 & 2.82  $\pm$ 0.16  $\pm$ 0.06 & 1.17  $\pm$ 0.15  $\pm$ 0.25 \\ \cline{1-5}
  4.525 & 4.9  $\pm$ 0.5  $\pm$ 0.8 & 0.91  $\pm$ 0.11  $\pm$ 0.06 & 2.81  $\pm$ 0.16  $\pm$ 0.06 & 1.17  $\pm$ 0.15  $\pm$ 0.25 \\ \cline{1-5}
  4.575 & 4.9  $\pm$ 0.5  $\pm$ 0.8 & 0.91  $\pm$ 0.11  $\pm$ 0.06 & 2.80  $\pm$ 0.16  $\pm$ 0.06 & 1.16  $\pm$ 0.15  $\pm$ 0.25 \\ \cline{1-5}
  4.625 & 4.9  $\pm$ 0.5  $\pm$ 0.8 & 0.90  $\pm$ 0.11  $\pm$ 0.06 & 2.79  $\pm$ 0.16  $\pm$ 0.06 & 1.15  $\pm$ 0.15  $\pm$ 0.25 \\ \cline{1-5}
  4.675 & 4.9  $\pm$ 0.5  $\pm$ 0.8 & 0.90  $\pm$ 0.11  $\pm$ 0.06 & 2.78  $\pm$ 0.16  $\pm$ 0.06 & 1.15  $\pm$ 0.15  $\pm$ 0.25 \\ \cline{1-5}
  4.725 & 4.9  $\pm$ 0.5  $\pm$ 0.8 & 0.90  $\pm$ 0.11  $\pm$ 0.06 & 2.76  $\pm$ 0.16  $\pm$ 0.06 & 1.14  $\pm$ 0.15  $\pm$ 0.24 \\ \cline{1-5}
  4.775 & 4.9  $\pm$ 0.5  $\pm$ 0.8 & 0.90  $\pm$ 0.11  $\pm$ 0.06 & 2.75  $\pm$ 0.15  $\pm$ 0.06 & 1.14  $\pm$ 0.15  $\pm$ 0.24 \\ \cline{1-5}
  4.825 & 4.9  $\pm$ 0.5  $\pm$ 0.8 & 0.89  $\pm$ 0.11  $\pm$ 0.06 & 2.74  $\pm$ 0.15  $\pm$ 0.06 & 1.13  $\pm$ 0.15  $\pm$ 0.24 \\ \cline{1-5}
  4.875 & 4.9  $\pm$ 0.5  $\pm$ 0.8 & 0.89  $\pm$ 0.11  $\pm$ 0.06 & 2.73  $\pm$ 0.15  $\pm$ 0.06 & 1.13  $\pm$ 0.15  $\pm$ 0.24 \\ \cline{1-5}
  4.925 & 4.9  $\pm$ 0.5  $\pm$ 0.8 & 0.89  $\pm$ 0.11  $\pm$ 0.06 & 2.72  $\pm$ 0.15  $\pm$ 0.06 & 1.12  $\pm$ 0.15  $\pm$ 0.24 \\ \cline{1-5}
  4.975 & 4.9  $\pm$ 0.5  $\pm$ 0.8 & 0.89  $\pm$ 0.11  $\pm$ 0.06 & 2.71  $\pm$ 0.15  $\pm$ 0.06 & 1.11  $\pm$ 0.14  $\pm$ 0.24 \\ \cline{1-5}
  5.025 & 4.9  $\pm$ 0.5  $\pm$ 0.8 & 0.88  $\pm$ 0.11  $\pm$ 0.06 & 2.70  $\pm$ 0.15  $\pm$ 0.06 & 1.11  $\pm$ 0.14  $\pm$ 0.24 \\ \cline{1-5}
  5.075 & 4.9  $\pm$ 0.5  $\pm$ 0.8 & 0.88  $\pm$ 0.11  $\pm$ 0.06 & 2.69  $\pm$ 0.15  $\pm$ 0.06 & 1.10  $\pm$ 0.14  $\pm$ 0.24 \\ \cline{1-5}
  5.125 & 4.9  $\pm$ 0.5  $\pm$ 0.8 & 0.88  $\pm$ 0.11  $\pm$ 0.06 & 2.68  $\pm$ 0.15  $\pm$ 0.06 & 1.10  $\pm$ 0.14  $\pm$ 0.23 \\ \cline{1-5}
  5.275 & 4.9  $\pm$ 0.5  $\pm$ 0.8 & 0.87  $\pm$ 0.10  $\pm$ 0.06 & 2.65  $\pm$ 0.15  $\pm$ 0.06 & 1.08  $\pm$ 0.14  $\pm$ 0.23 \\ \cline{1-5}
  5.325 & 4.9  $\pm$ 0.5  $\pm$ 0.8 & 0.87  $\pm$ 0.10  $\pm$ 0.06 & 2.64  $\pm$ 0.15  $\pm$ 0.06 & 1.08  $\pm$ 0.14  $\pm$ 0.23 \\ \cline{1-5}
  5.375 & 4.9  $\pm$ 0.5  $\pm$ 0.8 & 0.87  $\pm$ 0.10  $\pm$ 0.06 & 2.64  $\pm$ 0.15  $\pm$ 0.06 & 1.07  $\pm$ 0.14  $\pm$ 0.23 \\ \cline{1-5}
  5.475 & 4.9  $\pm$ 0.5  $\pm$ 0.8 & 0.87  $\pm$ 0.10  $\pm$ 0.06 & 2.62  $\pm$ 0.15  $\pm$ 0.06 & 1.07  $\pm$ 0.14  $\pm$ 0.23 \\ \cline{1-5}
  5.525 & 4.8  $\pm$ 0.5  $\pm$ 0.8 & 0.86  $\pm$ 0.10  $\pm$ 0.06 & 2.61  $\pm$ 0.15  $\pm$ 0.06 & 1.06  $\pm$ 0.14  $\pm$ 0.23 \\ \cline{1-5}
  5.625 & 4.8  $\pm$ 0.5  $\pm$ 0.8 & 0.86  $\pm$ 0.10  $\pm$ 0.06 & 2.59  $\pm$ 0.15  $\pm$ 0.06 & 1.05  $\pm$ 0.14  $\pm$ 0.22 \\ \cline{1-5}
  5.675 & 4.8  $\pm$ 0.5  $\pm$ 0.8 & 0.86  $\pm$ 0.10  $\pm$ 0.06 & 2.59  $\pm$ 0.15  $\pm$ 0.06 & 1.05  $\pm$ 0.14  $\pm$ 0.22 \\ \cline{1-5}
  5.725 & 4.8  $\pm$ 0.5  $\pm$ 0.8 & 0.86  $\pm$ 0.10  $\pm$ 0.06 & 2.58  $\pm$ 0.14  $\pm$ 0.06 & 1.04  $\pm$ 0.14  $\pm$ 0.22 \\ \cline{1-5}
  5.955 & 4.8  $\pm$ 0.5  $\pm$ 0.8 & 0.85  $\pm$ 0.10  $\pm$ 0.06 & 2.54  $\pm$ 0.14  $\pm$ 0.06 & 1.03  $\pm$ 0.13  $\pm$ 0.22 \\ \cline{1-5}
  6.915 & 4.7  $\pm$ 0.5  $\pm$ 0.8 & 0.82  $\pm$ 0.10  $\pm$ 0.06 & 2.42  $\pm$ 0.14  $\pm$ 0.06 & 0.96  $\pm$ 0.12  $\pm$ 0.21 \\ \cline{1-5}
  7.267 & 4.7  $\pm$ 0.5  $\pm$ 0.8 & 0.81  $\pm$ 0.10  $\pm$ 0.05 & 2.38  $\pm$ 0.13  $\pm$ 0.05 & 0.94  $\pm$ 0.12  $\pm$ 0.20 \\ \cline{1-5}
  7.630 & 4.7  $\pm$ 0.5  $\pm$ 0.8 & 0.80  $\pm$ 0.10  $\pm$ 0.05 & 2.34  $\pm$ 0.13  $\pm$ 0.05 & 0.92  $\pm$ 0.12  $\pm$ 0.20 \\ \cline{1-5}
  8.021 & 4.7  $\pm$ 0.5  $\pm$ 0.8 & 0.79  $\pm$ 0.10  $\pm$ 0.05 & 2.31  $\pm$ 0.13  $\pm$ 0.05 & 0.91  $\pm$ 0.12  $\pm$ 0.19 \\ \cline{1-5}
  8.847 & 4.6  $\pm$ 0.5  $\pm$ 0.8 & 0.78  $\pm$ 0.09  $\pm$ 0.05 & 2.24  $\pm$ 0.13  $\pm$ 0.05 & 0.87  $\pm$ 0.11  $\pm$ 0.19 \\ \cline{1-5}
  9.775 & 4.6  $\pm$ 0.5  $\pm$ 0.8 & 0.76  $\pm$ 0.09  $\pm$ 0.05 & 2.17  $\pm$ 0.12  $\pm$ 0.05 & 0.84  $\pm$ 0.11  $\pm$ 0.18 \\ \cline{1-5}
 10.267 & 4.6  $\pm$ 0.5  $\pm$ 0.8 & 0.76  $\pm$ 0.09  $\pm$ 0.05 & 2.14  $\pm$ 0.12  $\pm$ 0.05 & 0.83  $\pm$ 0.11  $\pm$ 0.18 \\ \cline{1-5}
 10.762 & 4.6  $\pm$ 0.5  $\pm$ 0.8 & 0.75  $\pm$ 0.09  $\pm$ 0.05 & 2.12  $\pm$ 0.12  $\pm$ 0.05 & 0.81  $\pm$ 0.11  $\pm$ 0.17 \\ \cline{1-5}
 11.344 & 4.5  $\pm$ 0.5  $\pm$ 0.8 & 0.74  $\pm$ 0.09  $\pm$ 0.05 & 2.08  $\pm$ 0.12  $\pm$ 0.05 & 0.80  $\pm$ 0.10  $\pm$ 0.17 \\ \cline{1-5}
 12.580 & 4.5  $\pm$ 0.5  $\pm$ 0.7 & 0.73  $\pm$ 0.09  $\pm$ 0.05 & 2.03  $\pm$ 0.11  $\pm$ 0.05 & 0.77  $\pm$ 0.10  $\pm$ 0.16 \\ \cline{1-5}
 13.238 & 4.5  $\pm$ 0.5  $\pm$ 0.7 & 0.72  $\pm$ 0.09  $\pm$ 0.05 & 2.00  $\pm$ 0.11  $\pm$ 0.05 & 0.76  $\pm$ 0.10  $\pm$ 0.16 \\ \cline{1-5}
 14.689 & 4.4  $\pm$ 0.5  $\pm$ 0.7 & 0.71  $\pm$ 0.08  $\pm$ 0.05 & 1.95  $\pm$ 0.11  $\pm$ 0.05 & 0.73  $\pm$ 0.09  $\pm$ 0.16 \\ \cline{1-5}
 17.108 & 4.4  $\pm$ 0.5  $\pm$ 0.7 & 0.69  $\pm$ 0.08  $\pm$ 0.05 & 1.87  $\pm$ 0.11  $\pm$ 0.05 & 0.70  $\pm$ 0.09  $\pm$ 0.15 \\ \cline{1-5}
 19.072 & 4.3  $\pm$ 0.5  $\pm$ 0.7 & 0.67  $\pm$ 0.08  $\pm$ 0.05 & 1.82  $\pm$ 0.10  $\pm$ 0.05 & 0.67  $\pm$ 0.09  $\pm$ 0.14 \\ \cline{1-5}
 20.108 & 4.3  $\pm$ 0.5  $\pm$ 0.7 & 0.67  $\pm$ 0.08  $\pm$ 0.04 & 1.80  $\pm$ 0.10  $\pm$ 0.04 & 0.66  $\pm$ 0.09  $\pm$ 0.14 \\ \cline{1-5}
 21.097 & 4.3  $\pm$ 0.5  $\pm$ 0.7 & 0.66  $\pm$ 0.08  $\pm$ 0.04 & 1.78  $\pm$ 0.10  $\pm$ 0.04 & 0.66  $\pm$ 0.09  $\pm$ 0.14 \\ \cline{1-5}
 24.259 & 4.3  $\pm$ 0.5  $\pm$ 0.7 & 0.65  $\pm$ 0.08  $\pm$ 0.04 & 1.73  $\pm$ 0.10  $\pm$ 0.04 & 0.63  $\pm$ 0.08  $\pm$ 0.13 \\ \cline{1-5}
 26.680 & 4.2  $\pm$ 0.5  $\pm$ 0.7 & 0.64  $\pm$ 0.08  $\pm$ 0.04 & 1.69  $\pm$ 0.09  $\pm$ 0.04 & 0.61  $\pm$ 0.08  $\pm$ 0.13 \\ \cline{1-5}
 32.500 & 4.2  $\pm$ 0.4  $\pm$ 0.7 & 0.62  $\pm$ 0.07  $\pm$ 0.04 & 1.62  $\pm$ 0.09  $\pm$ 0.04 & 0.58  $\pm$ 0.08  $\pm$ 0.12 \\ \cline{1-5}
 34.932 & 4.2  $\pm$ 0.4  $\pm$ 0.7 & 0.61  $\pm$ 0.07  $\pm$ 0.04 & 1.60  $\pm$ 0.09  $\pm$ 0.04 & 0.57  $\pm$ 0.07  $\pm$ 0.12 \\ \cline{1-5}
 36.750 & 4.1  $\pm$ 0.4  $\pm$ 0.7 & 0.61  $\pm$ 0.07  $\pm$ 0.04 & 1.58  $\pm$ 0.09  $\pm$ 0.04 & 0.57  $\pm$ 0.07  $\pm$ 0.12 \\ \cline{1-5}
 43.970 & 4.1  $\pm$ 0.4  $\pm$ 0.7 & 0.60  $\pm$ 0.07  $\pm$ 0.04 & 1.53  $\pm$ 0.09  $\pm$ 0.04 & 0.54  $\pm$ 0.07  $\pm$ 0.12 \\ \cline{1-5}
 47.440 & 4.1  $\pm$ 0.4  $\pm$ 0.7 & 0.59  $\pm$ 0.07  $\pm$ 0.04 & 1.51  $\pm$ 0.08  $\pm$ 0.04 & 0.54  $\pm$ 0.07  $\pm$ 0.11 \\ \cline{1-5}
 64.270 & 4.0  $\pm$ 0.4  $\pm$ 0.7 & 0.57  $\pm$ 0.07  $\pm$ 0.04 & 1.43  $\pm$ 0.08  $\pm$ 0.04 & 0.50  $\pm$ 0.06  $\pm$ 0.11 \\ \cline{1-5}
 75.000 & 4.0  $\pm$ 0.4  $\pm$ 0.7 & 0.55  $\pm$ 0.07  $\pm$ 0.04 & 1.39  $\pm$ 0.08  $\pm$ 0.04 & 0.48  $\pm$ 0.06  $\pm$ 0.10 \\ \cline{1-5}
 86.000 & 3.9  $\pm$ 0.4  $\pm$ 0.7 & 0.55  $\pm$ 0.07  $\pm$ 0.04 & 1.36  $\pm$ 0.08  $\pm$ 0.04 & 0.47  $\pm$ 0.06  $\pm$ 0.10 \\ \cline{1-5}
 97.690 & 3.9  $\pm$ 0.4  $\pm$ 0.7 & 0.54  $\pm$ 0.06  $\pm$ 0.04 & 1.33  $\pm$ 0.07  $\pm$ 0.04 & 0.46  $\pm$ 0.06  $\pm$ 0.10 \\ \cline{1-5}
\end{tabular}
}
\end{center}
\end{table*}

Low-order moments of the isovector $u - d$ quark distributions have been recently calculated 
in both unquenched \cite{Dolgov} and quenched \cite{Goeckeler} lattice QCD. Our results for 
the nonsinglet LT moments $M_n^{p-n}$ can be easily compared with those of the isovector 
moments $( \langle x^{n-1} \rangle_u - \langle x^{n-1} \rangle_d )$, taking into account that for even values of the 
order $n$ one has
\begin{eqnarray}
 M_n^{p-n}(Q^2) = \frac{1}{3} C_n^{NS} \left( \langle x^{n-1} \rangle_u - 
 \langle x^{n-1} \rangle_d  \right) ~ , 
 \label{eq:NS}
\end{eqnarray}
where $C_n^{NS}$ is the ($n-1$)-th moment of the quark coefficient function. At the scale 
$Q^2 = 4$~GeV$^2$ and in the $\overline{MS}$ renormalization scheme at NLO, which are 
the scale and the scheme adopted in Refs.~\cite{Dolgov,Goeckeler}, one gets $C_2^{NS} 
\simeq 1.01$ and $C_4^{NS} \simeq 1.15$.

The quark masses accessible in present lattice calculations are still relatively high 
and thus the extrapolation to the physical point is one of the open issues in lattice 
applications to light hadron phenomenology. In Refs.~\cite{Dolgov,Goeckeler} the 
lattice data are extrapolated linearly in the quark masses up to the physical point, 
while in Ref.~\cite{Detmold} the extrapolation includes the effects of meson loops and 
intermediate $\Delta(1232)$ resonance which are known to provide important non-analytic 
terms present in chiral perturbation theory. 

Using Eq.~(\ref{eq:NS}) the results of Refs.~\cite{Dolgov,Goeckeler,Detmold} are translated
in terms of nonsinglet $p - n$ moments, which are reported in Table~\ref{table:NS} together 
with our corresponding results from Table~\ref{table:pn_ltw}. It can be seen that the 
results of Refs.~\cite{Dolgov,Goeckeler}, which adopt a naive linear extrapolation in the 
quark masses up to the chiral point, are larger than our findings both at $n = 2$ and 
$n = 4$. Note that, by combining all the uncertainties, the deviations between our data 
and the lattice results of Refs.~\cite{Dolgov,Goeckeler} are $\approx 3.3 \sigma$ at $n = 2$, 
but only $\approx 1.6 \sigma$ at $n = 4$. The results of Ref.~\cite{Detmold}, which includes 
the effects of meson loops and intermediate $\Delta(1232)$ resonance in the chiral extrapolation, 
are in excellent agreement with our numbers for both $n = 2$ and $n = 4$. It should be 
mentioned that the role of finite volume effects at small quark masses still remains to 
be investigated.

\begin{table*}[!ht]
\begin{center}
\caption{\label{table:NS} \it \small
Comparison of the nonsinglet leading twist moments of the $F_2$ structure function 
given in Table~\ref{table:pn_ltw} with the lattice predictions from 
Refs.~\cite{Dolgov,Goeckeler,Detmold} at the scale $Q^2 = 4$~GeV$^2$.} 
\vspace{5mm}
\begin{tabular}{||c|c||c|c|c||} \hline
$n$ & This work 
    & Ref.~\cite{Dolgov}
    & Ref.~\cite{Goeckeler}
    & Ref.~\cite{Detmold} \\ \hline
 2  & 0.050~~~(5)~(8) & 0.091~~(8) & 0.082~(3) & 0.059~(8) \\ \hline
 4  & 0.0094~(11)~(6) & 0.030~(16) & 0.023~(7) & 0.009~(3) \\ \hline
\end{tabular}
\end{center}
\end{table*}

\section{\label{sec:ht_isospin}Isospin dependence of the Higher Twist contribution}

Once the leading twist is settled and checked, one can study the isospin dependence
of the higher twist (HT) contribution. Previous studies were performed in $x$-space
in Refs.~\cite{HT_NMC,HT_Kulagin}. These analyses showed that in the region
$x \leq 0.7$ HT terms are approximately the same for the proton and the deuteron, while 
at larger $x$ some isospin dependence was seen in Ref.~\cite{HT_Kulagin}, but not in 
Ref.~\cite{HT_NMC} (although the experimental errors in Ref.~\cite{HT_NMC}
were much larger than the difference between the HTs in the proton and in
the deuteron).

In the present paper we consider the total HT contribution in moment space, 
simply defined as the difference between the total and leading twist moments of 
$F_2$, 
\begin{equation}
 {\rm HT}_n^N(Q^2) = M_n^{N (\rm tot)}(Q^2) - M_n^N(Q^2)\ ,
\end{equation}
where the leading twist moment $M_n^N(Q^2)$ is defined as in Eq.~(\ref{eq:momdef_CN}).

Let us briefly recall that our approach is characterized by some relevant
features which are essential for a reliable extraction of the HT
contribution.
In particular:
  (i) the target mass corrections are removed from the total moments through the
      use of the Nachtmann definition;
 (ii) effective HT anomalous dimensions are extracted phenomenologically;
(iii) the effects of Sudakov logarithms are taken into account through 
      Soft Gluon Resummation; and
 (iv) the running coupling constant $\alpha_s(M_Z^2)$ is consistently
      extracted from the proton moments at high $Q^2$~\cite{alphas},
      giving the value
      $\alpha_s(M_Z^2) = 0.1188 \pm 0.0010_{\rm stat} \pm 0.0014_{\rm syst}$,
      which is consistent with the latest world average value
      $\alpha_s(M_Z^2) = 0.1187 \pm 0.0020$~\cite{PDG}.
As a result the total higher twist contribution ${\rm HT}_n^N(Q^2)$ is
extracted reliably and with good precision both for the proton and
deuteron~\cite{osipenko_f2p,osipenko_f2d} for $Q^2 \gtrsim 1$~GeV$^2$. 

\begin{figure}[!ht]
\begin{center}
\includegraphics[bb=0cm 5cm 20cm 25cm, scale=0.365]{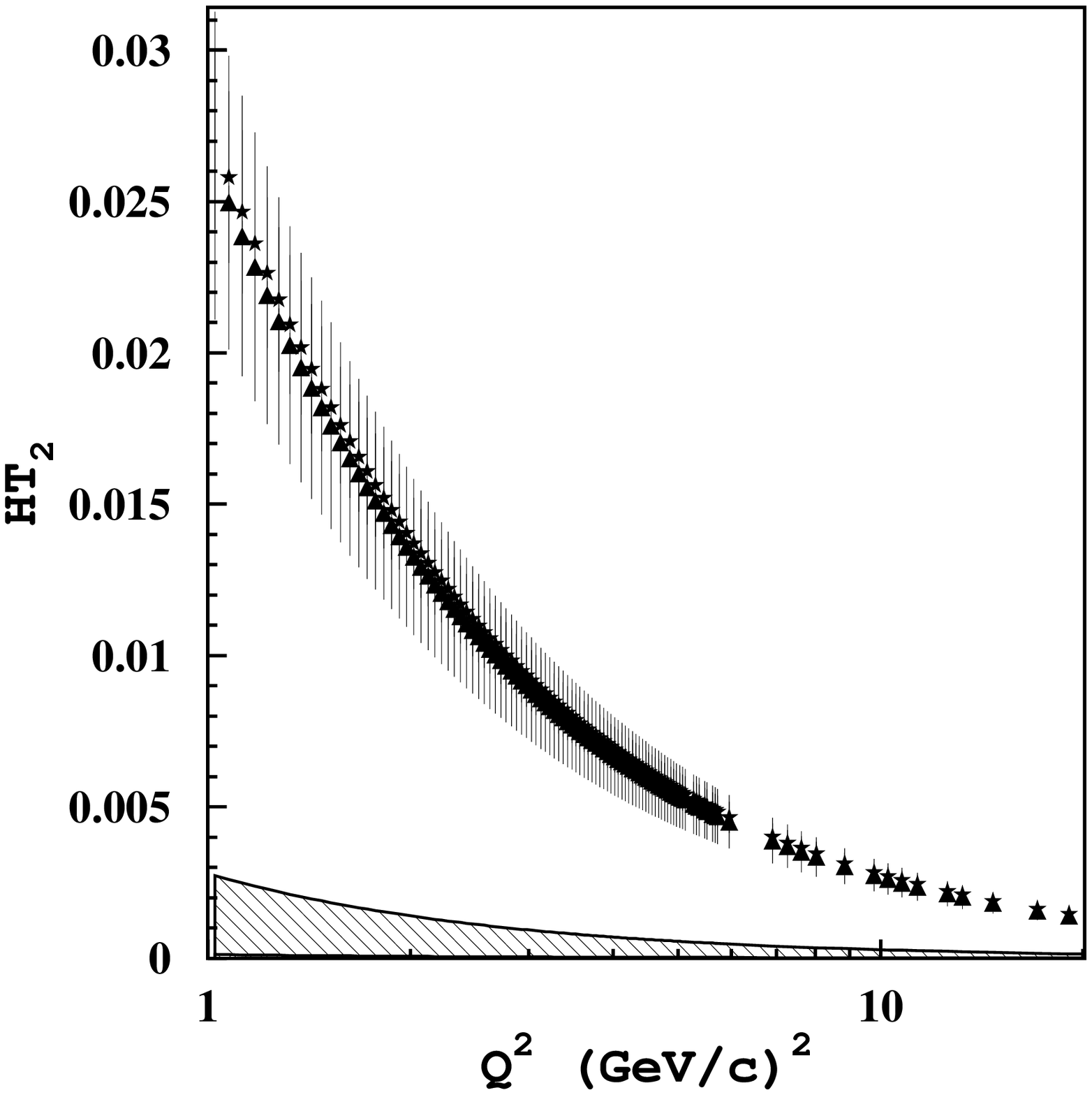}~%
\includegraphics[bb=0cm 5cm 20cm 25cm, scale=0.365]{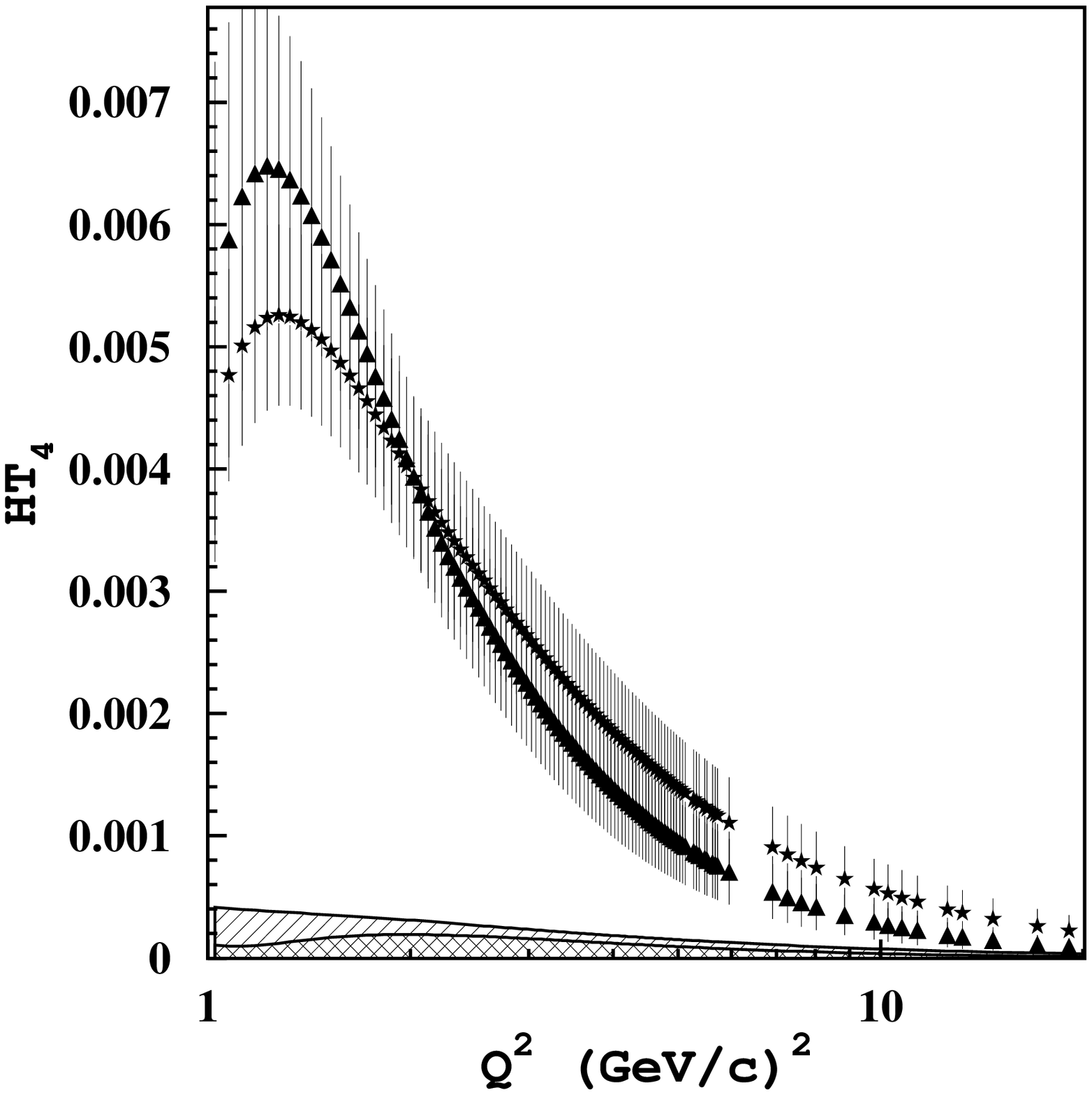}
\includegraphics[bb=0cm 5cm 20cm 25cm, scale=0.365]{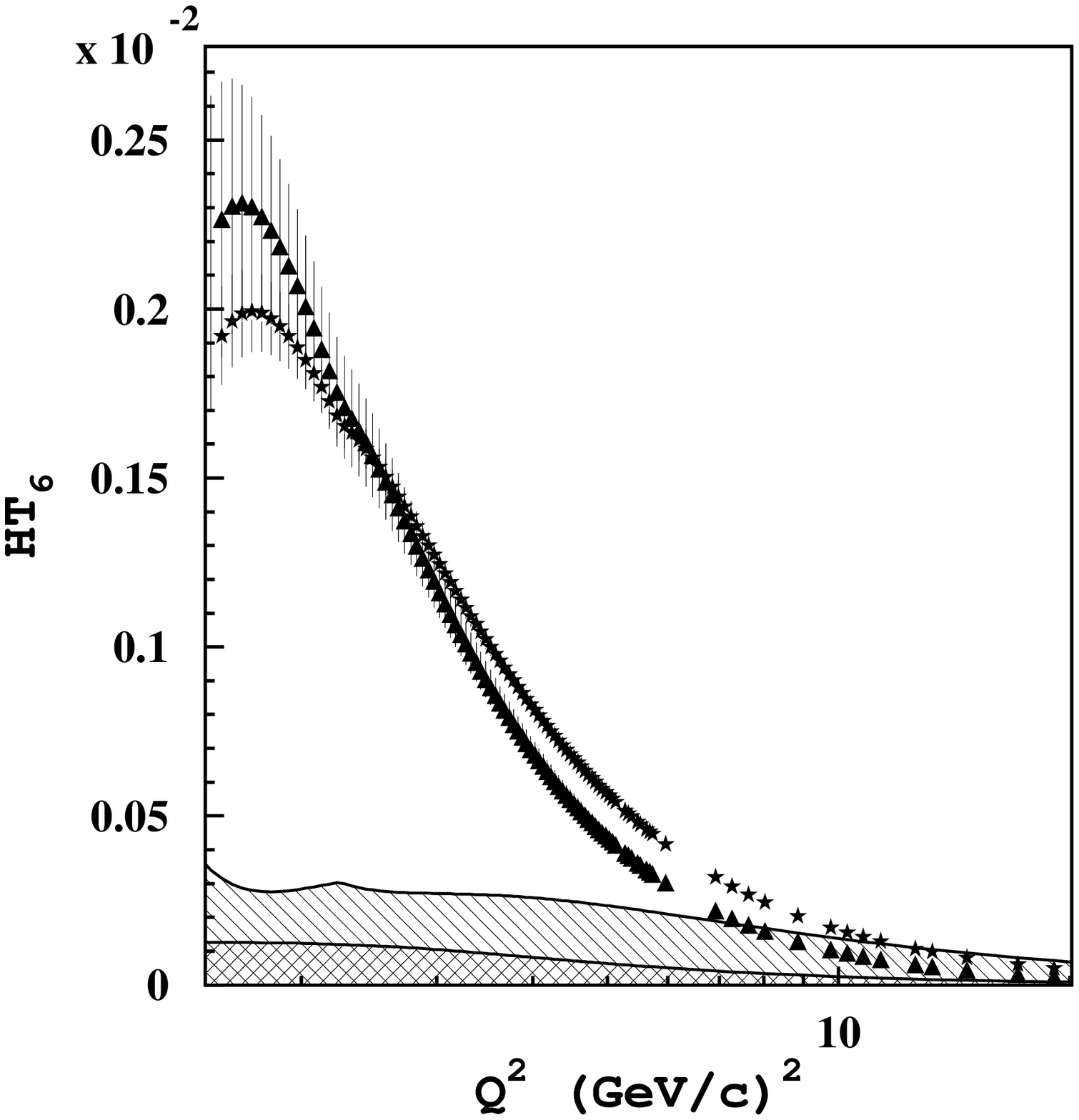}~%
\includegraphics[bb=0cm 5cm 20cm 25cm, scale=0.365]{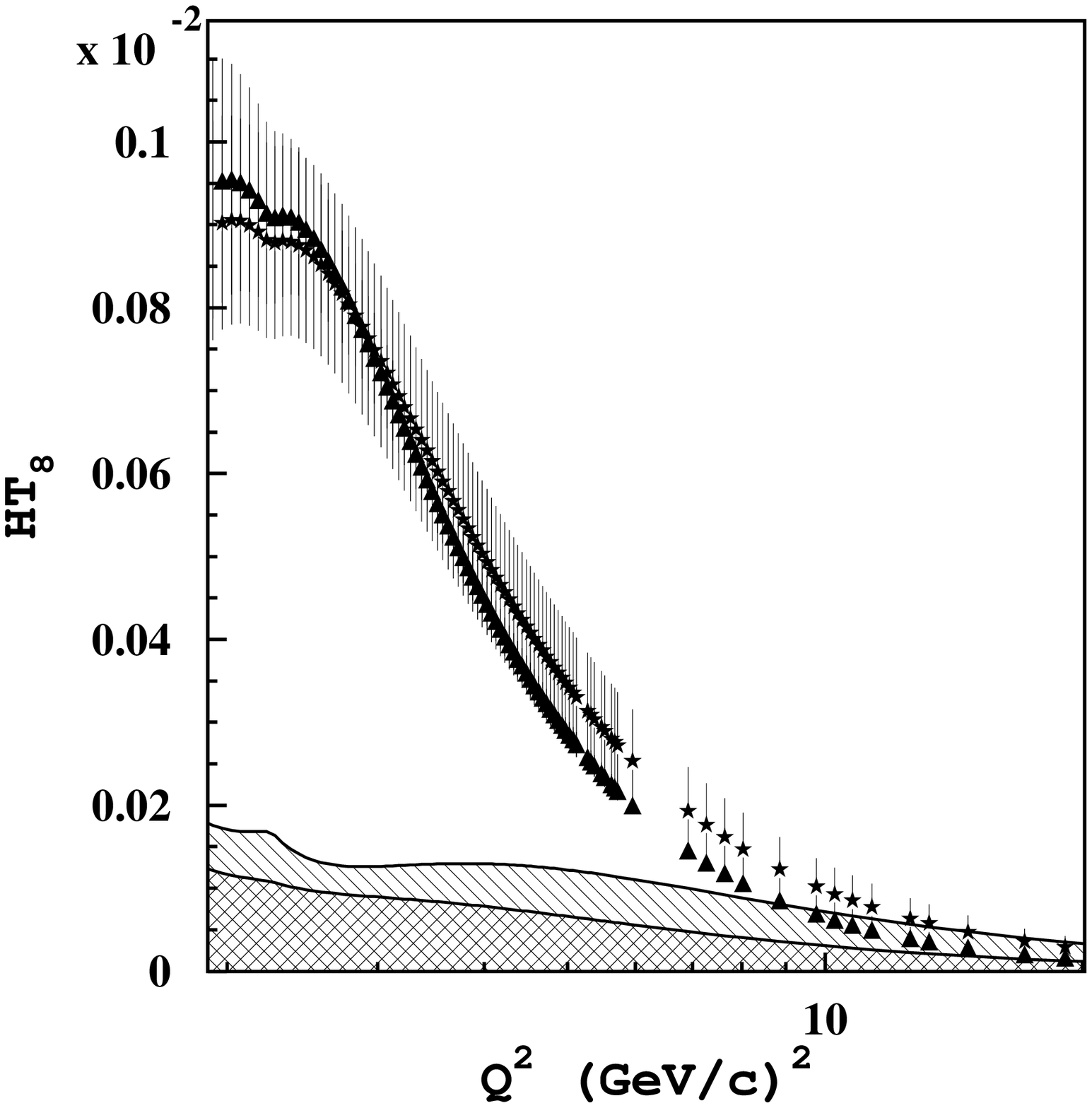}
\caption{\label{fig:ht_np_comp} \it \small 
Total higher twist contribution to the moments with $n = 2, 4 ,6, 8$ together 
with their statistical (bars) and systematic (left- and across-hatched areas) 
uncertainties: triangles show the higher twist contribution in the proton moments 
and the stars represent the same contribution in the deuteron moments corrected 
by the Fermi motion factor ${\cal{F}}_n^D$ given in Eq.~(\ref{eq:fz_mom}). The 
left- and cross-hatched areas refer to the case of proton and deuteron, 
respectively.}
\end{center}
\end{figure}

In Fig.~\ref{fig:ht_np_comp} the total HT term in the proton moments, ${\rm HT}_n^p(Q^2)$, 
is compared with that in the deuteron, ${\rm HT}_n^D(Q^2)$, after applying to the latter 
the same nuclear corrections used for the leading twist, namely ${\cal{F}}_n^D$ of 
Eq.~(\ref{eq:fz_mom}). Within the quoted uncertainties the higher twists appear 
to be the same in the proton and deuteron (per nucleon) moments. 
Assuming that possible additional nuclear effects are small with respect to the moment 
uncertainties (at least for $Q^2 \gtrsim 1$ GeV$^2$), the total HT contribution appears 
to be isospin-independent. This means that for the isovector combination $p - n$ our 
total HT term is consistent with zero within the uncertainties.

Based on the above results we now speculate about two possible scenarios
in which our findings can be interpreted: 

\begin{enumerate}

\item isospin-independent $q-q$ correlations.

\item dominance of $u-d$ correlations with respect to $u-u$ and $d-d$ ones.

\end{enumerate}

\noindent Assuming that HTs are dominated by coupling of the virtual boson with 
correlated $qq$ pairs, the neutron to proton ratio of HTs is expected to be
\begin{equation}
\frac{HT^n}{HT^p}=\frac{2(ud)\frac{1}{9}+(dd)\frac{4}{9}}{2(ud)\frac{1}{9}+(uu)\frac{16}{9}}
\label{eq:ratio_HT}
\end{equation}
\noindent If $q-q$ correlations are isospin-independent, the HT ratio (\ref{eq:ratio_HT}) 
becomes equal to $1 / 3$, a value which is not supported by the present analysis.

The second scenario can be better understood in terms of a diquark model (see \cite{diquark} 
and references therein), which represent a simple, natural way to account for the so-called 
lack of missing nucleon resonances. The correlation between two quarks in a $0^+$ color 
antitriplet state is expected to be of particular importance~\cite{diquark,jw}. Indeed, such a 
configuration is very likely in the nucleon because the correlation between two quarks in a color 
antitriplet state forming a scalar, isoscalar diquark is as strong as the corresponding $\bar{q} q$ 
correlation in the pion, as is known from both QCD sum rules~\cite{stech} and instanton calculations~\cite{shuryak}. 
The most dramatic effect of scalar, isoscalar diquark formation occurs in strangeness changing 
weak decays at low energies, leading to the explanation of the huge $\Delta I = 1/2$ enhancement 
present in these processes \cite{huge}.

In the quark-diquark picture of the nucleon diquarks are expected to be spatially extended objects and 
therefore it is natural to consider that HT terms in lepton-nucleon DIS originate mainly from the 
elastic scattering of the virtual boson off diquarks, representing correlated pairs of quarks. 
Many applications of the quark-diquark model to nucleon observables indicate that scalar, isoscalar 
diquarks are more abundant and have a size smaller than the one of vector, isovector diquarks~\cite{diquark}. 
This means that $u-d$ correlations are expected to dominate over $u-u$ and $d-d$ ones, and therefore the HT 
ratio (\ref{eq:ratio_HT}) becomes equal to $1$, a value which is fully consistent with our findings. A clear-cut 
determination of the spin of the dominant $u - d$ diquark requires the simultaneous investigation of the 
HTs in the longitudinal channel.

For completeness we should mention that the dominance of isoscalar over 
isovector higher-twist contributions in unpolarized structure functions is 
qualitatively predicted by the $1 / N_c$ expansion of QCD, as was recently 
pointed out in the context of a calculation of twist-4 matrix elements in 
the chiral quark-soliton model \cite{Dressler}.

\begin{figure}[!ht]
\begin{center}
\includegraphics[bb=0cm 5cm 20cm 25cm, scale=0.365]{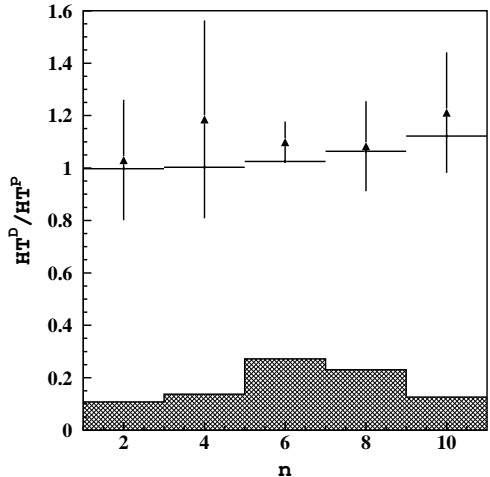}
\caption{\label{fig:ht_np_ratio} \it \small 
Ratio of the deuteron to the proton higher twist contribution at $Q^2 = 3$ GeV$^2$ 
with its statistical and systematic uncertainties represented by bars and hatched area, 
respectively. The solid curve represents the nuclear correction factor ${\cal{F}}_n^D$ 
defined in Eq.~(\ref{eq:fz_mom}).}
\end{center}
\end{figure}

Before closing this Section, we want to point out that, if HTs are isospin-independent,
then the nuclear correction can be obtained directly from the ratio between higher twist 
contributions in the deuteron and proton moments. In Fig.~\ref{fig:ht_np_ratio} such a ratio 
is compared with the nuclear correction factor given by Eq.~(\ref{eq:fz_mom}). The agreement 
is reasonable within the statistical and systematic error bars.

\section{\label{sec:con} Conclusions and future directions}

We have extracted the leading twist contribution to first few moments ($n = 2-10$) 
of the neutron structure function $F_2$ in the $Q^2$ range from 1 up to 
100 GeV$^2$, by combining proton and deuteron measurements over a huge range of 
kinematics. We have paid particular attention to the issue of nuclear effects in the
deuteron, which are found to be increasingly important for the higher 
moments. The model dependence arising from our incomplete knowledge of the nuclear
corrections, including the high-momentum tail of the deuteron wave function, 
as well as nucleon off-shell effects, were estimated and included in the systematic 
errors. From our analysis the following conclusions can be made: 

\begin{itemize}
\item
the extracted moments of the neutron structure function $F_2$ are consistent 
at large $Q^2$ with the predictions of several PDFs available in the literature;
\item
the ratio of neutron to proton structure functions is determined from the ratio of 
the corresponding moments up to $x \simeq 0.70$; we find $F_2^n / F_2^p(x = 0.70) = 
0.34 \pm 0.12_{\rm stat} \pm 0.13_{\rm syst}$. Our results are consistent with the 
asymptotic limit $1 / 4$ at $x = 1$, which originates from the dominance of soft, 
non-perturbative physics at large $x$. Nevertheless the alternative limiting value 
of $3 / 7$, derived from helicity conservation arguments, is not excluded at all, 
but the possible enhancement of the $d$-quark distribution should be confined to 
the region $x \gtrsim 0.7$;
\item
the nonsinglet, isovector $p -n$ combination of the leading twist moments is constructed 
and compared with available lattice results. We have found that our data are larger than 
the lattice values of Refs.~\cite{Dolgov,Goeckeler}, extrapolated linearly in the quark 
masses down to the physical point, by $\approx 3.3 \sigma$ for the second moment, but only 
by $\approx 1.6 \sigma$ in case of the fourth moment. Our results are in excellent agreement 
with those of Ref.~\cite{Detmold}, where the chiral extrapolation is performed taking into 
account the effects of non-analytic terms due to meson loops and intermediate $\Delta(1232)$ 
resonance;
\item
the total contribution of higher twists is found to be isospin independent, which 
implies that in the isovector combination $p - n$ the higher twists are consistent 
with zero within the uncertainties. The isospin independence is expected in a 
quark-diquark picture of the nucleon assuming the dominance of isoscalar $u-d$ 
diquarks.
\end{itemize}

The determination of the large-$x$ behavior of the $d$-quark distribution in the 
nucleon is still one of the main unsolved problems in partonic physics. The use of
deuteron inclusive DIS data to extract the neutron to proton structure 
function ratio produces unavoidably a model dependence. The strategy used 
in this paper has its own practical limitations which are due to the 
increasing uncertainties of the nuclear corrections as the order of 
the moments considered increases. It appears that moments of order $n = 12$ 
may be still accessible by improving measurements of the structure function 
$F_2$ at large $x$. As we have shown, at $n = 12$ one can probe values of 
the Bjorken-$x$ as large as $0.75$, which represents therefore an upper limit. 

Future directions for the experimental determination of the $d$-quark 
distribution at high $x$ have been recently discussed in the 
literature~\cite{Strikman,SIM96,Wally,SSS}. One of the strategies, which is 
currently under experimental investigation at Jefferson Lab~\cite{Bonus}, 
is the use of the tagged semi-inclusive process off the 
deuteron~\cite{Strikman,SIM96}, which requires the detection of a low 
momentum proton ($p \lesssim 150 MeV/c$). An alternative method which has 
been discussed, tries to exploit the mirror symmetry of $A = 3$ nuclei 
to reduce nuclear effects in the extraction of the $F_2^n / F_2^p$ 
ratio~\cite{Wally,SSS}.

In the semi-inclusive deuteron method one can {\em tag} the momentum of the 
struck neutron by detecting the slow recoiling proton; in this way it is 
possible to select initial deuteron configurations in which the two 
nucleons are far apart, so that the struck nucleon can be considered as 
free and the uncertainties due to the low-momentum part of the deuteron wave 
function are minimal~\cite{Strikman,SIM96}. The neutron structure function 
can then be extracted directly from the semi-inclusive deuteron cross section.

We want to point out that an important improvement of the {\em tagged} strategy
may be achieved by adding the detection of slow recoiling neutrons. 
In this way the $n / p$ structure function ratio can be directly obtained from 
the ratio of the corresponding semi-inclusive cross sections. Most of the 
nuclear corrections, e.g.~Fermi motion, off-shellness and FSIs, are expected 
to largely cancel out in the ratio~\cite{Strikman,SIM96}. Thus, from the 
measurements of both semi-inclusive processes $D(e, e'p)X$ and $D(e, e'n)X$ 
the shape of the $d$-quark distribution may be experimentally investigated 
up to and hopefully beyond $x \approx 0.75$.

\section*{Acknowledgements}
The work of W.M.~was supported by the U.S. Department of Energy contract 
DE-AC05-84ER40150, under which the Southeastern Universities 
Research Association (SURA) operates the Thomas Jefferson National 
Accelerator Facility (Jefferson Lab). The work of S.K.~was partially 
supported by the Russian Foundation for Basic Research grant 03-02-17177 
and INTAS project 03-51-4007.


\end{document}